\documentclass[aoas,preprint]{imsart}

\RequirePackage{amsthm,amsmath,amsfonts,amssymb,mathtools}
\RequirePackage[authoryear]{natbib}
\RequirePackage[colorlinks,citecolor=blue,urlcolor=blue,backref=page]{hyperref}
\usepackage[ruled,vlined,linesnumbered]{algorithm2e}
\usepackage{bbm}
\newcommand\given[1][]{\:#1\vert\:}
\newcommand{\dd}{\mathop{}\!\mathrm{d}}
\RequirePackage{graphicx}
\usepackage{multirow}
\usepackage{float}
\usepackage{forest}
\usepackage{threeparttable}
\usepackage{caption}
\usepackage{subcaption}
\usepackage{etoolbox}
\usepackage{anyfontsize}
\usepackage{comment}
\AtBeginEnvironment{algorithm}{\linespread{0.8}\selectfont}
\pubyear{2024}
\volume{TBA}
\issue{TBA}
\firstpage{1}
\lastpage{1}

\startlocaldefs
\newcommand*\samethanks[1][\value{footnote}]{\footnotemark[#1]}
\theoremstyle{plain}

\theoremstyle{definition}

\theoremstyle{remark}

\endlocaldefs

\begin{document}

\begin{frontmatter}
\title{Seemingly unrelated Bayesian additive regression trees for cost-effectiveness analyses in healthcare}
\runtitle{Seemingly unrelated BART}

\begin{aug}
\author[A]{\fnms{\textbf{Jonas}}~\snm{\textbf{Esser}}\thanks{\textbf{M. Maia} and \textbf{J. Esser} contributed equally to this article and share first-authorship.}{\footnotesize\textsuperscript{\!,\!}}\ead[label=e1]{j.esser@vu.nl}\orcid{0009-0005-4964-5525}},
\author[B]{\fnms{\textbf{Mateus}}~\snm{\textbf{Maia}}\samethanks{\footnotesize\textsuperscript{\!,\!}}\ead[label=e2]{mateus.maiamarques@glasgow.ac.uk}\orcid{0000-0001-7056-386X}},
\author[C]{\fnms{Andrew C.}~\snm{Parnell}\ead[label=e3]{andrew.parnell1@ucd.ie}\orcid{0000-0001-7956-7939}},
\author[A]{\fnms{\linebreak{}Judith E.}~\snm{Bosmans}\ead[label=e4]{j.e.bosmans@vu.nl}\orcid{0000-0002-1443-1026}},
\author[A]{\fnms{Johanna Maria}~\snm{van Dongen}\ead[label=e5]{j.m.van.dongen@vu.nl}\orcid{0000-0002-1606-8742}},
\author[D]{\fnms{Thomas}~\snm{Klausch}\ead[label=e6]{t.klausch@amsterdamumc.nl}},
\and
\author[E]{\fnms{Keefe}~\snm{Murphy}\ead[label=e7]{keefe.murphy@mu.ie}\orcid{0000-0002-7709-3159}}

\address[A]{Faculty of Science, Health Economics and Health Technology Assessment, Vrije Universiteit Amsterdam,\linebreak{}\printead[presep={}]{e1,e4,e5}}

\address[B]{School of Mathematics and Statistics, University of Glasgow,\printead[presep={}]{e2}}

\address[C]{School of Mathematics and Statistics, Insight Centre for Data Analytics, University College Dublin,\printead[presep={}]{e3}}

\address[D]{Department of Epidemiology and Data Science, Amsterdam University Medical Center,\printead[presep={}]{e6}}

\address[E]{Hamilton Institute and Department of Mathematics and Statistics, Maynooth University,\printead[presep={}]{e7}}

\runauthor{Esser, Maia, et al.}

\end{aug}

\begin{abstract}
In recent years, theoretical results and simulation evidence have shown Bayesian additive regression trees to be a highly-effective method for nonparametric regression. Motivated by cost-effectiveness analyses in health economics, where interest lies in jointly modelling the costs of healthcare treatments and the associated health-related quality of life experienced by a patient, we propose a multivariate extension of BART which is applicable in regression analyses with several dependent outcome variables.
Our framework allows for continuous or binary outcomes and overcomes some key limitations of existing multivariate BART models by allowing each individual response to be associated with different ensembles of trees, while still handling dependencies between the outcomes. In the case of continuous outcomes, our model is essentially a nonparametric version of seemingly unrelated regression. Likewise, our proposal for binary outcomes is a nonparametric generalisation of the multivariate probit model. We give suggestions for easily interpretable prior distributions, which allow specification of both informative and uninformative priors.
We provide detailed discussions of MCMC sampling methods to conduct posterior inference. Our methods are implemented in the \textsf{R} package \texttt{subart}. We showcase their performance through extensive simulation experiments and an application to an empirical case study from health economics. By also accommodating propensity scores in a manner befitting a causal analysis, we find substantial evidence for a novel trauma care intervention's cost-effectiveness.
\end{abstract}


\end{frontmatter}

\section{Introduction}
Many research questions in health economics are concerned with trading off the costs and benefits of a medical intervention. The most prominent examples are in \emph{cost-effectiveness analysis} (CEA), where we wish to decide whether a new innovative treatment is worth the associated increase in costs compared to usual care. We therefore need to estimate the average treatment effects on both costs and health. However, in order to get coherent measures of uncertainty, the two treatment effects must be estimated jointly  in order to account for the correlation between them \citep{baio2012bayesian}. Furthermore, if the CEA is performed with observational data, where the treatment assignment is not randomised, we additionally have to adjust for confounding bias in the analysis. These points are elaborated in Section \ref{sec:application}.

In this paper, we aim to estimate the cost-effectiveness of a novel treatment for physical trauma rehabilitation, called the \emph{transmural trauma care model} (TTCM), using data gathered under a study by \citet{wiertsema2019cost} which we will henceforth refer to as the \emph{TTCM data}. The treatment assignment is not randomised and the number of potential confounders is large relative to the sample size. Of the multiple cost and effectiveness outcomes the authors investigated, we focus on healthcare-related costs and health-related quality of life. It is of interest to jointly estimate both outcomes, and reasonable to assume both that the outcomes are non-linearly related to the available predictors and that each outcome may depend on different subsets of predictors, which may in turn interact in complex ways. The challenge of CEA under these circumstances motivated the development of our novel methodology, though we also anticipate its use in other CEA studies and broader healthcare settings.

In the causal inference literature, there is wide agreement that flexible nonparametric methods, which do not impose strong parametric assumptions on the regression functions, are the best tools for estimating treatment effects with observational data \citep{dorie2019automated,rudolph2023all}. It is not straightforward, however, to apply this knowledge in the context of CEAs. Seemingly unrelated regression  models \citep[SUR;][]{zellner1962efficient}, the most recommended statistical method for CEAs \citep{willan2004regression,el2022scoping}, impose strong linearity assumptions, which can bias the inferences if there are strong non-linear relationships between the variables of interest. On the other hand, there is a distinct lack of nonparametric regression methods which can handle multivariate outcomes. We adopt a Bayesian perspective and fill this gap by developing a nonparametric version of SUR. The idea is to replace the linear predictors in the SUR model by sums of regression trees. In the univariate case, this regression method has become known as \emph{Bayesian additive regression trees} (BART). BART has already demonstrated competitive performance for univariate responses \citep{dorie2019automated,rudolph2023all} and it seems plausible that this efficacy will extend to situations with multiple outcomes of interest. However, by embedding BART in the SUR framework, we also seek to overcome some limitations of existing multivariate BART extensions.

\citet{chipman2010bart} introduced BART as an ensemble method, where each learner is a tree following the Bayesian CART approach previously proposed by the same authors \citep{chipman1998bayesian}. Considering a univariate response vector $\mathbf{y} \in \mathbb{R}^{n}$ and a set of predictors $\mathbf{X} \in \mathbb{R}^{n\times p}$, which may be of mixed type, the main objective of regression modelling is to estimate the conditional expectation $\mathbb{E}\lbrack y_i\given \mathbf{x}_i\rbrack$, also called the regression function. As a nonparametric model, BART offers high flexibility in estimating this conditional expectation. The tendency of decision trees to overfit is mitigated by the Bayesian underpinnings of the method, allowing informative prior distributions to impose regularisation in a principled and transparent fashion, as well as the additive nature of the ensemble. These features facilitate better generalisation, in a similar vein to the gradient boosting approach of \citet{friedman2001greedy}. The success of BART is widely reported in the literature across a broad spectrum of applications \citep{janizadeh2021novel,sarti2023bayesian,yee2023evaluating}. In addition, theoretical work has demonstrated the frequentist optimality of BART under certain conditions \citep{linero2018bayesian,rockova2020semi,rockova2020posterior}.

The BART model was originally developed for univariate responses but subsequent formulations emerged to adapt BART to multivariate responses. Examples include the Bayesian additive vector autoregressive tree \citep{florian2022vabart} for multivariate time-series and formulations by \citet{peruzzi2022spatial} for multivariate spatial data. Other applied work involves an extension of BART to forecast the tails of multivariate responses \citep{clark2023tail}. \citet{um2023bayesian} adapted BART to cover not only multivariate responses but also the assumption of a skew-normal distribution; throughout this paper, we refer to the variant without skewness as mvBART. Furthermore, \citet{mcjames2024bayesian} proposed an extension of Bayesian causal forests \citep[BCF;][]{hahn2020bayesian} for multivariate responses. All  aforementioned approaches share the limitation that the tree structure must be identical for each component of the outcome vector. It is easy to envisage situations for which this is inappropriate: a specific covariate may be strongly associated with one outcome but independent of another; e.g., factors which govern costs may be unrelated to quality of life, and \emph{vice versa}.

In this study, we propose a novel variant of BART termed seemingly unrelated BART (suBART) which is designed to handle multivariate continuous responses and address this key limitation. \citet{chipman2010bart} previously drew parallels to SUR models \citep{zellner1962efficient} and alluded to the potential for extending BART in this direction. Our framework differs from the aforementioned multivariate BART extensions, which assume a single set of trees with correlated multivariate terminal node parameters. Instead, we jointly fit individual ensembles of trees for each outcome and model their interdependence through correlated error terms. Thus, suBART also differs from merely applying univariate BART models to each outcome independently. Motivated by our investigation into the cost-effectiveness of the TTCM intervention, we further extend suBART to incorporate propensity scores, in the spirit of \citet{hahn2020bayesian}, as befits causal analyses. Beyond CEA settings, we also develop probit suBART, an extension of suBART to accommodate multivariate binary outcomes. While not the main focus of this paper, we envision this version of the model being useful in marketing applications with correlated binary outcomes; see the examples in \citet{rossi2005bayesian}. Our approach is similar to that of \citet{chakraborty2016bayesian}, who offers a version of seemingly unrelated BART for exclusively continuous outcomes with an adaptive number of trees. However, it is not specifically tailored to causal inference objectives typical of CEA and lacks an available open-source software implementation. Moreover, the non-informative inverse-Wishart prior adopted by \citet{chakraborty2016bayesian} on the covariance matrix of the error terms fails to properly calibrate the prior on the residual variance parameters, which is crucial for regularising the contributions of the trees in the ensemble. We address these gaps by providing a comprehensive framework, which covers either continuous or binary outcomes, with appropriately specified priors, and a practical implementation through an \textsf{R} package named \texttt{subart}\footnote{Implemented using \texttt{Rcpp} and available at:     \url{https://github.com/MateusMaiaDS/subart}.}.

This article proceeds as follows: Section \ref{sec:application} provides background theory on CEA and motivates the development of the suBART model in the context of the application to the TTCM data from \citet{wiertsema2019cost}. Section \ref{sec:suBART} then elaborates on the theoretical underpinnings of the suBART methodology and Section \ref{sec:posterior} discusses the posterior inference for both the multivariate continuous and multivariate binary outcome settings. Section \ref{sec:simstudies} considers different simulation scenarios to compare the predictive performance of suBART against standard competitors. The empirical findings of our application of suBART to the TTCM data are presented in Section \ref{sec:case_study}, along with an additional simulation experiment based on these data which assesses suBART's estimation of treatment effects in the CEA setting. Finally, Section \ref{sec:conclusion} summarises the proposed methodologies, highlighting both their limitations and potential for further extension, and presents conclusions regarding the health economic application. Additional results, comparisons, and findings are included in the appendices.

\section{CEA and the suBART model}\label{sec:application}

Our main motivation in developing suBART is its potential applicability in cost-effectiveness analyses of healthcare treatments, particularly in cases where the treatment assignment is not randomised. When estimating treatment effects from such observational data, simple parametric models often lead to severe bias and flexible nonparametric models are preferable \citep{hernan2024causal}. 
Naturally, such concerns extend to CEA settings, where we want to estimate two treatment effects simultaneously. However, there is a lack of statistical methods fit for such purposes. Given that BART has proven to be very useful for univariate causal inference tasks \citep{hill2011bayesian,dorie2019automated,rudolph2023all}, we consider it a promising basis for multivariate CEA.

We continue by sketching some relevant ideas from health economics and causal inference in Section \ref{sec:CEA_theory}, drawing on the detailed treatments found in \citet{gabrio2019bayesian} and \citet{li2023bayesian}. Subsequently, we give a detailed presentation of the TTCM data in Section \ref{sec:TTCM_intro}.

\subsection{Cost-effectiveness analyses with observational data}\label{sec:CEA_theory}

The fundamental problem of cost-effectiveness analyses in health economics is to determine which of two competing healthcare treatments --- usually, but not always, for the same disease --- should be implemented. Often one is both more effective and more expensive than the other, which raises the question whether the increase in health is worth the added expenses. Henceforth, we let $c_i$ and $q_i$ respectively denote the healthcare costs and the health-related quality of life associated with a patient. We also suppose that there are two different treatments of interest, $t = 0$ and $t = 1$. Using the usual potential outcomes notation, we let $c_i(t)$ denote the costs associated with patient $i$, had they received treatment $t$. We do likewise for $q_i(t)$. We furthermore suppose that there is some vector of baseline characteristics $\mathbf{x}_i$ which have an effect on both the outcomes $c_i$ and $q_i$, as well as the treatment indicator $t_i$.

Given a sample of $n$ observations, we wish to estimate the mixed average treatment effect (MATE)\footnote{The MATE is closely related to the population average treatment effect (PATE), although the terminology for treatment effects is not consistent across the literature. We elect to use the terminology of \citet{li2023bayesian}, according to whom the PATE requires a generative model for the covariates $\mathbf{x}_i$ and thus necessitates additional assumptions. We work with the MATE in Equation \eqref{eq:li_MATE} to simplify the analyses. Note that this common approach actually corresponds to what is called the PATE by \citet{imbens2015pate} and other authors.} on the costs in this sample, which we define as 
\begin{equation}\Delta_c \coloneqq \frac{1}{n} \sum^n_{i = 1} \mathbb{E}\left\lbrack c_i(1)\given \mathbf{x}_i\right\rbrack -  \mathbb{E}\left\lbrack c_i(0)\given \mathbf{x}_i\right\rbrack.\label{eq:li_MATE}
\end{equation}
We now make the assumption of \emph{ignorability}, which means that conditional on the baseline covariates $\mathbf{x}_i$, the treatment $t_i$ is independent of the potential outcomes $c_i(0)$ and $c_i(1)$. Under this assumption, we may rewrite our treatment effect as 
\[\Delta_c = \frac{1}{n} \sum^n_{i = 1} \mathbb{E}\left\lbrack c_i\given t_i = 1, \mathbf{x}_i\right\rbrack -  \mathbb{E}\left\lbrack c_i\given t_i = 0, \mathbf{x}_i\right\rbrack.\]
Thus, $\Delta_c$ is completely specified by the conditional expectations $\mathbb{E}\lbrack c_i\given t_i, \mathbf{x}_i\rbrack$. 
We proceed in the same manner for $\Delta_q$, the MATE on the patient's quality of life. It is then customary to combine the two treatment effects into a utility function, the incremental net benefit (INB):
\[\text{INB}_{\lambda} \coloneqq \lambda \Delta_q - \Delta_c.\]
The scalar parameter $\lambda$ is called the \emph{willingness-to-pay}. Roughly speaking, $\lambda$ quantifies how much cost (in the given currency) a decision-maker is willing to trade for a one-unit increase in healthcare-related quality of life for one patient. The decision rule is then simple: if the INB is at most zero, we say that treatment $1$ is not cost-effective, and treatment $0$ should be implemented. If the INB is larger than zero, we consider treatment $1$ to be cost-effective and worthy of being implemented. Throughout this paper, we give $\lambda$ values in units of €1,000.

To illustrate the importance of modelling the two outcomes $c$ and $q$ jointly, let us assume for simplicity that the joint distribution of $\Delta_c$ and $\Delta_q$ is bivariate normal. Then
\begin{align}\Pr\left(\text{INB}_{\lambda} > 0\right) &= \Phi \left(\frac{\mathbb{E} \left\lbrack \text{INB}_{\lambda} \right\rbrack}{\sqrt{\text{Var}\left\lbrack \text{INB}_{\lambda} \right\rbrack}} \right)\nonumber \\
&= \Phi \left(\frac{\lambda \mathbb{E} \left\lbrack \Delta_q \right\rbrack - \mathbb{E} \left\lbrack \Delta_c \right\rbrack}{\left(\lambda^2 \text{Var}\left\lbrack \Delta_q \right\rbrack + \text{Var}\left\lbrack \Delta_c \right\rbrack - 2 \lambda \text{Cov}\left\lbrack \Delta_q, \Delta_c \right\rbrack \right)^{1/2}} \right).\label{eq:Pr_Inb}
\end{align}
Consequently, the probability of cost-effectiveness depends on the covariance of $\Delta_c$ and $\Delta_q$. Without the normality assumption, this probability can usually not be found explicitly, but the same principle applies nonetheless: the probability of cost-effectiveness depends on the joint distribution of $\Delta_c$ and $\Delta_q$ \citep{lothgren2000definition, gabrio2019bayesian}. It follows that we must model $c_i$ and $q_i$ jointly, as modelling them separately would enforce the unrealistic assumption that the treatment effects $\Delta_c$ and $\Delta_q$ are independent.  This belief is seldom appropriate, since empirical cost and health data are often strongly correlated \citep{willan2004regression}. We expound on these points through simulation experiments in Appendix \ref{app:indBART}.

We hence use the suBART model developed below to jointly estimate the conditional expectations $\mathbb{E}\lbrack c\given t, \mathbf{x}_i\rbrack$ and $\mathbb{E}\lbrack q\given t, \mathbf{x}_i\rbrack$. The treatment effects and INB can then be obtained as functions of these estimates. Our approach mirrors that of \citet{hahn2020bayesian}: we first estimate propensity scores (using probit BART), and then condition the suBART model on all covariates, the treatment indicator, and the estimated propensity scores.

\subsection{Transmural Trauma Care Model (TTCM)}\label{sec:TTCM_intro}

Our data of interest, pertaining to $n=140$ patients suffering from traumatic injuries, were originally collected and  analysed by \citet{wiertsema2019cost}. The two treatment options are usual care and the novel Transmural Trauma Care Model (TTCM), denoted by $t=0$ and $t=1$, respectively. The treatment assignment was not randomised. The outcomes we use, for $c_i$ and $q_i$ respectively, are the costs from the healthcare perspective and generic healthcare-related quality of life. As is usual in CEAs, the former is an aggregate measure: $c_i$ comprises the total costs acquired from hospital records as well as several questionnaires conducted over the course of nine months following treatment, in which patients were surveyed on their use of various healthcare resources. The responses --- which relate to issues such as hospital stays, medication use, and surgeries --- were then converted to costs. Conversely, the effectiveness outcome $q_i$ was calculated from one single survey administered nine months after treatment using the EQ-5D-3L instrument \citep{lamers2006dutch}. 
The data also include $p=11$ baseline covariates, with respective sample sizes of $83$ and $57$ in the two treatment groups. The ratio of covariates to observations is thus reasonably large. We reproduce the table of baseline variables in Table \ref{tab:baseline_data}. 

The original dataset had some missing observations --- for survey items related to the outcome variables only --- which \citet{wiertsema2019cost} dealt with through multiple imputation. $17\%$ of patients did not complete any follow-up questionnaires, and hence were missing all information on $q_i$ and some survey information on $c_i$ (though hospital records were available for all patients). Additionally, $39\%$ and $7\%$ of respondents were missing some (but not all) survey items related to $c$ and $q$, respectively. As missing data is not the subject of this paper, we avoid this complication by simply using \emph{one} imputed dataset, obtained through predictive mean matching \citep{vink2014predictive}, and treating that as complete data. Specifically, the imputation is applied to the missing survey items prior to the calculation of $c$ and $q$. The imputation was done using the \texttt{mice} package in \textsf{R} \citep{van2011mice}, with default settings for everything other than setting the imputation method to predictive mean matching.
It follows that our analyses of these data are not directly comparable to the original one in \citep{wiertsema2019cost} and we do not claim that ours are more valid in this regard. We discuss this issue further in Section \ref{sec:conclusion}.

The statistical analysis in \citet{wiertsema2019cost} is based on simple linear models\footnote{More specifically: the SUR model, as presented in Section \ref{sec:suBART_continuous}.}, without any interaction terms or other transformations of the covariates. This is imposes rather strong assumptions on the regression functions, which we find difficult to justify --- and as alluded to in the introduction, violations of these assumptions can lead to severely biased results. We hence found it necessary to develop the more flexible suBART model, which allows us to jointly estimate both treatment effects without strong functional assumptions.
\begin{table}[H]
\caption{Baseline data from \citet{wiertsema2019cost}.}
\label{tab:baseline_data}
\centering
\begin{threeparttable}
\begin{tabular}{lll}
\textbf{Characteristics}    & \multicolumn{2}{c}{\textbf{Mean (SD) \emph{or} frequency (\%)}}\\
\cline{2-3}
&Intervention group ($t=1$)  & Control group ($t=0$)\\
\hline
$n$                & $83$                                & $57$                                      \\\\[-2.5ex]
Age                & $43.4~(15.6)$                       & $50.5~(17.9)$                             \\\\[-2.5ex]
Gender (M/F)       & $39/44~(47/53\%)$                   & $26/31~(46/54\%)$                         \\\\[-2.5ex]
\emph{Education level}    &                              &                                           \\
Low                & $7~(8.4\%)$                         & $6~(11.1\%)$                              \\
Middle             & $19~(22.9\%)$                       & $16~(29.6\%)$                             \\
High               & $57~(68.7\%)$                       & $32~(59.3\%)$                             \\\\[-2.5ex]
\emph{Medical history}    &                              &                                           \\
None               & $53~(63.9\%)$                       & $30~(52.6\%)$                             \\
Chronic            & $14~(16.9\%)$                       & $13~(22.8\%)$                             \\
Musculoskeletal    & $16~(19.3\%)$                       & $14~(24.6\%)$                             \\\\[-2.5ex]
\emph{Trauma type}        &                              &                                           \\
Traffic            & $44~(53.0\%)$                       & $25~(43.9\%)$                             \\
Work related       & $0~(0.0\%)$                         & $2~(3.5\%)$                               \\
Fall               & $27~(32.5\%)$                       & $17~(29.8\%)$                             \\
Sports             & $11~(13.3\%)$                       & $9~(15.8\%)$                              \\
Other              & $1~(1.2\%)$                         & $4~(7.0\%)$                               \\\\[-2.5ex]
\emph{Fracture region}    &                              &                                           \\
Upper extremity    & $31~(37.3\%)$                       & $25~(43.9\%)$                             \\
Lower extremity    & $41~(49.4\%)$                       & $19~(33.0\%)$                             \\
Vertebral          & $7~(8.4\%)$                         & $1~(1.8\%)$                               \\
Multitrauma        & $4~(4.8\%)$                         & $12~(21.1\%)$                             \\\\[-2.5ex]
Injury severity score                & $7.9~(4.4)$                         & $8.6~(6.3)$                               \\\\[-2.5ex]
Hospital admission & $62~(75\%)$                         & $29~(51\%)$                               \\\\[-2.5ex]
Length of hospital stay (days)     & $7.1~(6.1)$                         & $10.0~(11.4)$                             \\\\[-2.5ex]
Surgery            & $53~(64\%)$                         & $21~(37\%)$                               \\\\[-2.5ex]
TTO\tnote{b}        & $24.3~(14.3)$                       & $14.6~(14.7)$\\[0.75ex]
\hline
\end{tabular}
\begin{tablenotes}
    \item[b] Days between trauma and first outpatient consultation.
\end{tablenotes}
\end{threeparttable}
\end{table}

\section{The suBART models for continuous and binary responses}\label{sec:suBART}
We start by reviewing the original BART model in the univariate setting in Section \ref{sec:review_BART}, in order to provide context for what is to follow. We then present our novel extensions to the multivariate continuous outcome setting in Section \ref{sec:suBART_continuous} and the multivariate binary outcome setting in Section \ref{sec:suBART_probit}. Specific details regarding posterior inference for suBART models are deferred to Section \ref{sec:posterior}.

\subsection{A review of univariate BART}\label{sec:review_BART} 
\begin{figure}[H]
\centering
\begin{subfigure}[t]{0.4\textwidth}
\centering
        \begin{forest}
    for tree={
        grow=south, draw, minimum size=3ex, 
        inner sep=3pt, 
        s sep=7mm,
        l sep=6mm
    }
    [$x \le 0.3$,
        [$0.2$, edge label={node[midway, left, font=\footnotesize]{TRUE\:}}, circle,]
        [$x \le 0.6$, edge label={node[midway, right, font=\footnotesize]{\:FALSE}},
            [$x \le 0.5$, edge label={node[midway, left, font=\footnotesize]{}},
                [0.9, edge label={node[midway, left, font=\footnotesize]{}}, circle,]
                [0.4, edge label={node[midway, right, font=\footnotesize]{}}, circle,]
            ]
            [0.6, edge label={node[midway, right, font=\footnotesize]{}}, circle,]
        ]
    ]
\end{forest}
\end{subfigure}
\hspace*{\fill}
\begin{subfigure}[t]{0.5\textwidth}
\centering
     \includegraphics[width=0.8\columnwidth]{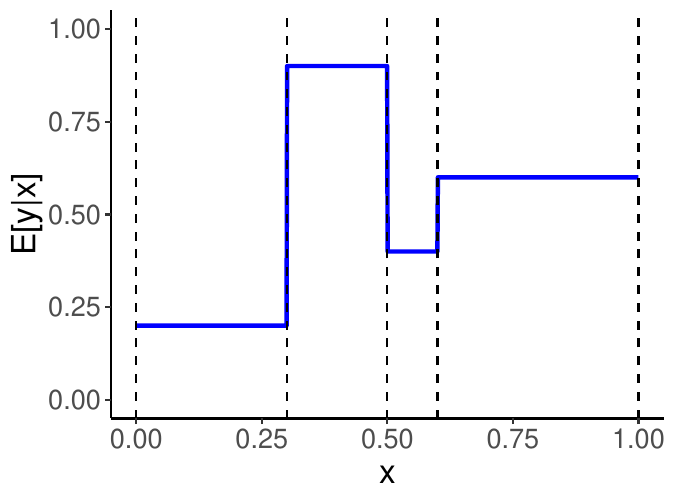}
\end{subfigure}
\caption{Regression tree (left) and implied regression function (right).}
\label{fig:tree_example}
\end{figure}

BART was designed to solve the classic regression problem of the form
$y_i = \mathbb{E}\left\lbrack y_i\given \mathbf{x}_i\right\rbrack + \varepsilon_i$,
where $y_i$ is a univariate response variable for observation $i=1,\ldots,n$, $\mathbf{x}_i$ is a $p-$dimensional predictor, and $\varepsilon_i \sim \text{N}(0,\sigma^2)$. The idea is to find a flexible approximation for the conditional expectation $\mathbb{E}\lbrack y_i\given \mathbf{x}_i\rbrack$ by expressing it as a sum of regression trees.
A regression tree consists of two components:
\begin{enumerate}
    \item A binary tree $\mathcal{T}$, which defines a finite partition $\{\mathcal{A}_1,\ldots,\mathcal{A}_h\}$ of $\mathbb{R}^p$ based on the feature space of $\mathbf{X}$, using the available predictors 
    to form splitting rules. In other words, $\mathcal{A}_1,\ldots,\mathcal{A}_h$ are subsets of $\mathbb{R}^p$ such that any $\mathbf{x}_i \in \mathbb{R}^p$ is contained in exactly one $\mathcal{A}_\ell$.
    \item A collection of scalar parameters $\mathcal{M} = (\mu_1,\ldots,\mu_h)$, called leaf nodes, with each component being associated with the corresponding subset in the partition.
\end{enumerate}
We now define a function $\mathbf{x}_i,\mathcal{T},\mathcal{M} \mapsto g(\mathbf{x}_i,\mathcal{T},\mathcal{M})$ as follows: $\mathbf{x}_i \in \mathcal{A}_\ell$ for exactly one $\ell$; then $g(\mathbf{x}_i,\mathcal{T},\mathcal{M}) \coloneqq \mu_\ell$. In the case of a single regression tree, we may then define the regression function $\mathbf{x}_i \mapsto \mathbb{E}\lbrack y_i\given \mathbf{x}_i\rbrack$ by $\mathbb{E}\lbrack y_i\given \mathbf{x}_i\rbrack \coloneqq g(\mathbf{x}_i,\mathcal{T},\mathcal{M})$. Figure \ref{fig:tree_example} shows an illustration of a simple regression tree, including the regression function it implies, for the case $p = 1$.
\vspace{-1.25em}%
\noindent To extend this idea to \emph{additive} regression trees, we consider not just one tree, but multiple trees $\mathcal{T}_1,\ldots,\mathcal{T}_m$, each with their own corresponding partitions and leaf nodes. Then we let 
$\mathbb{E}\lbrack y_i\given \mathbf{x}_i\rbrack \coloneqq
\sum^m_{t=1}  g(\mathbf{x}_i,\mathcal{T}_t,\mathcal{M}_t)$.
The definition of the statistical model now becomes 
\[y_i = \sum^m_{t=1} g\left(\mathbf{x}_i,\mathcal{T}_t,\mathcal{M}_t\right) + \varepsilon_i,\]
which is the basic BART model presented in \citet{chipman2010bart}. The model is typically not identified, since different sets of trees can lead to the same regression function. However, this is not a problem, since the individual trees are rarely of direct interest. 

The sum of trees framework can also be used to model the conditional expectation of a binary response $\mathbf{y}$, which takes values in $\{0,1\}$. This is the probit BART model, again proposed originally in \citet{chipman2010bart}. The model is easier to present and analyse when it is cast in terms of a continuous latent variable. Suppose $\mathbf{z}$ is such that
\begin{equation}
y_i = 
    \begin{cases}
        1 & \text{ if } z_i > 0 \\
        0 & \text{ otherwise}.
    \end{cases}
    \label{eq:latent_z}
\end{equation}
As before, we model $z_i$ as 
$z_i = \sum^m_{t=1} g\left(\mathbf{x}_i,\mathcal{T}_t,\mathcal{M}_t\right) + \varepsilon_i$,
with $\varepsilon_i \sim \text{N}(0,1)$. 
This then implies
\begin{align*}
    \mathbb{E}\left\lbrack y_i\given \mathbf{x}_i\right\rbrack = \Pr\left(y_i = 1\given \mathbf{x}_i\right) = \Pr\left(z_i > 0\given \mathbf{x}_i\right) = \Phi\left( \sum_{t=1}^{m}g\left(\mathbf{x}_i, \mathcal{T}_t, \mathcal{M}_t\right)\right),
\end{align*}
where $\Phi(\cdot)$ is the cumulative distribution function of the standard normal distribution. 

For both types of outcome, given a sample of size $n$ and a univariate outcome vector $\mathbf{y} \in \mathbb{R}^{n}$ associated with a predictor matrix $\mathbf{X} \in \mathbb{R}^{n\times p}$, we are interested in sampling from the joint posterior distribution $\pi(\boldsymbol{\mathcal{T},\mathcal{M}} \given \mathbf{y}, \mathbf{X})$ where $\boldsymbol{\mathcal{T}} = (\mathcal{T}_{1}, \ldots, \mathcal{T}_{m})$ denotes the collection of all trees and $\boldsymbol{\mathcal{{M}}} = (\mathcal{M}_{1}, \ldots, \mathcal{M}_{m})$ denotes their corresponding mean parameters. To obtain the posterior, it is necessary to define priors for both the trees and the terminal node parameters. Assuming the independence of leaf parameters conditional on the tree structures, \citet{chipman2010bart} define the joint prior distribution as
\begin{align*}
\pi\left(\boldsymbol{\mathcal{T}},\boldsymbol{\mathcal{M}}, \sigma^{2} \right) &= \left\lbrack\prod_{t=1}^{m}\pi\left(\mathcal{T}_{t},\mathcal{M}_{t}\right)\right\rbrack\times \pi\left(\sigma^{2}\right)  = \left\lbrack\prod_{t=1}^{m}\pi\left(\mathcal{M}_{t} \given{\mathcal{T}_{t}}\right) \times \pi\left(\mathcal{T}_{t}\right)\right\rbrack\times\pi\left(\sigma^{2}\right)\\
& = \left\lbrack\prod_{t=1}^{m}\prod_{\ell=1}^{h_{t}}\pi\left(\mathcal{\mu}_{t\ell} \given{\mathcal{T}_{t}}\right) \times \pi\left(\mathcal{T}_{t}\right)\right\rbrack\times\pi\left(\sigma^{2}\right).
\end{align*}
To achieve conjugacy, it is typically assumed that the residual variance parameter $\sigma^{2}$ follows an inverse-gamma distribution and $\mu_{t\ell} \sim N(0,\sigma_{\mu}^{2})$, with $\sigma_{\mu}^{2} \coloneqq 1/(4\kappa^{2}m)$ being proportional to the number of trees in order to regularise the contribution of each tree. The definition of $\pi(T_{t})$ includes specifying the probability of a non-terminal node as $\alpha (1+\gamma_{t\ell})^{-\beta}$ where $\gamma_{t\ell}$ denotes the depth of that node. The hyperparameters $\alpha$ and $\beta$ take the default values suggested in \citet{chipman2010bart} of $0.95$ and $2$, respectively, to favour shallow trees.

Once the prior is defined, a sampler for the conditional posterior of the trees and leaves, given $\sigma^2$, can be obtained. Referring to $\boldsymbol{\mathcal{T}}_{(-t)} \coloneqq \boldsymbol{\mathcal{T}} \setminus \{\mathcal{T}_{t}\}$ and $\boldsymbol{\mathcal{M}}_{(-t)} \coloneqq \boldsymbol{\mathcal{M}} \setminus \{\mathcal{M}_{t}\}$, an MCMC sampler is built by sequentially sampling from $\pi(\mathcal{T}_{t} \given \boldsymbol{\mathcal{T}}_{(-t)},\boldsymbol{\mathcal{M}},\mathbf{y},\mathbf{X},\sigma^{2})$ and $\pi(\mathcal{M}_{t} \given \boldsymbol{\mathcal{T}}, \boldsymbol{\mathcal{M}}_{(-t)}, \mathbf{y}, \mathbf{X}, \sigma^{2})$. It can be shown that $\mathcal{T}_{t}$ and $\mathcal{M}_{t}$ depend on $(\boldsymbol{\mathcal{T}}_{(-t)},\mathbf{y})$ only through the partial residuals $\mathbf{r}_{t} \coloneqq \mathbf{y} - \sum_{j\neq t}^{m}g(\mathbf{X},\mathcal{T}_{j},\mathcal{M}_{j})$. This can be used to construct a Bayesian back-fitting algorithm \citep{hastie2000bayesian}. The successive draws become
\begin{center}
$\pi\left(\mathcal{T}_{t}\given \mathbf{r}_{t},\sigma^{2}\right)$ \\ 
$\pi\left(\mathcal{M}_{t}\given \mathcal{T}_{t},\mathbf{r}_{t},\sigma^{2}\right)$, 
\end{center}
where new trees and splitting rules are sampled through a Metropolis-Hastings step calculated using the integrated-likelihood for the tree $\mathcal{T}_{t}$ over the leaf parameters $\mathcal{M}_{t}$. See \citet{chipman2010bart}, \citet{kapelner2016bartmachine}, and \citet{tan2019bayesian} for further details of the model and additional information the algorithmic implementation on which that of the suBART models is based.

\subsection{The suBART model for continuous outcomes}\label{sec:suBART_continuous}
We consider a regression problem of the form
\begin{equation}
\begin{pmatrix}  y_i^{(1)} \\ \vdots \\ y_i^{(d)} \end{pmatrix} = 
\begin{pmatrix}  \mathbb{E}\left\lbrack y_i^{(1)}\given \mathbf{x}_i\right\rbrack \\ \vdots \\ \mathbb{E}\left\lbrack y_i^{(d)}\given \mathbf{x}_i\right\rbrack \end{pmatrix} +
\begin{pmatrix}  \varepsilon_i^{(1)} \\ \vdots \\ \varepsilon_i^{(d)} \end{pmatrix}
\label{eq:basic_SUR}
\end{equation}
where $\mathbf{y}^{(j)}$ represents the $j$-th component of a $d$-variate outcome, and $(\varepsilon_i^{(1)},\ldots,\varepsilon_i^{(d)})^\top \sim \text{MVN}_d(\mathbf{0}_d, \boldsymbol{\Sigma})$. In principle, different models can be used for each conditional expectation in Equation \eqref{eq:basic_SUR}.
 If the conditional expectations are all assumed to be linear in $\mathbf{x}_i$, we obtain the classic SUR model  \citep{zellner1962efficient}\footnote{A more modern presentation from a Bayesian perspective, closer in spirit to the present work, may be found in \citet{greenberg2012introduction}, Chapter 9.}
. We instead want to allow the possibility that the conditional expectations are non-linear. Given that the BART model has been shown to be a viable model in the one-dimensional setting, it seems reasonable to expect this viability to extend to the multivariate case. We therefore proceed by assigning an ensemble of regression trees to each $\mathbb{E}\lbrack y_i^{(j)}\given \mathbf{x}_i\rbrack$ as follows:
\begin{equation}
\begin{pmatrix}  y_i^{(1)} \\ \vdots \\ y_i^{(d)} \end{pmatrix} = 
\begin{pmatrix} 
\sum_{t=1}^{m}g\left(\mathbf{x}_i, \mathcal{T}^{(1)}_t, \mathcal{M}^{(1)}_t\right) \\  \vdots \\
\sum_{t=1}^{m}g\left(\mathbf{x}_i, \mathcal{T}^{(d)}_t, \mathcal{M}^{(d)}_t\right)
\end{pmatrix} + 
\begin{pmatrix}  \varepsilon_i^{(1)} \\ \vdots \\ \varepsilon_i^{(d)} \end{pmatrix},
\label{eq:basic_suBART}
\end{equation}
where
\begin{itemize}

\item $\mathcal{T}^{(j)}_{t}$ is a binary tree which defines a finite partition $\{\mathcal{A}^{(j)}_{t\ell}: 1 \leq \ell \leq h^{(j)}_t\}$ of $\mathbb{R}^p$. Note that $h^{(j)}_t$ is the number of leaf nodes of the tree $\mathcal{T}^{(j)}_{t}$. The collection of all trees pertaining to the $j$-th outcome is denoted by $\boldsymbol{\mathcal{T}}^{(j)} = (\mathcal{T}_{1}^{(j)},\dots, \mathcal{T}_{m}^{(j)})$.

\item $\mathcal{M}^{(j)}_t = (\mu^{(j)}_{t1},\ldots,\mu^{(j)}_{th^{(j)}_{t}})$ is the vector of leaf parameters associated with the tree $\mathcal{T}_t^{(j)}$. The collection of all leaf parameters related to the $j$-th outcome is similarly denoted by $\boldsymbol{\mathcal{M}}^{(j)}$.

\item 
$(\varepsilon_i^{(1)},\ldots,\varepsilon_i^{(d)})^\top \sim \text{MVN}_d(\mathbf{0}_d, \boldsymbol{\Sigma})$, with  $\boldsymbol{\Sigma}$ being a $d \times d$ covariance matrix. 
We write $\sigma^2_j \coloneqq \Sigma_{jj}$ for the diagonal elements, such that $\sigma^2_j$ is the variance of the error term $\boldsymbol{\varepsilon}^{(j)}$. 
We further write $\rho_{jk} \coloneqq \text{Cor}(\boldsymbol{\varepsilon}^{(j)},\boldsymbol{\varepsilon}^{(k)})\:\forall\:j \neq k$. Note that for any $j,k$, we have 
\begin{equation*}
    \rho_{jk} = 
        \frac{\text{Cov}\left\lbrack\boldsymbol{\varepsilon}^{(j)},\boldsymbol{\varepsilon}^{(k)}\right\rbrack}
            {\sqrt{\text{Var}\left\lbrack\boldsymbol{\varepsilon}^{(j)}\right\rbrack\text{Var}\left\lbrack\boldsymbol{\varepsilon}^{(k)}\right\rbrack}} 
            = \frac{\Sigma_{jk}}
                {\sqrt{\Sigma_{jj} \Sigma_{kk}}} 
        = \frac{\Sigma_{jk}}
                {\sigma_j\sigma_k}.
\end{equation*}
\end{itemize}
The model setup is straightforward, and indeed similar to the original BART model in \citet{chipman2010bart}. 
Our model is comprised of $d$ univariate BART models, which are linked through correlated error terms $\boldsymbol{\varepsilon}^{(j)}$. Although allowing an outcome-specific number of trees $m_j$ is possible, we assume that the number of trees $m$ is common across all $d$ outcomes for simplicity. However, we stress that each outcome is assigned its \emph{own} separate collection of $m$ trees. Thus, there are $m\times d$ trees in total. This is a key difference compared to other multivariate BART models \citep{mcjames2024bayesian, um2023bayesian}, which typically assume that all outcomes share the same $m$ trees and that the leaf parameters associated with each tree are vectors $\boldsymbol{\mu}_{t\ell}$ which follow a $d$-dimensional multivariate normal distribution. Although using $m \times d$ trees in suBART may be computationally prohibitive when $d$ is large, this is rarely the case in causal inference settings, and indeed we have only $d=2$ in the TTCM application.

A consequence of this assumption in existing multivariate BART versions is the implication that the $p$-dimensional predictor $\mathbf{x}_i$ is the same for all $d$ outcomes. However, in systems of equations such as \eqref{eq:basic_SUR} and \eqref{eq:basic_suBART}, the set of predictors $\mathbf{x}_i$ need not be of common dimension $p$ for each outcome. Indeed, our \texttt{subart} software implementation allows different subsets of the predictors to be used for each constituent univariate BART model. Though we will henceforth assume that the predictors are the same for all outcomes, for simplicity, it is important to note that having the same set of predictors $\mathbf{x}_i$ available as candidates to form splitting rules in each set of trees does \emph{not} imply that the trees for different outcomes will split on the same components of $\mathbf{x}_i$ at the same cutoff values. Unlike the standard SUR, imposing restrictions on the $\mathbf{x}_i$ for different outcomes is not required to ensure that different outcomes depend on different covariates under suBART, as different sets of trees will tend to form different partitions on different subsets of the feature space $\mathbf{X}$ anyway, by the inherent nature of the tree-generating process in BART. Although practitioners can impose such restrictions nonetheless, to strictly guarantee that the trees for a given outcome do not depend on certain predictors, this is an appealing property, in the sense that suBART minimises the need to pre-specify the parametric form of the models for each conditional expectation.

We assume that the $m$ regression trees for each outcome are all independent of each other and of the covariance matrix \emph{a priori}, as per \citet{chipman2010bart}; i.e,
\begin{align*}    \pi\left(\left(\boldsymbol{\mathcal{T}}^{(1)},\boldsymbol{\mathcal{M}}^{(1)}\right),\ldots,\left(\boldsymbol{\mathcal{T}}^{(d)},\boldsymbol{\mathcal{M}}^{(d)}\right),\boldsymbol{\Sigma}\right)
    = &\left(  \prod^{m}_{t=1} \prod^{d}_{j=1} \pi\left(\mathcal{T}^{(j)}_{t},\mathcal{M}^{(j)}_{t}\right) \right) \times  \pi\left(\boldsymbol{\Sigma}\right) \\
    = &\left( \prod^{m}_{t=1} \prod^{d}_{j=1} \pi\left(\mathcal{M}^{(j)}_{t}\given \mathcal{T}^{(j)}_{t}\right)\pi\left(\mathcal{T}^{(j)}_{t}\right) \right)  \times \pi\left(\boldsymbol{\Sigma}\right).
\end{align*}
We also assume that the leaves of a tree are conditionally independent given the tree structure, i.e.,\vspace{-1em}
\begin{equation*}
\pi\left(\mathcal{M}^{(j)}_{t}\given \mathcal{T}^{(j)}_t\right) = \prod^{h^{(j)}_{t}}_{\ell=1} \pi\left(\mu^{(j)}_{t\ell}\given \mathcal{T}^{(j)}_j\right).
\end{equation*}
With this setup, prior distributions for $\mathcal{T}^{(j)}_{t}, \mu^{(j)}_{t\ell}$, and $\boldsymbol{\Sigma}$ are sufficient to specify the joint prior distribution of all model parameters. The tree structure $\mathcal{T}^{(j)}_{t}$ is assigned the same prior as in the original work by \citet{chipman2010bart}, with default hyperparameters $\alpha = 0.95$ and $\beta = 2$ to favour shallow trees and avoid overfitting. 

The shrinkage prior used for the leaf node parameters is also aligned with the approach of standard BART. As noted earlier, this prior is formulated conditionally on the tree $\mathcal{T}^{(j)}_t$. We assume that each outcome component $\mathbf{y}^{(j)}$ is re-scaled such that $y^{(j)}_i \in \lbrack -0.5, 0.5\rbrack$, using its respective observed maximum and minimum values, as outlined by \cite{chipman2010bart}. This enables us to specify the prior probability that $\mathbb{E}\lbrack y^{(j)}_i\given\mathbf{x}_i\rbrack$ lies within the rescaled interval. Consequently, the prior on the leaf node parameters is then
\begin{equation}\mu^{(j)}_{t\ell}\given \mathcal{T}^{(j)}_{t} \sim \text{N}\left(0, {\sigma^{(j)}_{\mu}}^{2}\right),\label{eq:mu_prior}
\end{equation}
where $\sigma^{(j)}_{\mu} = 1/(2\kappa\sqrt{m})$. We suggest $\kappa = 2$ as a default choice, which assigns a prior probability of $0.95$ to the event $\{\mathbb{E}\lbrack y_i^{(j)}\given \mathbf{x}_i\rbrack\in \lbrack-0.5, 0.5\rbrack\}$. 
As per the standard BART, having the prior variance of the leaf nodes be inversely proportional to $m$ helps to prevent overfitting. Crucially, the regularisation of each set of $m$ trees per outcome does not depend on the total number of outcomes $d$. Hence, $d$ being large does not incur any additional risk of overfitting.

In the univariate BART model, the error variance is typically assigned an inverse-gamma prior, which is conditionally conjugate and can thus be easily incorporated into a Gibbs sampler. Furthermore, by choosing the hyperparameters accordingly, it is possible to put an informative prior on the error variance. The recommended approach of \citet{chipman2010bart} involves obtaining a `data-based overestimate' $\hat{\sigma}^2$ of the error variance $\sigma^2$ by fitting a linear model via least squares and calculating the standard deviation of the sample residuals\footnote{If the number of observations is smaller than the dimension of $\mathbf{X}$, we instead estimate $\hat{\sigma}$ using the LASSO in the \texttt{glmnet} \textsf{R} package \citep{glmnetR}.}. Presumably, the true value of  $\sigma^2$ is smaller than $\hat{\sigma}^2$, since more of the variation in $\mathbf{y}$ can be explained by the flexible BART model than the linear model. Therefore, a large prior probability $\alpha_\sigma$ is assigned to the event $\{\sigma^2 < \hat{\sigma}^2\}$ (for example, $\alpha_\sigma=0.95$).

We now wish to generalise this idea to the multivariate setting: for all components $j$ of the outcome vector $\left(\mathbf{y}^{(1)},\ldots,\mathbf{y}^{(d)}\right)$, we obtain an overestimate $\hat{\sigma}^2_{j}$ of $\sigma^{2}_{j}$ in a similar fashion and want to assign the large prior probability $\alpha_\sigma$ to the event $\{\sigma^2_j < \hat{\sigma}^2_j\}$. Additionally, we also would like to control the prior on the correlations $\rho_{jk}$, where $j \neq k$. Since there is usually not strong prior information about these correlations, it is preferable for the prior to not be too informative in any direction. Finally, the chosen solution should be computationally tractable and readily incorporated into MCMC samplers. One such candidate is the inverse-Wishart distribution, which is often used as a prior for covariance matrices. As it is a straightforward multivariate generalisation of the inverse-gamma distribution, it facilitates easy computations. However, although it has the desirable property that the prior on $\Sigma$ is invariant to the ordering of the outcome vector $\mathbf{y}_i$, the inverse-Wishart prior is unfortunately not well-suited to meet our other aforementioned goals; among other problems, it imposes a strong prior dependency between the variances and the correlations and, when the true variances are small, overestimates variances and induces downward bias towards zero on the correlations \citep{alvarez2014bayesian}.
    
In order to have more control over the variance parameters, we thus instead adapt an approach by \citet{huang2013simple} and parameterise the covariance matrix $\boldsymbol{\Sigma}$ in a hierarchical fashion as follows:
    \begin{itemize}
        \item $a_{j} \sim \text{Inv-Gamma}(1/2, 1/A^2_j)$, where $A_j > 0$ is a fixed scale hyperparameter.
        \item $\boldsymbol{\Sigma} \given  a_{1},\ldots,a_{d} \sim \text{Inv-Wishart}_{d} (\nu + d - 1, \mathbf{S}_0)$, where \begin{equation}
        \mathbf{S}_0\coloneqq  2\nu \times \mbox{diag}\left(1/a_{1}, \ldots,1/a_{d} \right).\label{eq:s0}
        \end{equation}
    \end{itemize}
As this prior is based on the inverse-Wishart, it shares the property of invariance to the ordering of the outcome vector. A similar prior was proposed by \citet{giannone2015prior} in the context of vector autoregressions. Under our prior, the implied prior distribution for the correlations is given by
\begin{equation}
    \pi\left(\rho_{jk}\right) = \left(1 - \rho_{jk}^2\right)^{\frac{\nu}{2} - 1},\quad\rho_{jk} \in \left(-1,1\right),\label{eq:implied_prior}
\end{equation}
which matches the prior implied under the standard inverse-Wishart prior \citep{barnard2000modeling}. Crucially, it does not depend on $A_1,\ldots,A_d$. It is uniform if and only if $\nu = 2$. For higher values, the prior increasingly concentrates around zero. Moreover, $\nu=1$ leads to a strongly informative and rarely reasonable prior on the correlations, which favours extreme values near $\pm 1$. Thus, we consider $\nu = 2$ a reasonable default choice, since we usually do not have any strong prior information on the correlations. It is also worth~recalling from Section \ref{sec:application} that the response vector is typically bivariate in the CEA setting~which motivated the development of suBART, such that there is only one such correlation parameter.

From the previous definitions, the induced prior for the standard deviations is 
$\sigma_{j} \sim \text{Half-$t$}(\nu, A_{j})$,
which in turn implies a scaled $F$-distribution as the prior for $\sigma_j^2$. Conversely, an inverse-Wishart prior would imply that the variances have inverse-gamma distributions. As half-$t$ priors on standard deviations are recommended over inverse-gamma priors on variances in univariate settings by \citet{gelman2006prior} and others, the construction due to \citet{huang2013simple} is preferable to the inverse-Wishart prior in multivariate settings for similar reasons. Using this construction, we can specify an uninformative prior on each $\sigma_{j}$, by setting a very large value for the associated hyperparameter $A_j$. However, in the spirit of the standard BART approach, we prefer to specify informative, data-dependent priors, by choosing each $A_j$ in such a way that $\Pr(\sigma_j < \hat{\sigma}_j)=\alpha_\sigma$ is ensured \emph{a priori}.

As well as ensuring that Equation \eqref{eq:implied_prior} is uniform, another appealing property of specifying $\nu=2$ is the heaviness of the tails of the induced half-$t$ prior on $\sigma_j$, which makes the prior more robust to the choice of the scale $A_j$ and thereby reduces sensitivity to the data-based overestimate $\hat{\sigma}_j$ used for prior calibration. Although \citet{gelman2006prior} highlights the special case of the half-Cauchy prior with $\nu=1$ and even heavier tails, we encountered numerical difficulties in our simulation experiments and the TTCM application with $\nu=1$, particularly in situations where the sample size is small, the dimension of $\mathbf{X}$ is large, and the variability of the multivariate outcome is almost entirely explained by $\mathbf{X}$.

We calibrate the prior for each $\sigma_j$ by setting up the following equation for each outcome
\begin{equation}
\alpha_{\sigma} =   \frac{2\Gamma(\frac{\nu + 1}{2})}{\Gamma(\frac{\nu}{2})\sqrt{\nu \pi A^2_{j}}} \int_0^{\hat{\sigma}_{j}}
    {\left(1 + \frac{x^2}{\nu A^2_{j}}\right)}^{-\frac{\nu + 1}{2}}
    \dd x\label{eq:half_t}
\end{equation}
and solving it for $A_{j}$, which can be done through numerical root-finding. This expression is the cumulative distribution function of a $\text{Half-$t$}$-distributed random variable with degrees of freedom $\nu$ and scale parameter $A_j$,
evaluated at $\hat{\sigma}_j$. By rewriting Equation \eqref{eq:half_t} in terms of regularised incomplete beta functions, it can be shown that this expression is continuous as well as strictly decreasing in $A_j$ and approaches $1$ and $0$ as $A_j$ approaches $0$ and $\infty$, respectively. Thus, the solution for $A_j$ exists and is unique. 

\subsection{Probit suBART}\label{sec:suBART_probit}

While the previous model was designed to jointly model multiple continuous outcome variables, we now turn our attention to binary outcomes. We will present a generalisation of the linear multivariate probit model \citep{chib1998analysis}, where the linear predictors are replaced by sums of regression trees. Alternatively, it can also be seen as a multivariate generalisation of the probit BART model.

Suppose that we have observed some predictor variables $\mathbf{x}_i$ and an associated collection of binary outcome vectors $\mathbf{y}_i \in \{0,1\}^d$ whose dependence on $\mathbf{x}_i$ we want to model. In particular, we do not want to assume that the components of $\mathbf{y}_i$ are conditionally independent, given $\mathbf{x}_i$; there may be some leftover correlation which is not explained by $\mathbf{x}_i$. As in the basic probit BART model, we cast the multivariate version in terms of latent variables $\mathbf{z}_i = (z_i^{(1)},\ldots,z_i^{(d)})$; the construction is exactly as per Equation \eqref{eq:latent_z} for each outcome $j$.
Then, the probit suBART model is given by
\begin{equation}
\begin{pmatrix}  z_i^{(1)} \\ \vdots \\ z_i^{(d)} \end{pmatrix}  = 
\begin{pmatrix} 
\sum_{t=1}^{m}g\left(\mathbf{x}_i, \mathcal{T}^{(1)}_t, \mathcal{M}^{(1)}_t\right) \\  \vdots \\
\sum_{t=1}^{m}g\left(\mathbf{x}_i, \mathcal{T}^{(d)}_t, \mathcal{M}^{(d)}_t\right)
\end{pmatrix} + 
\begin{pmatrix}  \varepsilon_i^{(1)} \\ \vdots \\ \varepsilon_i^{(d)} \end{pmatrix},
\end{equation}
where
\begin{itemize}
\item $g(\cdot)$, $\mathcal{T}^{(j)}_{t}$, and $\mathcal{M}^{(j)}_t$ are defined as before.

\item 
$(\varepsilon_i^{(1)},\ldots,\varepsilon_i^{(d)})^\top \sim \text{MVN}_d(\mathbf{0}_d, \boldsymbol{\Sigma})$, with  $\boldsymbol{\Sigma}$ being a $d \times d$ correlation matrix. Again writing $\rho_{jk} \coloneqq \text{Corr}(\boldsymbol{\varepsilon}^{(j)},\boldsymbol{\varepsilon}^{(k)})$ for $j \neq k$, we have 
$$\boldsymbol{\Sigma}_{jk} = \begin{cases}
    1 & \text{if $j = k$} \\
    \rho_{jk} & \text{if $j \neq k$}.  
\end{cases}$$
\end{itemize}
Conditional on the latent variables $\mathbf{z}_i$, the model is essentially the same as the suBART model presented earlier. The only difference is that for each error term $\boldsymbol{\varepsilon}^{(j)}$, we fix the variance at $1$. Without doing this, the model would be unidentified. This issue is not specific to probit suBART, but also arises in the linear multivariate probit model. See \citet{chib1998analysis} for details. It follows that $\boldsymbol{\Sigma}$, the covariance matrix of the error terms, is equal to $1$ in each diagonal entry, and hence must be a correlation matrix. 

The dependence structure of the trees, leaves, and covariance matrix is the same as for the suBART model. The priors for the trees are also exactly the same. Although the other priors are broadly similar, there are some differences that are worth briefly highlighting.
The prior for the terminal node parameters $\mu_{t\ell}^{(j)}$ again follows Equation \eqref{eq:mu_prior}, though the calibration of the variance hyperparameter requires more care. Taking ${\sigma^{(j)}_{\mu}} = q_z/(\kappa\sqrt{m})$, with $\kappa = 2$ as a default choice, we  assign a prior probability of $0.95$ to the event $\{\mathbb{E}\lbrack z_i^{(j)}\given \mathbf{x}_i\rbrack \in \lbrack -q_z,q_z\rbrack\}$. On the probability scale, this means that $\{\Pr(y_i^{(j)} = 1 \given \mathbf{x}_i) \in \lbrack\Phi(-q_z), \Phi(q_z)\rbrack\}$. For example, when taking $q_z = 3$ as per \citet{chipman2010bart}, we assign a prior probability of $0.95$ to the~event $\{\Pr(y_i^{(j)} = 1\given \mathbf{x}_i) \in \lbrack0.0013,0.9987\rbrack\}$.  This is reasonable for many applications, since extremely small or large probabilities are uncommon. 

In the probit setting, $\boldsymbol{\Sigma}$ is a correlation matrix and hence must be positive definite and have all diagonal entries equal to $1$. These restrictions make it difficult to choose a prior for $\boldsymbol{\Sigma}$ which has desirable properties \emph{and} facilitates easy sampling. \citet{chib1998analysis} present a prior (and related sampling strategy) which we found to be extremely inefficient in our application. We thus instead adapt an approach by \citet{zhang2020parameter} (see also \citet{barnard2000modeling}, where most of the following results originate, and \citet{zhang2006sampling}). We introduce an auxiliary parameter $\mathbf{D}$, which is a $d \times d$ diagonal matrix. We then define $\mathbf{W} \coloneqq \mathbf{D}^{\frac{1}{2}} \boldsymbol{\Sigma} \mathbf{D}^{\frac{1}{2}}$, and assume the prior $\mathbf{W} \sim \text{Inv-Wishart}_d (\nu + d - 1, \mathbf{I}_d)$, where $\mathbf{I}_d$ is the $d$-dimensional identity matrix. As before, the induced prior on $\boldsymbol{\Sigma}$ is invariant to the ordering of the outcome vector.
    
Given $\mathbf{W}$, we can recover $\mathbf{D}$ and $\boldsymbol{\Sigma}$ thanks to the identities $\mathbf{D} = \text{diag}(\mathbf{W})$ and $\boldsymbol{\Sigma}=  \mathbf{D}^{-\frac{1}{2}} \mathbf{W} \mathbf{D}^{-\frac{1}{2}}$. The induced marginal prior density of $\boldsymbol{\Sigma}$ is 

$$\pi\left(\boldsymbol{\Sigma}\right) \propto \left(\det \boldsymbol{\Sigma}\right)^{\frac{1}{2}(\nu+d-1)(d-1)-1} \left(\prod_{j=1}^{d} \det \left\lbrack\boldsymbol{\Sigma}\right\rbrack_{jj} \right)^{-\frac{\nu + d - 1}{2}},$$
where $\lbrack\boldsymbol{\Sigma}\rbrack_{jj}$ is the $j$-th principle submatrix of $\boldsymbol{\Sigma}$. It can be shown that the marginal prior density for the correlations is again given by Equation \eqref{eq:implied_prior}, as per the suBART model for continuous outcomes above, despite the different priors assumed for $\boldsymbol{\Sigma}$. The hyperparameter $\nu$ plays a similar role as before and we again consider $\nu=2$ to be a reasonable default choice.

\section{Posterior inference}\label{sec:posterior}

This section describes strategies and algorithmic details for conducting posterior inference under suBART and probit suBART for multivariate continuous and multivariate binary outcomes, respectively. For both frameworks, sampling is performed using a Metropolis-within-Gibbs sampler based on their respective priors and model specifications. Given an observed sample $\mathbf{y}_{1}, \ldots, \mathbf{y}_n$, with $\mathbf{y}_{i} \in \mathbb{R}^{d}$, all computations are carried out conditionally on the covariates $\mathbf{X} \in \mathbb{R}^{n\times p}$. Thus, for the sake of readability, we will consider $\mathbf{X}$ fixed and not condition on it explicitly. Additionally, the expression $\pi(\theta\given \boldsymbol{\Theta}\setminus\{\ldots\})$ denotes the conditional distribution of $\theta$ with respect to all parameters \emph{except} $\theta$ itself and the ones listed within the braces. For example, $\pi(\mathcal{T}^{(j)}_t\given \boldsymbol{\Theta}\setminus\{\mathcal{M}^{(j)}_t\})$ refers to the distribution of tree $\mathcal{T}^{(j)}_t$ given all parameters except $\mathcal{T}^{(j)}_t$ and $\mathcal{M}^{(j)}_t$. This slightly unusual notation reduces the complexity of the expressions which follow. Given that we routinely condition on all but two parameters, we find it clearer to highlight what is \emph{not} being conditioned on.

\subsection{suBART continuous}\label{sec:post_subart}

For brevity, we define the estimate for $y_{i}^{(j)}$ as $\hat{y}^{(j)}_{i} \coloneqq \linebreak{}\sum_{t=1}^{m}g(\mathbf{x}_i, \mathcal{T}^{(j)}_{t}, \mathcal{M}^{(j)}_{t})$ and $\mathbf{\hat{y}}_{i}=(\hat{y}_{i}^{(1)},\ldots,\hat{y}^{(d)}_{i})^{\top}$, since there are now multiple components of $\mathbf{y}_{i}$. Analogously, we define the residuals from a tree $t$ associated with the $j$-th component as $\mathbf{r}^{(j)}_t \coloneqq \{r^{(j)}_{t1}, \ldots, r^{(j)}_{tn}\}$, where $r^{(j)}_{ti} \coloneqq y^{(j)}_{i} - \sum_{k \neq t}^{m}g(\mathbf{x}_{i}, \mathcal{T}^{(j)}_{k}, \mathcal{M}^{(j)}_{k})$. As per the standard BART, we are interested in sampling from the posterior distribution 
$$\pi(\boldsymbol{\Theta}\given \mathbf{Y}) = \pi\left(\left(\boldsymbol{\mathcal{T}}^{(1)},\boldsymbol{\mathcal{M}}^{(1)}\right),\ldots,\left(\boldsymbol{\mathcal{T}}^{(d)},\boldsymbol{\mathcal{M}}^{(d)}\right),\boldsymbol{\Sigma}, a_1,\ldots,a_d\given \mathbf{Y}\right),$$
which, due to the back-fitting algorithm \citep{hastie2000bayesian} and properties of the multivariate normal distribution, can be obtained from sequential draws from a collection of conditional distributions.
In the multivariate continuous outcomes setting, we have that 
$\mathbf{y}_i \sim \text{MVN}_d \left(\hat{\mathbf{y}}_i,  \boldsymbol{\Sigma}\right) 
$. For the following, we will also need the conditional distribution of any component $y^{(j)}_i$ given all other components $\mathbf{y}_{i}^{(-j)}$.
Using a well-known result (see e.g., \citet{bierens2004introduction}, Section 5.3), this can be found in closed form:
\begin{align}
y^{(j)}_i\given \mathbf{y}^{(-j)}_i, \boldsymbol{\Theta} &\sim \text{N} \left(\hat{y}_i^{(j)} + 
\boldsymbol{\Sigma}_{j(-j)} 
\boldsymbol{\Sigma}_{(-j)(-j)}^{-1} \left(\mathbf{y}^{(-j)}_{i} - \hat{\mathbf{y}}_{i}^{(-j)} \right),\right.\nonumber\\ \phantom{y^{(j)}_i\given \mathbf{y}^{(-j)}_i, \boldsymbol{\Theta}} & \phantom{\sim \text{N}}\left.\quad\!{\Sigma}_{jj} -  
\boldsymbol{\Sigma}_{j(-j)} 
\boldsymbol{\Sigma}_{(-j)(-j)}^{-1} 
\boldsymbol{\Sigma}_{(-j)j} \right), 
\label{eq:marginal_normal}
\end{align}
 where $\boldsymbol{\Sigma}_{(-j)(-j)}$ is the submatrix obtained by excluding the $j$-th row and column from $\boldsymbol{\Sigma}$. Analogously, $\boldsymbol{\Sigma}_{j(-j)}$ is the vector obtained by selecting the $j$-th row and excluding~the $j$-th column. Using the conditional distribution in Equation \eqref{eq:marginal_normal}, Metropolis-Hastings draws from the posterior distribution of trees $\pi(\mathcal{T}^{(j)}_{t}\given \mathbf{r}_{t}^{(j)}, \boldsymbol{\Theta}\setminus\{\mathcal{M}_{t}^{(j)}\})$ can be obtained in essentially the same way as for the univariate BART model; see  Appendix A of \citet{kapelner2016bartmachine} for details.  Ultimately, as described in Section \ref{sec:review_BART}, the sampler for the joint posterior distribution of the trees $\mathcal{T}^{(j)}_{t}$ and their parameters $\mathcal{M}_{t}^{(j)}$, for the $j$-th component of $\mathbf{y}^{(1)},\ldots,\mathbf{y}^{(d)}$, is given by successively drawing from
 $\mathcal{T}^{(j)}_{t} \given  \mathbf{r}^{(j)}_{t},\boldsymbol{\Theta}\setminus\{\mathcal{M}_{t}^{(j)}\}$ and 
    $\mathcal{M}^{(j)}_{t} \given \mathbf{r}^{(j)}_{t},\boldsymbol{\Theta}$.
    
Notably, the algorithm reduces to that of the standard BART approach when $d=1$. Indeed, each draw above can also be viewed as univariate BART --- albeit with distinct mean and variance parameters --- since it is conditioned on the values of all other components in $\mathbf{Y}$, as illustrated by Equation \eqref{eq:marginal_normal}. The full structure of the suBART sampler is given in Algorithm \ref{alg:pseudocode}, but we first describe the remaining required posterior conditional distributions.

Firstly, the posterior distribution for $\mu^{(j)}_{t\ell}$ is given by
\begin{equation}
\mu^{(j)}_{t\ell} \given  \boldsymbol{\Theta} \sim \text{N} \left(\left( \frac{{\sigma_{\mu}^{(j)}}^{2}}{v^{(j)}+n_{t\ell}^{(j)}{\sigma_{\mu}^{(j)}}^{2}} \right)\times\left(\sum_{i=1}^{n_{t\ell}^{(j)}}r_{i}^{(j)} - \sum_{i=1}^{n_{t\ell}^{(j)}}u_{i}^{(j)}\right), \frac{v^{(j)}{\sigma_{\mu}^{(j)}}^{2}}{v^{(j)}+n_{t\ell}^{(j)}{\sigma_{\mu}^{(j)}}^{2}}  \right),
\label{eq:full_mu_post}
\end{equation}
where $u_{i}^{(j)}  \coloneqq \boldsymbol{\Sigma}_{j(-j)} 
\boldsymbol{\Sigma}_{(-j)(-j)}^{-1} (\mathbf{y}^{(-j)}_{i} - \hat{\mathbf{y}}_{i}^{(-j)})$, $v^{(j)}\coloneqq \Sigma_{jj} - \boldsymbol{\Sigma}_{j(-j)} 
\boldsymbol{\Sigma}_{(-j)(-j)}^{-1} 
\boldsymbol{\Sigma}_{(-j)j}$, and $\smash{n_{t\ell}^{(j)}}$ denotes the number of observations in the given terminal node. The derivation again closely follows that for the univariate BART model; see Appendix B.1 of \citet{tan2019bayesian} for details. Indeed, when $d=1$, these definitions imply that the $u_i^{(j)}$ term vanishes and $v^{(j)}=\Sigma_{jj}=\sigma^{2}$, resulting in an expression identical to the original BART formulation. Secondly, the conditional posteriors of $\boldsymbol{\Sigma}$ and the auxiliary parameters $a_1,\ldots,a_d$ take simple forms, due to the conditional conjugacy in the construction of \citet{huang2013simple}. We have
\begin{equation}
a_j\given \boldsymbol{\Theta}  \sim \text{Inv-Gamma} \left(\frac{\nu + d}{2}, \frac{1}{A^2_{j}} + \nu\left(\boldsymbol{\Sigma}^{-1}\right)_{jj}\right),
\label{eq:a_full_post}
\end{equation}
where $(\boldsymbol{\Sigma}^{-1})_{jj}$ denotes the $j$-th entry along the diagonal of $\boldsymbol{\Sigma}^{-1}$, and
\begin{equation}
\boldsymbol{\Sigma}\given \boldsymbol{\Theta} \sim 
\text{Inv-Wishart}_d \left(\nu + d - 1 + n, \mathbf{S}_0 + \mathbf{S}\right),
\label{eq:sigma_full_post}
\end{equation}
where $\mathbf{S} = \sum_{i=1}^{n} (\mathbf{y}_i - \hat{\mathbf{y}}_i) (\mathbf{y}_i - \hat{\mathbf{y}}_i)^\top$ and $\mathbf{S}_0$ is as given in Equation \eqref{eq:s0}.\vspace{-1em}
\begin{figure}[H] 
\begin{algorithm}[H]
\caption{suBART sampling algorithm.\label{alg:pseudocode}}
\SetAlgoLined
\SetKwInput{KwInput}{Input}
\SetKwInput{KwOutput}{Output}
\SetKwInput{KwInit}{Initialise}
\KwInput{$\mathbf{X}$, $\mathbf{Y}$, $m$, $N_{\text{MCMC}}$, $N_{\text{burn-in}}$, and all hyper-parameters of the priors.}
 \KwInit{$\boldsymbol{\mathcal{T}}^{(1)}, \ldots, \boldsymbol{\mathcal{T}}^{(d)}$ tree stumps, $\boldsymbol{\Sigma}$, $\mu^{(j)}_{t\ell} = 0 \:\forall\:(t,j)$.}
 \For{iterations $k$ from $1$ to $N_{\text{MCMC}}$}{
 \For{dimension $j$ from $1$ to $d$}{
  \For{trees $t$ from $1$ to $m$}{
   Calculate the partial residuals $\mathbf{r}^{(j)}_t$;\\
   Propose a new tree ${\mathcal{T}^{(j)}_{t}}^{\star}$ by a grow, prune, or change move\protect\footnote{See \citet{kapelner2016bartmachine} for more details on these tree proposal steps and transition probabilities $q(\cdot)$.};\\
   Accept and update $\mathcal{T}^{(j)}_{t}=
   {\mathcal{T}^{(j)}_{t}}^{\star}$ with probability 
   \[\gamma^{\star}\left(\mathcal{T}^{(j)}_{t},{\mathcal{T}_{t}^{(j)}}^{\star}\right) = 
   \min \left\{1, \frac{\pi \left(\mathbf{r}^{(j)}_{t} \given  {\mathcal{T}_{t}^{(j)}}^{\star}, \boldsymbol{\Theta}\setminus\left\{M_{t}^{(j)}\right\} \right) \pi\left({\mathcal{T}_{t}^{(j)}}^{\star}\right)q\left({\mathcal{T}_{t}^{(j)}}^{\star} \rightarrow \mathcal{T}^{(j)}_{t}\right)}{\pi \left(\mathbf{r}_{t}^{(j)} \given  {\mathcal{T}_{t}^{(j)}}, \boldsymbol{\Theta}\setminus\left\{M_{t}^{(j)}\right\} \right) \pi\left({\mathcal{T}_{t}^{(j)}}\right)q\left({\mathcal{T}_{t}^{(j)}} \rightarrow {\mathcal{T}^{(j)}_{t}}^{\star}\right)}\right\}.\]
   
   \For{terminal nodes $\ell$ from $1$ to $b^{(j)}_{t}$}{
        Update $\mu^{(j)}_{t \ell} \given \mathbf{r}_{t}^{(j)},\boldsymbol{\Theta}$ using Equation \eqref{eq:full_mu_post}.
   }
   }
   }{\For{ $j$ from $1$ to $d$}{
         Update $a_{j} \given \boldsymbol{\Theta}$ using Equation \eqref{eq:a_full_post}.
    }}
     Update $\boldsymbol{\Sigma} \given \boldsymbol{\Theta}$ using Equation \eqref{eq:sigma_full_post}.}
\end{algorithm}
\end{figure}

\subsection{Probit suBART}

In the multivariate binary setting, the goal is to draw samples from the similar posterior $$\pi(\boldsymbol{\Theta}\given \mathbf{Y}) = \pi\left(\left(\boldsymbol{\mathcal{T}}^{(1)},\boldsymbol{\mathcal{M}}^{(1)}\right),\ldots,\left(\boldsymbol{\mathcal{T}}^{(d)},\boldsymbol{\mathcal{M}}^{(d)}\right),\boldsymbol{\Sigma},\mathbf{D}\given \mathbf{Y}\right).$$
Note that $\mathbf{W}$ is a deterministic function of $\boldsymbol{\Sigma}$ and $\mathbf{D}$, and is hence omitted from the above distribution. However, it will be more convenient to work with the joint posterior of the parameters and the latent variables
$$\pi\left(\boldsymbol{\Theta},\mathbf{Z}\given \mathbf{Y}\right) = \pi\left(\left(\boldsymbol{\mathcal{T}}^{(1)},\boldsymbol{\mathcal{M}}^{(1)}\right),\ldots,\left(\boldsymbol{\mathcal{T}}^{(d)},\boldsymbol{\mathcal{M}}^{(d)}\right),\boldsymbol{\Sigma},\mathbf{D},\mathbf{Z}\given \mathbf{Y}\right),$$
for which the sampling algorithm is very similar to the previously presented Algorithm \ref{alg:pseudocode} in the continuous setting. For brevity, we discuss only the required modifications to Algorithm \ref{alg:pseudocode} without presenting a new algorithm in full.

The updates for the trees and leaf nodes stay essentially the same, with the one difference being that the latent variables $\mathbf{Z}$ replace the data $\mathbf{Y}$. An important additional step is that the latent variables for each component $j$ should be updated after line 10 in Algorithm \ref{alg:pseudocode}. In a similar manner to Equation \eqref{eq:marginal_normal}, the marginal distribution of $z_i^{(j)}$ can be obtained via
\begin{align}
z^{(j)}_i\given \mathbf{z}^{(-j)}_i, \boldsymbol{\Theta} &\sim \text{N} \left(\hat{z}_i^{(j)} + 
\boldsymbol{\Sigma}_{j(-j)} 
\boldsymbol{\Sigma}_{(-j)(-j)}^{-1} \left(\mathbf{z}^{(-j)}_{i} - \hat{\mathbf{z}}_{i}^{(-j)} \right),\right.\nonumber\\ \phantom{z^{(j)}_i\given \mathbf{z}^{(-j)}_i, \boldsymbol{\Theta}} & \phantom{\sim \text{N}}\left.\quad\!1 -  
\boldsymbol{\Sigma}_{j(-j)} 
\boldsymbol{\Sigma}_{(-j)(-j)}^{-1} 
\boldsymbol{\Sigma}_{(-j)j} \right), 
\label{eq:marginal_normal_prob}
\end{align}
where we exploit the fact that $\Sigma_{jj}=1$, since $\boldsymbol{\Sigma}$ is a correlation matrix in the probit setting. However, sampling the latent variables also requires conditioning on $\mathbf{Y}$. In doing so, we find that $\pi(z^{(j)}_i \given \mathbf{z}^{(-j)}_i, \mathbf{Y}, \boldsymbol{\Theta})$ follows a truncated normal distribution with the same location and scale parameters as Equation \eqref{eq:marginal_normal_prob}. However, we encounter a case distinction for the support of this conditional posterior distribution based on the values of the associated response. If $y_i^{(j)}=0$, which implies that $z_i^{(j)} \le 0$, the support is truncated to $(-\infty, 0\rbrack$. Conversely, if $y_i^{(j)}=1$, which implies that $z_i^{(j)} > 0$, the support is truncated to $(0, \infty)$. In each case, we draw the sample through the method proposed by \citet{robert1995simulation}.

The other major difference for the probit suBART sampler concerns the update of $\boldsymbol{\Sigma}$.
Since the $a_j$ parameters only apply to $\boldsymbol{\Sigma}$ in the continuous setting, their updates are of course dropped. 
In the probit setting, accounting also for the auxiliary parameter $\mathbf{D}$, we can write the associated conditional posterior distribution as
\begin{align*}
    \pi\left(\boldsymbol{\Sigma},\mathbf{D}\given \boldsymbol{\Theta}, \mathbf{Z}\right) &\propto
    \pi\left(\boldsymbol{\Sigma},\mathbf{D}\right) \left(\det \mathbf{D}\right)^{\frac{d - 1}{2}}
    \pi\left(\mathbf{Z}\given \boldsymbol{\Theta}\right) \\
    &\propto \pi\left(\boldsymbol{\Sigma},\mathbf{D}\right)\left(\det \mathbf{D}\right)^{\frac{d - 1}{2}}
   \left(\det \boldsymbol{\Sigma}\right)^{-n/2} \exp\left(-\frac{1}{2} \sum_{i=1}^{n} \left(\mathbf{z}_i - \hat{\mathbf{z}}_i\right)^\top \boldsymbol{\Sigma}^{-1} \left(\mathbf{z}_i - \hat{\mathbf{z}}_i\right) \right),
\end{align*}
where $\pi(\boldsymbol{\Sigma},\mathbf{D})=\pi(\mathbf{W})$ is the aforementioned inverse-Wishart prior on $\mathbf{W}$. The term $(\det \mathbf{D})^{\frac{d - 1}{2}}$ is the Jacobian determinant which arises due to the change of variables $\mathbf{W} \mapsto \boldsymbol{\Sigma},\mathbf{D}$.
This distribution is not of known form and hence cannot be sampled from directly.
We instead proceed using the parameter-expanded Metropolis-Hastings (PX-MH) algorithm of \citet{zhang2020parameter}. This defines a proposal for $\mathbf{W}^{(k+1)} \given \mathbf{W}^{(k)} \sim 
 \text{Inv-Wishart}_{d}\left(\nu_{\text{prop}}, \nu_{\text{prop}} \times \mathbf{W}^{(k)} \right)$ where $k$ denotes the current MCMC iteration and $\nu_{\text{prop}}$ is a tuning parameter which can be fine-tuned to balance the trade-off between the the acceptance rate of the PX-MH algorithm and the autocorrelation in the resulting chains. Such tuning should account for the sample size and the dimension $d$, as outlined in prior studies \citep{zhang2006sampling}. We set $\nu_{\text{prop}}$ so to as to ensure an acceptance rate of 20-30\%.

\section{Simulation experiments to assess predictive performance}\label{sec:simstudies}

In this section, we evaluate the predictive performance of the proposed models through experiments with simulated data. Section \ref{sec:sim_continuous} and Section \ref{sec:sim_binary} are devoted to simulation designs in which the responses $\mathbf{y}^{(1)},\ldots,\mathbf{y}^{(d)}$ are all continuous and each outcome $\mathbf{y}^{(j)} \in \{0,1\}$ is binary, respectively. We undertake a comparative analysis of suBART against benchmark models, including the standard BART model applied independently to each response, the multivariate BART (mvBART) model, and a Bayesian linear seemingly unrelated regression (SUR) model. We also consider probit versions of each model, where available. This comparative study aims to explore various aspects of the models, including predictive performance and their ability to accommodate assumptions regarding correlation among responses and/or assumptions of linearity. Furthermore, our simulations aim to elucidate the primary distinctions between suBART and mvBART. For example, the splits generated by trees under the mvBART framework entail a splitting rule in all components of $\mathbf{Y}$, potentially deviating from an accurate representation of the true function $f(\mathbf{X})$ for some scenarios. Consequently, each response variable in our experiments is generated using a different subset of covariates. Additionally, a significant improvement in predictive performance capacity is anticipated when compared with the linear SUR model as, for the most part, the responses in our experiments are almost all assumed to be non-linearly related to the covariates.

These assumptions were tested over $100$ replications of each simulation scenario, using different sample sizes of $ n_{\text{train}} = n_{\text{test}} = \{250, 500, 1000
\}$ for training and test samples respectively. The metrics employed to assess differences in model performance included the root mean squared error (RMSE), the continuous ranked probability score \citep[CRPS;][]{raftery2007crps}, and the prediction interval (PI) coverage for the multivariate regression cases, while the logarithmic loss, the accuracy (ACC), and the credible interval (CI) coverage of the probabilities from $\Phi(z_{i}^{(j)})$ were used for the multivariate probit scenarios. The posterior means for $\hat{y}_{i}^{(j)}$, $\sigma_{j}$, and $\rho_{jk}$ are obtained by averaging the corresponding posterior samples, while the 50\% PIs in the case of continuous outcomes are estimated by drawing from the predictive distribution of $y_{i}^{(j)}$ and computing the 25-th and 75-th percentiles.

Our choice for the inverse-Wishart hyperparameter is $\nu = 2$, reflecting our lack of prior information about the correlation structure \citep{huang2013simple}.  
For the probit version of suBART, our selection of the proposal degrees of freedom $\nu_{\text{prop}}$ varies according to the dimension of the response and the sample size. 
We adopt the default values of $\nu_{\text{prop}}=n_{\text{train}}/10$ when $d=2$ and $\nu_{\text{prop}}=n_{\text{train}}/2$ when $d=3$. These settings appear to ensure sufficiently well-behaved sampling in our simulation experiments. For the MCMC settings for all methods under our experiments with binary responses, we set a total of $N_{\text{MCMC}} = 10000$ iterations, of which $N_{\text{burn-in}} = 2000$ draws are discarded as burn-in. Otherwise, for our continuous response experiments, $N_{\text{MCMC}}=5000$ and $N_{\text{burn-in}}=1000$ suffice for satisfactory convergence for all methods. The higher number of iterations for the probit version of suBART is required since the sampler for $\boldsymbol{\Sigma}$ is less efficient in this case. See \citet{zhang2006sampling} for details on these aspects of the sampler
and Appendix \ref{app:convergence} for a discussion of the satisfactory convergence of the continuous and probit versions of suBART for these simulation experiments.

Lastly, we note the software implementations for each model included in the comparisons. The suBART models are fitted using our own \texttt{subart} implementation and the BART models are fitted  using the \texttt{dbarts} package \citep{dorie2024package},  while different software is used to fit the linear Bayesian SUR models (henceforth BayesSUR) depending on the nature of the outcome variables. We use the \text{R} package \texttt{surbayes} \citep{alt2020surbayes} for the continuous response experiments in Section \ref{sec:sim_continuous}, which is based on the direct Monte Carlo scheme and default priors from \citet{zellner2010direct}. For the binary response experiments in Section \ref{sec:sim_binary}, we use the probabilistic programming language \texttt{Stan} \citep{stanusersguide}, through the \texttt{rstan} package \citep{rstanref} which provides an \textsf{R} interface for this library. The mvBART model was evaluated using the \texttt{skewBART} implementation provided by \citet{um2023bayesian}, specifically by setting the argument \texttt{do\_skew=FALSE} of the main \texttt{MultiskewBART()} function. None of our subsequent analyses use the sparsity-inducing Dirichlet prior on the splitting probabilities \citep{linero2018highdim} which is available as an option in the \texttt{skewBART} package, as the TTCM application and our simulated scenarios each have only a moderate number of covariates. We have not yet incorporated this prior into our \texttt{subart} software implementation for the same reason. 
Furthermore, it is worth noting that the \texttt{skewBART} package is currently limited to continuous scenarios with two dimensions, thereby results were constrained to such cases. The default hyperparameter settings were retained for all competing models with the exception of the number of trees for the tree-based models. We use $m=100$ trees \emph{per outcome} for suBART and independent BART and --- for the sake of fairness --- configure mvBART with $m \times d$ trees.

\subsection{Continuous response experiments}\label{sec:sim_continuous}

The simulation scenario described by the system of equations below was created to accommodate different types of complexity. In this experiment, the values of the response are non-linear functions modified from examples given in \citet{friedman1991multivariate} and \citet{breiman1996bagging} for a multivariate response scenario. The third response is exceptional in the sense that the generating function is purely linear. Note that correlated noise $(\varepsilon_i^{(1)}, \ldots, \varepsilon_i^{(d)})^\top \sim \text{MVN}_{d}(\mathbf{0}_d, \boldsymbol{\Sigma})$ is subsequently added to the $d$-dimensional response.\smallskip

\noindent\textbf{Friedman \#1}:
\begin{align*}
&x^{(1)}_{i}, \ldots, x_{i}^{(10)} \overset{\mathrm{iid}}{\sim} \text{Uniform}(0, 1) \\[-0.75ex]
&y_{i}^{(1)} = 10 \sin(x_{i}^{(1)} x_{i}^{(2)} \pi) + 20(x_{i}^{(3)} - 0.5)^2 \\[-0.75ex]
&y_{i}^{(2)} = 8 x_{i}^{(4)} + 20 \sin(x_{i}^{(1)} \pi) \\[-0.75ex]
&y_{i}^{(3)} = 10 x_{i}^{(5)} - 5 x_{i}^{(2)} - 5 x_{i}^{(4)}
\end{align*}
\indent The $p=10$ predictors are generated from a uniform distribution. It is notable that not all predictors are used to build the responses. However, all predictors are used for model fitting. It is anticipated that the tree-based models will be able to identify the uninformative noise variables. It is also essential to emphasise that each component of the outcome vector $\mathbf{Y}$ is derived from a distinct set of predictors. However, no restrictions are imposed on which predictors are associated with which response during model fitting. As previously mentioned, it is anticipated that mvBART may encounter challenges in accurately approximating the true generating functions in such cases. For each tree, the partitioning of the covariate space is shared across all responses, which may
be an unrealistically restrictive assumption.

\begin{figure}[H]
    \centering
    \includegraphics[width=\textwidth]{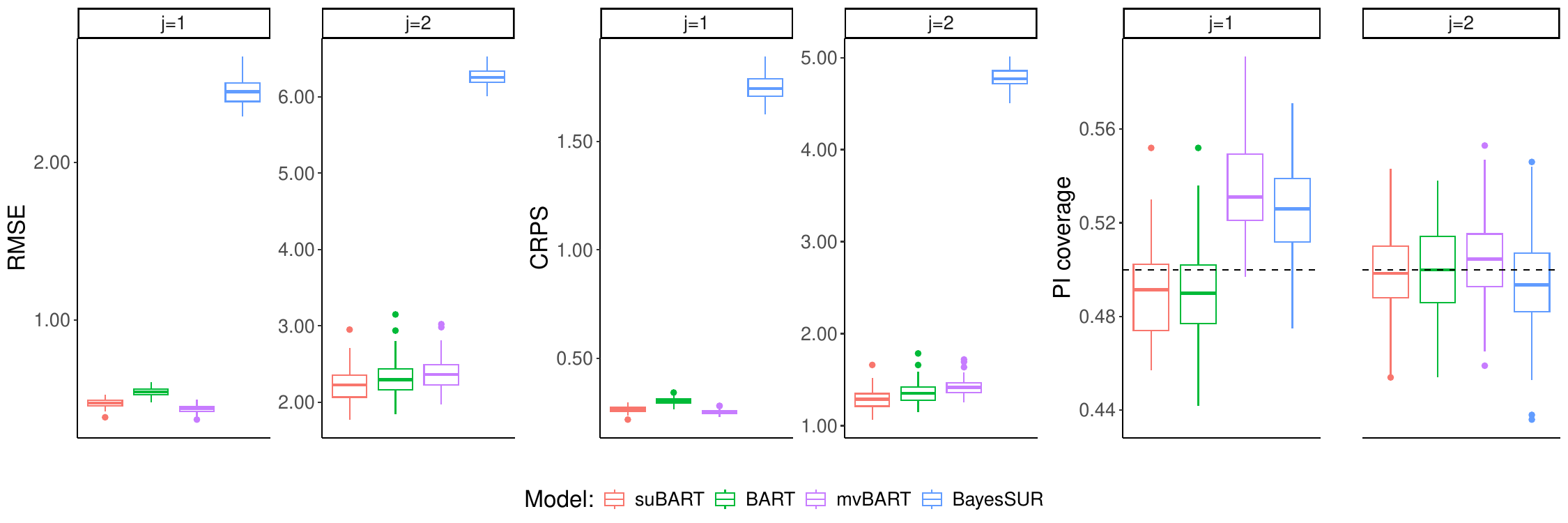}
    \caption{Simulation results for Friedman \#1 under the suBART, BART, mvBART, and BayesSUR models (from left to right) with $n_{\text{train}} = n_{\text{test}}=1000 $ and $d = 2$.}
    \label{fig:sim_reg_friedman_one_1000_2d}
\end{figure}

\begin{figure}[H]
    \centering
    \includegraphics[width=\textwidth]{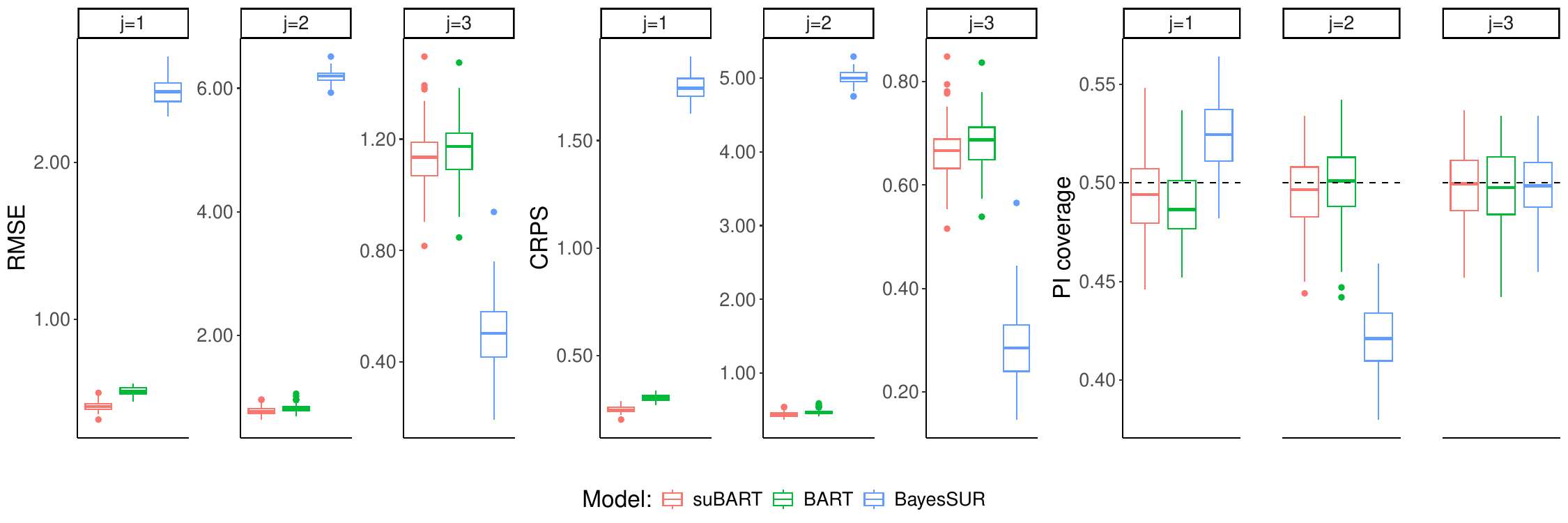}
    \caption{Simulation results for Friedman \#1 under the suBART, BART, and BayesSUR models (from left to right) with $n_{\text{train}} = n_{\text{test}}=1000 $ and $d = 3$.}
    \label{fig:sim_reg_friedman_one_1000_3d}
\end{figure}

We varied the dimension of the covariance error matrix within $d = \{2,3\}$ and defined $\boldsymbol{\Sigma}$ accordingly, with specific values assigned to each parameter $\sigma_{j}$ and each correlation parameter $\rho_{jk}$, for all $j\neq k$. The error covariance parameters were set as detailed in Table \ref{tab:friedman_reg_parameters} and it is the first two responses $\smash{y_i^{(1)}}$ and $\smash{y_i^{(2)}}$ which comprise the $d=2$ setting. The restriction to $d=2$ enables consideration of the mvBART model in the comparisons, owing to the aforementioned limitation of the \texttt{skewBART} software to bivariate outcome settings.
\begin{table}[H]
\centering
\caption{True parameters of $\boldsymbol{\Sigma}$ used for the Friedman \#1 simulation scenario.}\vspace{-1em} 
\label{tab:friedman_reg_parameters}
\begin{tabular}{ccccccc}
\multicolumn{7}{c}{}\\
\hline
$d$ & $\sigma_1$ & $\sigma_2$ & $\sigma_3$ & $\rho_{12}$ & $\rho_{13}$ & $\rho_{23}$ \\
\hline
$2$ &  $1.00$   &   $10.00$   &   ---   &   $0.75$   & --- &   ---   \\ 
$3$ &   $1.00$   &   $2.50$   &   $5.00$   &   $0.80$  & $0.50$  &   $0.25$   \\
\hline
\end{tabular}
\end{table}%
A comparison of results is depicted in the boxplots in Figure \ref{fig:sim_reg_friedman_one_1000_2d} and Figure \ref{fig:sim_reg_friedman_one_1000_3d} which confirm previous assumptions about suBART performance. These figures illustrate the results for \textit{Friedman \#1} with $d=2$ and $d=3$, respectively, with $n_{\text{train}} = n_{\text{test}} = 1000$ in each case. In general, suBART exhibits either slightly superior or competitive predictive performance compared to BART and mvBART, as evidenced by small average RMSE and CRPS values over the test samples. Furthermore, all tree-based methods exhibit a clear superiority over BayesSUR in estimating the non-linear responses. The primary discrepancy occurs when $j=3$, where BayesSUR has the best performance owing to the linearity of this response.
In terms of uncertainty estimation, Figure \ref{fig:sim_reg_friedman_one_1000_2d} shows that all methods exhibit reasonable coverage rates for the prediction intervals when $d=2$, except for the first outcome where the coverage for both mvBART and BayesSUR is notably high. Figure \ref{fig:sim_reg_friedman_one_1000_3d} also shows satisfactory coverage for both suBART and mvBART, while the coverage under BayesSUR is only close to the nominal rate when $j = 3$, which is again due to the linearity of this response.

For proper uncertainty quantification, it is essential to correctly estimate 
the elements of the
covariance matrix $\boldsymbol{\Sigma}$. Table \ref{tab:sim_corr_fried_one_combined_1000} displays the RMSE and coverage rates of 50\% credible intervals  for the correlation  parameters  $\rho_{jk}$ for all $j \neq k$ provided by suBART, mvBART, and BayesSUR. Notably, correlation values for the BART model are not provided as it assumes independence  (i.e., $\rho_{jk}= 0\:\forall\:j\ne k$), and the mvBART estimates are provided only when $d=2$, due to the aforementioned limitations of the \texttt{skewBART} package. Standard deviation parameters $\sigma_j$ are also given where applicable. From the results, it is clear that suBART outperforms BART, mvBART, and BayesSUR in terms of coverage, demonstrating its superior ability to estimate correlation structures. The coverage values of zero for BayesSUR are particularly notable and suggest an inability to accurately estimate correlations when  assuming linear regressions for non-linear responses. In terms of RMSE, suBART is superior to BART and BayesSUR and comparable to mvBART, albeit only in the $d=2$ setting.
\begin{table}[H] 
\small
\centering
\caption{RMSE and coverage of a 50\% CI for $\sigma_k$ and ${\rho}_{jk}$  for Friedman \#1 with $n_{\text{train}} = n_{\text{test}} = 1000$.}
\setlength{\tabcolsep}{3.5pt}
\label{tab:sim_corr_fried_one_combined_1000}
\begin{tabular}{lcccccccc}
\hline
\small
& \multicolumn{4}{c}{\textbf{RMSE}} & \multicolumn{4}{c}{\textbf{CI coverage}} \\
 & \textbf{suBART} & \textbf{BART} & \textbf{mvBART} & \textbf{BayesSUR} & \textbf{suBART} & \textbf{BART} & \textbf{mvBART} & \textbf{BayesSUR} \\
\hline
$d=2$ & & & & & & & & \\
\hline
{$\sigma_1$} & $0.02$ & $0.03$ & $0.03$ & $1.62$ & $0.43$ & $0.26$ & $0.31$ & $0.00$ \\
{$\sigma_2$} & $0.27$ & $0.33$ & $0.35$ & $1.78$ & $0.32$ & $0.24$ & $0.17$ & $0.00$ \\
{$\rho_{12}$} & $0.02$ & --- & $0.01$ & $0.31$ & $0.38$ & --- & $0.60$ & $0.00$ \\ 
\hline
$d=3$ & & & & & & & & \\
\hline
{$\sigma_1$} & $0.02$ & $0.04$ & --- & $1.62$ & $0.46$ & $0.28$ & --- & $0.00$ \\
{$\sigma_2$} & $0.07$ & $0.11$ & --- & $4.15$ & $0.34$ & $0.11$ & --- & $0.00$ \\
{$\sigma_3$} & $0.18$ & $0.19$ & --- & $0.10$ & $0.25$ & $0.22$ & --- & $0.41$ \\
{$\rho_{12}$} & $0.01$ & --- & --- & $0.34$ & $0.39$ & --- & --- & $0.00$ \\ 
{$\rho_{13}$} & $0.02$ & --- & --- & $0.31$ & $0.45$ & --- & --- & $0.00$ \\
{$\rho_{23}$} & $0.03$ & --- & --- & $0.16$ & $0.50$ & --- & --- & $0.00$ \\ 
\end{tabular}
\end{table}%
Equivalent figures and tables summarising the results for the remaining scenarios, with $d=3$ and/or different sample sizes, yield the same conclusions as above and have been deferred to Appendix \ref{app:friedman} for brevity. 

\subsection{Binary response experiments}\label{sec:sim_binary}

The experiments for binary responses are aligned with those from Section \ref{sec:sim_continuous}, wherein different sample sizes of $ n_{\text{train}} = n_{\text{test}} = \{250, 500, 1000
\}$ are used for training and test samples respectively. The simulation of the latent variables $z^{(j)}$ is described by the system of equations below. Other than when $j=3$, the values of the latent variables are non-linear functions. Recall that correlated noise $(\varepsilon_i^{(1)}, \ldots, \varepsilon_i^{(d)})^\top \sim \text{MVN}_{d}(\mathbf{0}_d, \boldsymbol{\Sigma})$ is subsequently added to the $d$-dimensional latent variable and that the generating process for each binary response follows Equation \eqref{eq:latent_z} thereafter. As the generative model is invariant to the values of the variances \citep{chib1998analysis}, we set them to $1$ for convenience, but the true correlations are set to the same values as in Table \ref{tab:friedman_reg_parameters}.\smallskip

\begin{figure}[H]
    \centering
    \includegraphics[width=\textwidth]{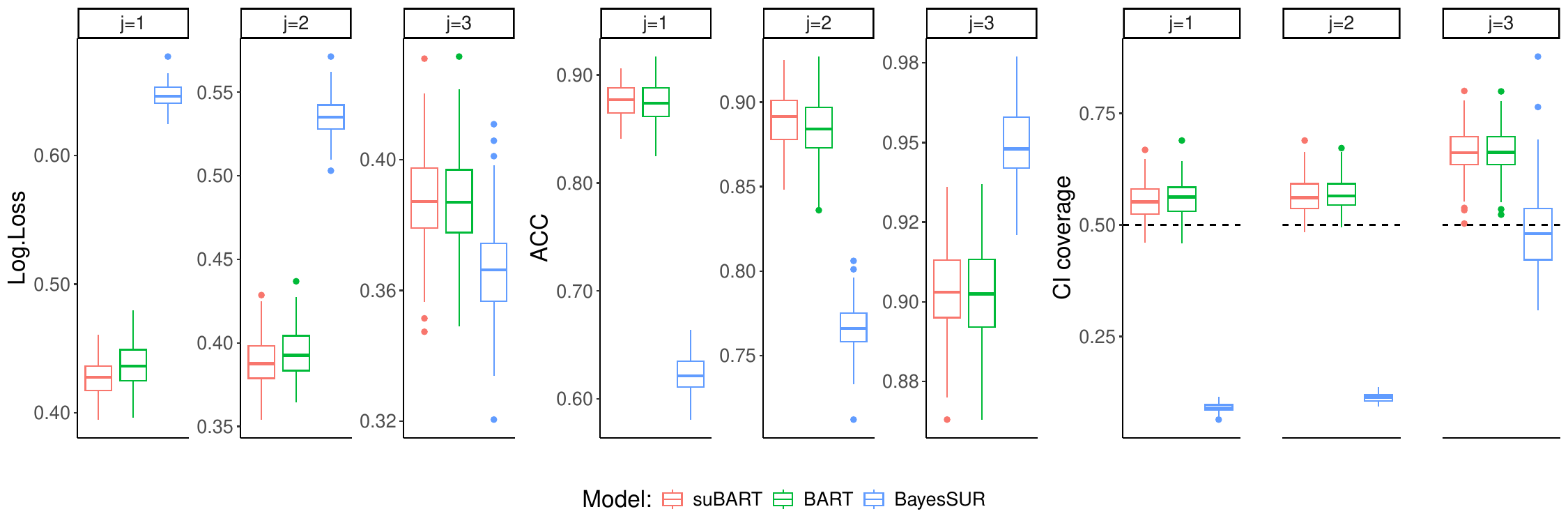}
    \caption{Simulation results Friedman \#2 with $n_{\text{train}} = n_{\text{test}}=1000 $ and $d = 3$.}
    \label{fig:sim_class_friedman_one_1000_3d}
\end{figure}

\noindent\textbf{Friedman \#2}:
\begin{align*}
&x_{i}^{(1)}, \ldots, x_{i}^{(10)} \overset{\mathrm{iid}}{\sim} \text{Uniform}(0, 1) \\[-0.75ex] 
&z_{i}^{(1)} = \sin(x_{i}^{(1)} x_{i}^{(2)} \pi) + {x_{i}^{(3)}}^{3} \\[-0.75ex]
&z_{i}^{(2)} = -1 + 2x_{i}^{(1)}x_{i}^{(4)} +e^{x_{i}^{(5)}}\\[-0.75ex]
&z_{i}^{(3)} = 0.5(x_{i}^{(2)} +  x_{i}^{(4)}) + x_{i}^{(5)}
\end{align*}
\indent The models evaluated for these experiments are the probit suBART and probit extensions of the standard BART and BayesSUR. The mvBART model is excluded, since the \texttt{skewBART} software does not offer a probit implementation of the model, for any dimensionality.
We persist in evaluating settings with varying dimension $d=\{2,3\}$ with the $d=2$ setting again comprising the first two responses.
The results for $d=3$, with $n_{\text{train}}=n_{\text{test}}=1000$, are summarised in 
Figure \ref{fig:sim_class_friedman_one_1000_3d} and  are consistent with the findings from Section \ref{sec:sim_continuous}. When logarithmic loss and ACC are considered as metrics for evaluating predictive performance, suBART either exhibits superior results or comparable averages. Both tree-based models outperform Bayesian SUR, with the exception of the linear third response in the $d=3$ setting, as per the continuous-outcome simulation experiments in Section \ref{sec:sim_continuous}.

Regarding calibration, it is evident that suBART and BART are satisfactory across all scenarios, showing coverage rates close to the nominal value. On the other hand, due to the inherent linearity of BayesSUR, its calibration performance is notably poorer, though the third linear response in Figure \ref{fig:sim_class_friedman_one_1000_3d} is again an exception in this regard, as per Section \ref{sec:sim_continuous}. 
As per the Friedman \#1 simulations in Section \ref{sec:sim_continuous}, the results for remaining sample sizes are deferred to Appendix \ref{app:friedman}, as they lead to similar conclusions.
The results regarding the estimation of the correlation parameters are presented in Table \ref{tab:sim_corr_class_fried_one_combined_1000}, which includes the RMSE and CI coverage for the correlation parameters $\rho_{jk}$ associated with binary responses when $n_{\text{train}}=1000$. As BART assumes independent errors, it is excluded from these results. Overall, these results are consistent with Table \ref{tab:sim_corr_fried_one_combined_1000} in clearly demonstrating superior performance for suBART in terms of 
estimating correlation structures.
\begin{table}[H]
\centering
\caption{RMSE and coverage of a 50\% CI for ${\rho}_{jk}$ for Friedman \#2 with $n_{\text{train}} = n_{\text{test}} = 1000$.}
\label{tab:sim_corr_class_fried_one_combined_1000}
\begin{tabular}{lcccccc}
\hline
\small
 & \multicolumn{3}{c}{\textbf{RMSE}} & \multicolumn{3}{c}{\textbf{CI coverage}} \\
 & \textbf{suBART} & \textbf{BayesSUR} & \textbf{suBART} & \textbf{BayesSUR} \\
\hline
$d=2$ & & & & \\
\hline
{$\rho_{12}$} & $0.04$ & $0.24$ & $0.39$ & $0.00$ \\ 
\hline
$d=3$ & & & & \\
\hline
{$\rho_{12}$} & $0.04$ & $0.26$ & $0.44$ & $0.00$ \\
{$\rho_{13}$} & $0.05$ & $0.09$ & $0.41$ & $0.00$ \\
{$\rho_{23}$} & $0.05$ & $0.06$ & $0.49$ & $0.00$ \\
\hline
\end{tabular}
\end{table}

\section{Analysis of the TTCM data}\label{sec:case_study}

We now apply the continuous suBART model to the TTCM data from \citet{wiertsema2019cost} described previously in Section \ref{sec:application}. We do this in two different ways. We first construct a simulation experiment in Section \ref{sec:TTCM_sim} using the baseline covariates contained in the TTCM data. We then apply six different models (detailed below) to 1000 datasets created through simulation. As the true treatment effects are known, this will allow us to evaluate how well each model estimates them. Subsequently, we apply the same six models to the full dataset from the TTCM study in Section \ref{sec:TTCM_real_V2}, and compare the results. 

In each case, we compare three methods for estimating the treatment effects: (1) \mbox{suBART}, (2) mvBART (using the aforementioned \textsf{R} package \texttt{skewBART}), and (3) BayesSUR~(\mbox{using} the aforementioned \textsf{R} package \texttt{surbayes}). As before, we use $m=100$ trees for each outcome under suBART and configure mvBART with $m\times d$ trees. Additionally, we do not impose any restrictions on the sets of covariates associated with each response, for any of these methods. This means that all sets of trees for suBART and the single set of multivariate trees for mvBART are \emph{allowed} to form splitting rules using all covariates in Table \ref{tab:baseline_data} and all linear regressions for BayesSUR also share all covariates. Following \citet{wiertsema2019cost}, we do not specify any interaction effects or non-linear terms in the linear predictors for BayesSUR, owing to the difficulty of pre-specifying appropriate functional forms in the presence of a large amount of covariates and the associated challenges in terms of model selection. In any case, \citet{dorie2019automated} found that linear models perform poorly in causal settings even when also including interactions, polynomial terms, and regularisation to avoid overfitting. Conversely, BART-based methods are well-equipped to automatically capture low-order interactions and non-linearities \citep{rockova2020semi,rockova2020posterior}.

In addition, we evaluate each method again with the set of predictors augmented using point estimates of the propensity scores obtained via probit BART. This procedure is inspired by the univariate ps-BART method proposed by \citet{hahn2020bayesian}, which induces a covariate-dependent prior on the regression function. This step 
can substantially reduce bias due to regularisation-induced confounding, thereby making the models less prone to attributing the effect of confounders to the treatment variable, and yield results which are more robust to model misspecification \citep{li2023bayesian}\footnote{There is some conceptual disagreement as to whether one should merely include a point estimate of the propensity score among the covariates, or condition on the propensity score; the latter implies a joint model for the treatment and outcomes.
\citet{hahn2020bayesian} argue convincingly in favour of the former; see in particular Section 8.3. 
For an opposing viewpoint arguing for joint modelling, see for example \citet{li2023bayesian}, Section 5.
Our adoption of the point estimate approach is also well-supported by extensive simulations conducted by \citet{dorie2019automated} which show that ps-BART performs well from a frequentist perspective.}. We first estimate propensity scores for each patient through probit BART, which give the probability that a patient received treatment $1$, conditional on their baseline characteristics, and then add the posterior mean propensity score estimates to the set of predictors $\mathbf{x}_i$ used to estimate the conditional expectations $\mathbb{E}\lbrack c\given t, \mathbf{x}_i\rbrack$ and $\mathbb{E}\lbrack q\given t, \mathbf{x}_i\rbrack$ via the chosen model. 
Thus, we expand the comparison to include what we refer to as ps-suBART, ps-mvBART, and ps-BayesSUR, which are straightforward adaptations of the univariate ps-BART method. Although the use of BART-based propensity scores in conjunction with linear SUR deviates from usual practice in the applied health economics literature, we nonetheless use the same set of propensity scores estimated via probit BART for each method, in order to ensure the comparison is fair in this regard. Expanding the comparison to include versions of each method with and without propensity scores will help to establish the extent to which differences in results are attributable to differences in model specification or due to the inclusion of propensity scores.

Given that there are several unordered categorical covariates with more than two levels in the TTCM dataset --- Table \ref{tab:baseline_data} reports \textit{Medical history}, \textit{Trauma type}, and \textit{Fracture region} --- such variables require particular attention. With probit BART (when estimating the propensity scores) and suBART, we treat such predictors by adapting the method of ordering categories proposed by \citet{breiman1984classification}.
More specifically, whenever such a variable is split on, we order the levels according to the mean values of the partial residuals and then treat the reordered variable in the same way as a continuous variable. This approach makes the tree-growing process more effective by allowing splitting rules which group multiple levels and can reduce the risk of overfitting by allowing for shallower trees. By contrast, both mvBART and BayesSUR accommodate these predictors as dummy variables.

\subsection{Simulation experiment to assess performance in CEA}\label{sec:TTCM_sim}
In this section, we present the results of a simulation experiment based on the TTCM data. We keep the baseline covariates fixed and define generative functions for the potential outcomes and propensity scores. We then simulate treatments and outcomes for each patient and use all six previously discussed models to estimate the treatment effects and $\text{INB}_\lambda$ at representative values of $\lambda=20\mathrm{k}$ and $\lambda=50\mathrm{k}$, where $\lambda=20\mathrm{k}$ corresponds to €20,000. By repeating this procedure $1000$ times, we get an idea of how well each method performs at recovering the true parameters used in the simulation. We summarise the performance in terms of bias, standard deviation (SD), and RMSE, all evaluated using the posterior mean as a point estimate. We further include the coverage rates of 50\% credible intervals (CI coverage), and the average width of 50\% credible intervals (CI width). We stress that the chosen $\lambda$ values are not indicative of the cost of the TTCM intervention, which averages €272 per patient \citep{wiertsema2019cost}. Rather, $\lambda=20\mathrm{k}$, for example, is indicative of a situation in which decision-makers would be willing to pay €20,000 \emph{per additional unit of healthcare-related quality of life}, which is much greater as the treatment effect of $\Delta_q$ is near-zero. In any case, both €20,000 and €50,000 are commonly used thresholds for determining whether an intervention represents value for money in the CEA literature \citep{drummond2015methods,gabrio2019bayesian}.

We first present the generative models used. The ideas are inspired by similar simulation experiments in the literature --- \citet{hill2011bayesian} and \citet{hahn2020bayesian} in particular --- but there are also important novelties. There is first the setting with two correlated outcomes, which is characteristic of the cost-effectiveness context for which the suBART model was developed. We also use somewhat larger variances for the error terms in the simulation (not necessarily in absolute terms, but relative to the variation in $\mathbb{E}\lbrack y_i\given \mathbf{x}_i\rbrack$). In our experience, this is more reflective of real cost-effectiveness datasets. 

Our simulation includes all covariates from Table \ref{tab:baseline_data}. We refer to the covariates in the~same order as listed in the table (so $ x^{(1)}_i$ refers to age,  $x^{(2)}_i$ refers to gender, and so on). Here, all continuous covariates are normalised prior to generating the outcomes. For the sake of clarity, we introduce the additional notation $\mu_c(\mathbf{x}_i) \coloneqq \mathbb{E}\left\lbrack c_i(0)\given \mathbf{x}_i\right\rbrack$ and $\tau_c(\mathbf{x}_i) \coloneqq \mathbb{E}\left\lbrack c_i(1)\given \mathbf{x}_i\right\rbrack - \mathbb{E}\left\lbrack c_i(0)\given \mathbf{x}_i\right\rbrack$, adopt analogous definitions for $q$, and further let $\overline{\mu_q(\mathbf{x}_i)}$ and $\text{sd}(\mu_q(\mathbf{x}_i))$ denote the corresponding sample mean and standard deviation, respectively.\smallskip

\noindent\textbf{Data-generating process for outcomes}:
\begin{align*}
\mu_c(\mathbf{x}_i) &= 2000 + 500x^{(1)}_i -200x^{(3)}_i + 500x^{(10)}_i&\mu_q(\mathbf{x}_i) &= 0.5 + 0.2(x^{(2)}_i + 1)\sin(x^{(1)}_i) \\
\tau_c(\mathbf{x}_i) &= 500&\tau_q(\mathbf{x}_i) &= -0.1 + 0.1\exp(-x^{(11)}_i) \\
c_i &= \mu_c(\mathbf{x}_i) + t_i \times \tau_c(\mathbf{x}_i) + \varepsilon^{(1)}_i  
&q_i &= \mu_q(\mathbf{x}_i) + t_i \times \tau_q(\mathbf{x}_i) + \varepsilon^{(2)}_i
\end{align*}
$$\begin{pmatrix}  \varepsilon_i^{(1)} \\ \varepsilon_i^{(2)} \end{pmatrix} \sim \text{MVN}_2 \left( \begin{pmatrix}  0 \\ 0 \end{pmatrix} , \begin{pmatrix}  500^2 & -0.25 \times 500 \times 0.05 \\ -0.25 \times 500 \times 0.05 & 0.05^2 \end{pmatrix}\right).$$
The generating functions for $c_i$ and $q_i$ each involve subsets of the available covariates which only partially overlap. The coefficients are chosen such that the simulated outcomes are reflective of values typically observed in CEA data. We emphasise that the data-generating process is linear for $c_i$ and non-linear for $q_i$. 
In particular, the treatment effect $\tau_q(\mathbf{x}_i)$ is heterogeneous, given its dependence on $\smash{x_i^{(11)}}$, while $\tau_c(\mathbf{x}_i)$ is constant. 
These settings imply true values for $\Delta_c, \Delta_q$, $\text{INB}_{20\mathrm{k}}$, and $\text{INB}_{50\mathrm{k}}$ of $500$, $0.0399$, $299.24$, and $1489.10$, respectively. We specify a negative correlation of $-0.25$ for the errors to model a common occurrence in CEA studies: patients who report lower well-being also have higher associated costs \citep{glick2014economic}.
The results of additional experiments using correlation values of $0$ and $-0.5$ --- considering only ps-suBART and independent BART models --- are deferred to Appendix \ref{app:indBART}.

We further define the propensity scores $\pi_i \coloneqq \Pr(t_i = 1 \given \mathbf{x}_i)$ as follows:
\begin{equation}
\pi_i = 0.9 \Phi\left(-0.5 + x^{(10)}_i -1.5\left(\frac{\mu_q(\mathbf{x}_i) - \overline{\mu_q(\mathbf{x}_i)}}{\text{sd}(\mu_q(\mathbf{x}_i))}\right)\right) + 0.05.\label{eq:ps_dgp}
\end{equation}
By design, the propensity scores are strongly dependent on $\mu_q(\mathbf{x}_i)$, which is the expected health outcome under treatment $0$. This is an example of \emph{targeted selection} \citep{hahn2020bayesian}, which induces strong confounding with respect to the outcome $q_i$, and reflects how treatments are assigned in reality: patients expected to have a bad outcome under treatment $0$ are more likely to be given treatment $1$ by medical professionals. Further confounding with respect to $c_i$ arises by including $\smash{x^{(10)}_i}$, on which $\mu_c(\mathbf{x}_i)$ also depends, in Equation \eqref{eq:ps_dgp}. The coefficients are chosen to ensure roughly balanced sample sizes for the two treatment groups.

The performance metrics for $\Delta_c$, $\Delta_q$, and $\text{INB}_\lambda$ (under all six models, across $1000$ simulation replications) are presented in Table \ref{tab:results_Delta_c}, Table \ref{tab:results_Delta_q}, and Table \ref{tab:results_INB}, respectively. We see that including propensity scores generally leads to smaller bias and improved coverage rates, while slightly increasing the SD of the estimates. Including propensity scores also lowers the RMSE in all but one case while increasing the average CI width for all methods. Across all estimands, the intervals from BayesSUR are wider than those from mvBART, which in turn are wider than those from suBART, with or without propensity scores. For $\Delta_c$, the BayesSUR models have the smallest bias, which is expected given the linearity of $\mu_c(\mathbf{x}_i)$~in the data-generating process. For $\Delta_q$, ps-suBART exhibits the lowest bias by far, and also the lowest RMSE by a reasonable margin. BayesSUR exhibits substantial bias for $\Delta_q$, although this is mitigated through the inclusion of propensity scores. For $\text{INB}_\lambda$, ps-suBART's superiority in terms of bias, RMSE, and coverage notably holds at both $\lambda=20\mathrm{k}$ and $\lambda=50\mathrm{k}$.

The results also show that the non-linear BART models perform only slightly worse than linear BayesSUR when the conditional expectation is actually linear ($\Delta_c$ in Table \ref{tab:results_Delta_c}), and much better when it is non-linear ($\Delta_q$ in Table \ref{tab:results_Delta_q}). The subpar performance of SUR in the second case can be improved significantly by the inclusion of propensity scores, which seem to absorb much of the non-linearity, although we emphasise that this is a somewhat unnatural model which is unlikely to be used by anyone in practice; if we estimate propensity scores through a non-linear model, it would seem appropriate to also use a non-linear outcome model. Comparing ps-suBART and ps-mvBART specifically, we see a better performance from ps-suBART across all estimands and metrics. Overall, ps-suBART exhibits the smallest bias and RMSE, while also being competitive in terms of interval width and coverage.\vspace{-1em} 
\begin{table}[H]
\caption{Performance metrics for $\Delta_c$, evaluated over 1000 simulation replications. \label{tab:results_Delta_c}}
\setlength{\tabcolsep}{4.9pt}
\begin{tabular}{lrrrrrr}
\textbf{Model} & Bias & SD & RMSE & CI coverage & CI width \\
\hline
\textbf{ps-suBART} & $-60$& $118$& $132$& $0.423$& $159$ \\
\textbf{suBART} & $-101$& $113$& $151$& $0.358$& $150$ \\
\textbf{ps-mvBART} & $-93$& $109$& $144$& $0.404$& $160$ \\
\textbf{mvBART} & $-110$& $107$& $154$& $0.343$& $154$ \\
\textbf{ps-BayesSUR} & $14$& $141$& $142$& $0.493$& $189$ \\
\textbf{BayesSUR} & $4$& $125$& $125$& $0.478$& $167$
\end{tabular}
\end{table}
\vspace{-1em}
\begin{table}[H]
\caption{Performance metrics for $\Delta_q$, evaluated over 1000 simulation replications. \label{tab:results_Delta_q}}
\setlength{\tabcolsep}{4.9pt}
\begin{tabular}{lrrrrrr}
\textbf{Model} & Bias & SD & RMSE & CI coverage & CI width \\
\hline
\textbf{ps-suBART} & $-0.0041$& $0.0160$& $0.0166$& $0.475$& $0.0207$ \\
\textbf{suBART} & $-0.0115$& $0.0155$& $0.0193$& $0.348$& $0.0191$\\
\textbf{ps-mvBART} & $-0.0088$& $0.0161$& $0.0184$& $0.522$& $0.0251$\\
\textbf{mvBART} & $-0.0162$& $0.0157$& $0.0225$& $0.371$& $0.0236$\\
\textbf{ps-BayesSUR} & $-0.0056$& $0.0230$& $0.0237$& $0.592$& $0.0385$\\
\textbf{BayesSUR} & $-0.0574$& $0.0227$& $0.0617$& $0.039$& $0.0360$
\end{tabular}
\end{table}

\begin{table}[H]
\caption{Performance metrics for $\text{INB}_\lambda$ with $\lambda = 20\mathrm{k}$ and $\lambda=50\mathrm{k}$, evaluated over 1000 simulation replications. \label{tab:results_INB}}
\setlength{\tabcolsep}{3.21pt}
\begin{tabular}{lrrrrrrrrrrr}
\multirow{2}{*}{\textbf{Model}} &\multicolumn{5}{c}{$\textbf{INB}_{\mathbf{20k}}$}&\multicolumn{5}{c}{$\textbf{INB}_{\mathbf{50k}}$}\\
\cline{2-11}
 & Bias & SD & RMSE & CI coverage & CI width & Bias & SD & RMSE & CI coverage & CI width \\
\hline
\textbf{ps-suBART} &$-22$& $353$& $354$& $0.466$& $467$& $-146$ & $823$& $836$& $0.473$& $1071$ \\
\textbf{suBART} & $-130$& $342$& $366$& $0.452$& $433$& $-475$& $795$& $926$& $0.392$& $991$\\
\textbf{ps-mvBART} & $-83$& $349$& $359$& $0.560$& $543$& $-349$& $823$& $894$ & $0.527$& $1282$\\
\textbf{mvBART} & $-212$& $341$& $401$& $0.482$& $514$& $-697$& $801$& $1062$& $0.422$& $1208$\\
\textbf{ps-BayesSUR} & $-127$& $501$& $517$& $0.586$& $815$& $-296$& $1180$& $1217$& $0.597$& $1957$\\
\textbf{BayesSUR} & $-1151$& $483$& $1249$& $0.053$& $757$& $-2872$& $1155$& $3096$& $0.042$& $1824$
\end{tabular}
\end{table}

\subsection{Results of the TTCM data analysis}\label{sec:TTCM_real_V2}
\begin{figure}[H]
    \centering
    \includegraphics[width=\textwidth]{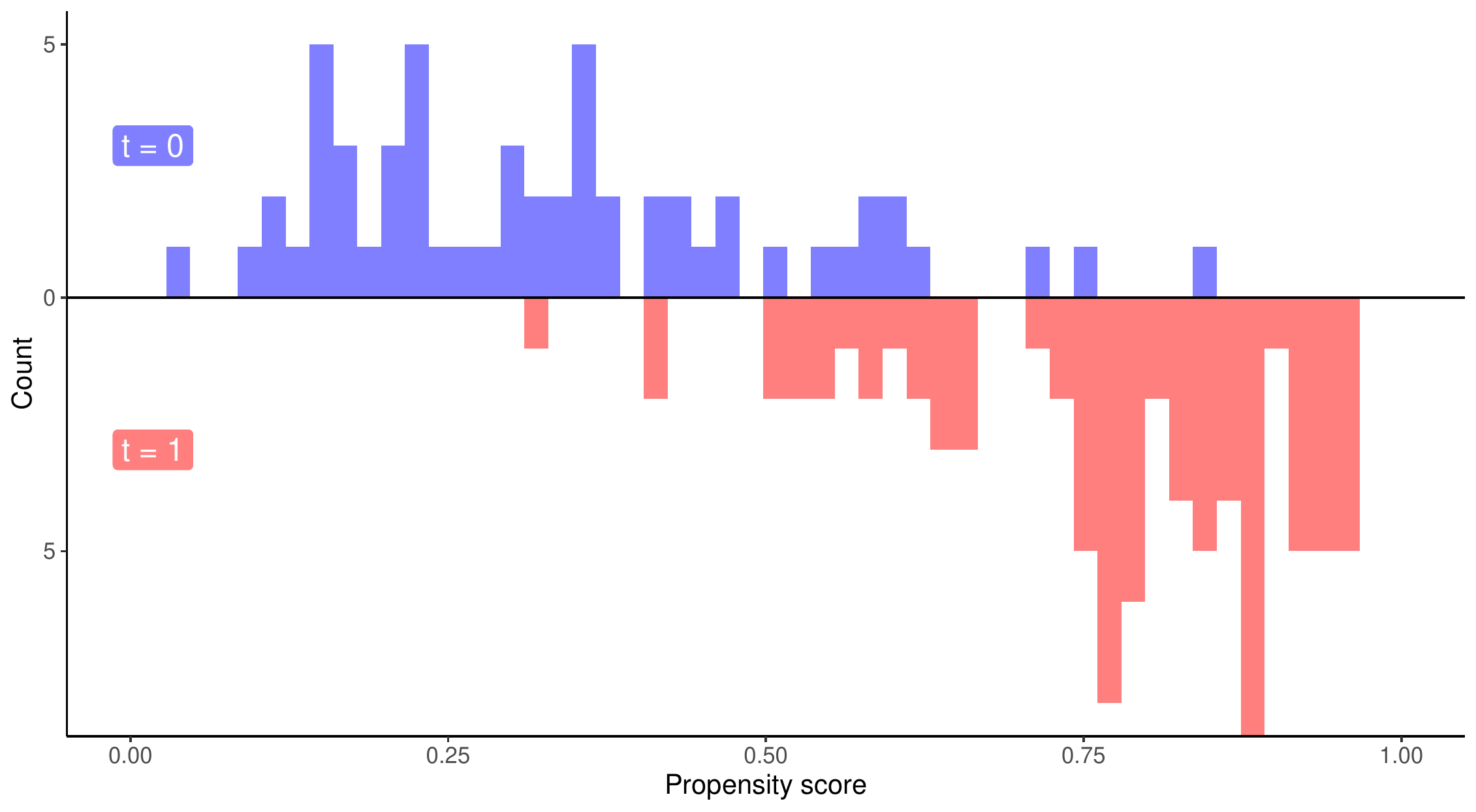}
    \caption{Propensity scores by treatment arm, estimated via probit BART, where $t=0$ corresponds to the control group.}
    \label{fig:PSplot}
\end{figure}

\begin{figure}[H]
    \centering
    \includegraphics[width=\textwidth]{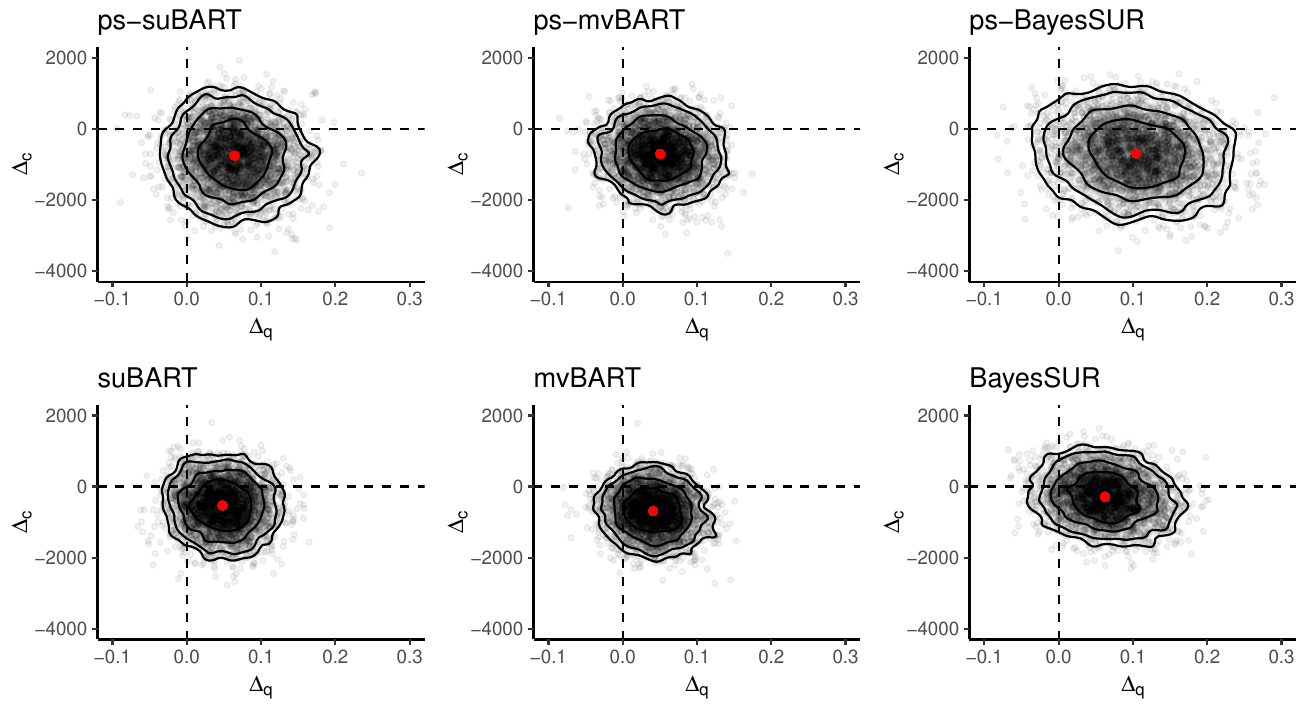}
    \caption{CEPs showing highest density regions of kernel density estimates of the posterior distributions of $\Delta_c$ and $\Delta_q$ according to each model, with and without propensity scores. Posterior means
    are indicated by red dots, individual draws of $\Delta_c$ and $\Delta_q$ are shown via grey points, and the contour lines correspond to probability levels of $0.5$, $0.75$, $0.9$, and $0.95$.} 
    \label{fig:CEplane_model_comparison}
\end{figure}

Figure \ref{fig:PSplot} shows the distribution of the estimated propensity scores. Although there is some overlap, it is evident that the treatment groups are quite imbalanced with respect to their baseline characteristics and that some form of covariate adjustment is necessary to avoid biased results. As already described, we reevaluate each model with estimated propensity scores included as an additional covariate. In Figure \ref{fig:CEplane_model_comparison}, we show highest density regions of kernel density estimates of the posterior distributions of $\Delta_c$ and $\Delta_q$ --- obtained through suBART, mvBART, and BayesSUR, as well as their counterpart models which also incorporate the estimated propensity scores as additional predictors --- in the form of a cost-effectiveness plane 
(CEP; see \citet{gabrio2019bayesian} for more details). 
In general, for both $\Delta_c$ and $\Delta_q$, the centers of the distributions are shifted further away from $0$ and posterior uncertainty is greater when the propensity scores are incorporated. These differences are most pronounced between ps-BayesSUR and BayesSUR.
A notable distinction between suBART and BayesSUR is that the latter appears to be more uncertain about $\Delta_q$.
The same applies to the comparison between ps-suBART and ps-BayesSUR.
\vspace{-1em}%
In Table \ref{tab:inb_ci}, we provide summary statistics for $\Delta_c$, $\Delta_q$, and the INB at the same representative values of $\lambda = 20\mathrm{k}$ and $\lambda=50\mathrm{k}$ adopted in Section \ref{sec:TTCM_sim}. 
We see some considerable differences between ps-suBART and the other methods. This suggests that there may be strong non-linear functional relationships between the covariates and the outcomes (such that suBART benefits from relaxing the linearity assumption of BayesSUR, without requiring pre-specification of the functional forms) and that those relationships may differ for the two outcomes $c_i$ and $q_i$ (such that suBART benefits from relaxing the mvBART assumption of a common tree structure). With the exception of ps-BayesSUR, the INB values from all models are much lower than those from ps-suBART.  For all models, the estimated treatment effects are consistently larger in absolute value for each method when propensity scores are included. The same is true for $\text{INB}_\lambda$ at both $\lambda$ values. 
Surprisingly, for both $\Delta_q$ and $\text{INB}_\lambda$, ps-BayesSUR produces far higher estimates than all other models, including ps-suBART.
We recall in this context that in our simulation study, the standard error of the $\Delta_q$ estimates was the highest by far for ps-BayesSUR. Hence this estimate tends to be rather volatile, and we are inclined to put more faith in the estimates from the other models.
\begin{table}[H]
\caption{Posterior means with 95\% credible intervals for $\Delta_c$, $\Delta_q$, and $\text{INB}_\lambda$ at representative values of $\lambda=20\mathrm{k}$ and $\lambda=50\mathrm{k}$ for each model, with and without the propensity scores.\label{tab:inb_ci}}
\setlength{\tabcolsep}{3.33pt}
\begin{tabular}{lrrrrrrrr}
\multirow{2}{*}{\textbf{Model}} &\multicolumn{6}{c}{\textbf{Mean and 95\% CI}} \\ \cline{2-9}
& \multicolumn{2}{c}{$\Delta_c$}  & \multicolumn{2}{c}{$\Delta_q$} & \multicolumn{2}{c}{$\text{INB}_{20\mathrm{k}}$}& \multicolumn{2}{c}{$\text{INB}_{50\mathrm{k}}$}\\
\hline
\textbf{ps-suBART}   & $-759$&$\lbrack-2321, 745\rbrack$ & $ 0.064$&  $\lbrack-0.013, 0.147\rbrack$ & $2045$&$\lbrack-217, 4317\rbrack$  &$3975$&$\lbrack-250, 8519\rbrack$  \\
\textbf{suBART}   & $-537$&$\lbrack-1752, 619\rbrack$ & $0.048$&$\lbrack-0.016, 0.114\rbrack$ & $1500$&$\lbrack-338, 3320\rbrack$  &$2943$&$\lbrack-627, 6497\rbrack$  \\
\textbf{ps-mvBART}   & $-706$&$\lbrack-1938, 518\rbrack$ & $0.051$&$\lbrack-0.022, 0.122\rbrack$ & $1717$&$\lbrack-245, 3631\rbrack$ &$3234$&$\lbrack-647, 7054\rbrack$  \\
\textbf{mvBART}   & $-692 $&$\lbrack-1781, 359\rbrack$ & $0.041$&$\lbrack-0.021, 0.104\rbrack$ & $1511$&$\lbrack-162, 3235\rbrack$ &$2740$&$\lbrack-588, 6235\rbrack$  \\
\textbf{ps-BayesSUR} & $-698$&$\lbrack-2253, 795\rbrack$ & $0.104$&$\lbrack-0.007, 0.214\rbrack$  & $2783$&$\lbrack-55, 5646\rbrack$ &$5912$&$\lbrack-80, 11735\rbrack$  \\
\textbf{BayesSUR} & $-286$&$\lbrack-1400, 827\rbrack$ & $0.063$&$\lbrack-0.016, 0.145\rbrack$  & $1537$&$\lbrack-538, 3723\rbrack$ &$3413$&$\lbrack-901, 7959\rbrack$
\end{tabular}
\end{table}%

\begin{figure}[H]
    \centering
    \includegraphics[width=\textwidth]{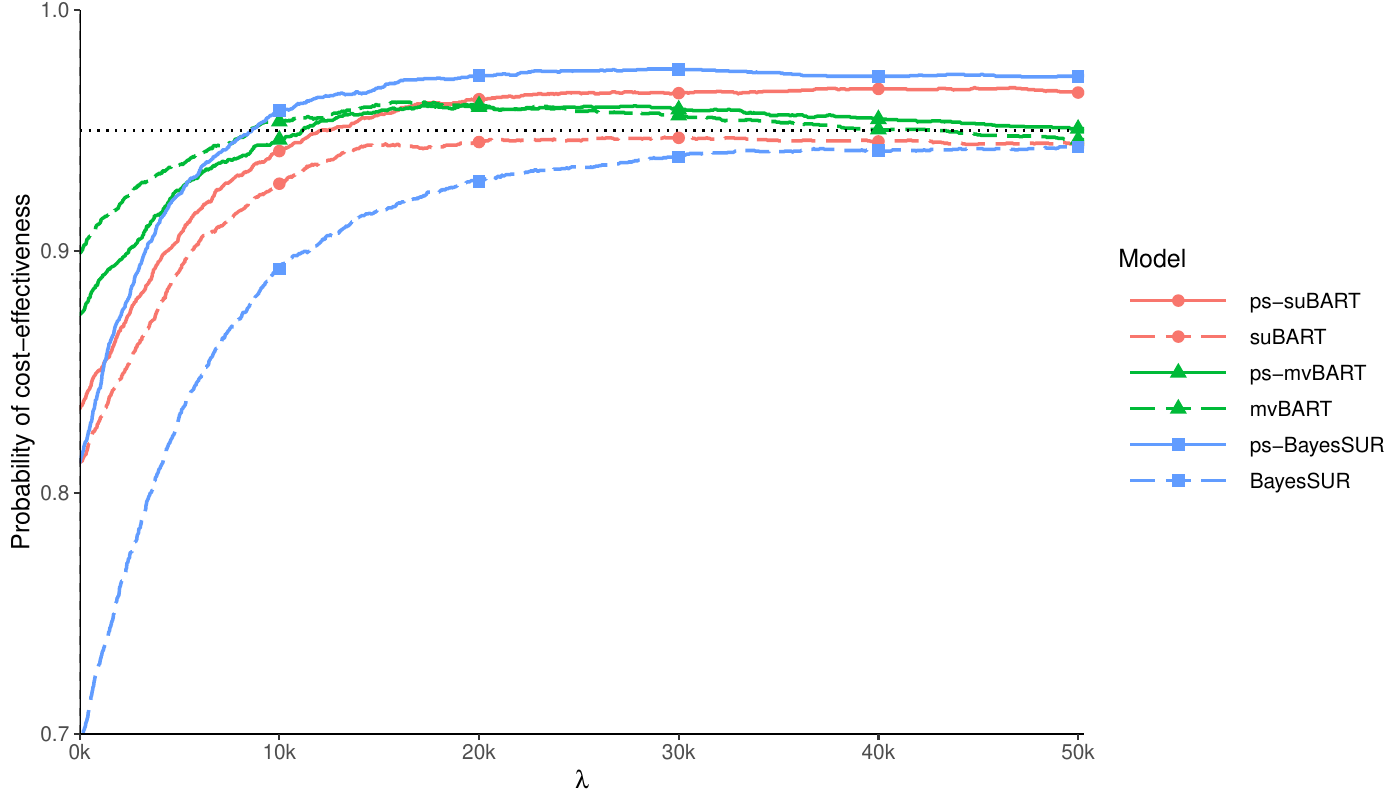}
    \caption{CEACs showing probabilities of cost-effectiveness as a function of $\lambda$ for each model: suBART (red, circles), mvBART (green, triangles), and BayesSUR (blue, squares). Each model is shown with propensity scores (solid lines) and without (dashed lines), with a horizontal dotted line at the $0.95$ reference level.} 
    \label{fig:CEAC_comparison}
\end{figure}

Rather than relying on a single $\lambda$ value, we also show the probability of cost-effectiveness as a function of the willingness-to-pay $\lambda$ in Figure \ref{fig:CEAC_comparison}. This plot, called a cost-effectiveness acceptability curve \citep[CEAC;][]{lothgren2000definition}, is a highly-important tool in guiding the decision of which medical intervention to implement. The probabilities are simply estimated by counting all posterior draws for which $\text{INB}_{\lambda} > 0$ and dividing this count by the total number of posterior draws. 

For suBART and BayesSUR, inclusion of the propensity scores leads to a uniformly higher probability of cost-effectiveness. Hence, these analyses give greater certainty to the  conclusion that the treatment is cost-effective. The greatest difference in this regard is observed for BayesSUR. Although mvBART does not exhibit the same effect, all three models involving propensity scores give similar results: if $\lambda$ exceeds a value of around $10000$, then the probability of cost-effectiveness exceeds $0.95$, a commonly used benchmark in CEAs.
However, we stress that the superiority of any one method cannot be established from these estimated CEACs, as the true $\text{INB}_{\lambda}$ is unknown for all values of $\lambda$. In light of the simulation study based on the TTCM data presented in Section \ref{sec:TTCM_sim}, which demonstrated the superiority of ps-suBART in recovering the true INB, we tend to put most trust in the estimates from this model.

In Appendix \ref{app:extraTTCM}, we sketch two additional analyses of the TTCM data, which also showcase some additional features of our computational implementation.
\begin{itemize}
    \item Appendix \ref{app:CATE} shows how we can estimate treatment effects for individual patients in the TTCM data, and how they vary depending on the baseline covariates.
    \item Appendix \ref{app:var_imp} shows variable importance plots, which were already used by \citet{chipman2010bart} and can be created for suBART in much the same way. 
\end{itemize}

\section{Discussion}\label{sec:conclusion}

In this paper, we introduce the suBART model for multivariate outcomes both as means of accounting for non-linearities and interactions in the seemingly unrelated regression framework and as a means of addressing the key limitation of existing multivariate BART approaches which assume a single set of trees, such that the entire response vector is partitioned in the same way by the splitting rules in the ensemble. By modelling each component of the outcome via univariate BART, suBART captures non-linearities in the relationship between the response and predictors while allowing for and detecting the different subsets of covariates --- along with the interactions between them --- associated with each response without enforcing common tree structures. Another defining feature of suBART is that it allows each individual outcome to be associated with a distinct ensemble of trees while handling dependencies between the outcomes through correlated error terms. In this respect, inference on $\boldsymbol{\Sigma}$ is key. Running independent BART models implicitly assumes that $\boldsymbol{\Sigma}$ is diagonal, which has been shown to harm performance in our analyses of the TTCM data (see Appendix \ref{app:indBART}). 

We further develop a probit version of the suBART model to handle multivariate binary outcomes. The effectiveness of suBART and probit suBART is demonstrated through extensive simulation experiments. In the simulation experiments in Section \ref{sec:simstudies}, it is shown that the model adequately captures non-linear responses, accurately estimates the covariance structure for multivariate responses, and generally outperforms its main competitors --- including other tree-based alternatives and the Bayesian linear SUR --- from the points of view of both predictive accuracy and uncertainty quantification. Notably, suBART is consistently superior to the strategy of applying the standard BART model independently to each response, which demonstrates the benefits of modelling the covariance between multivariate response under our suBART framework. Furthermore, the results show that the model exhibits enhanced flexibility compared to its direct multivariate counterpart, mvBART. This flexibility stems from suBART permitting variation in splitting rules across each response, by allowing the trees for each outcome component to differ rather than imposing common tree structures. This enables a more accurate representation, especially when different outcome components depend on distinct sets of predictors. As a cautionary note, we recall that BayesSUR exhibited competitive performance 
when estimating conditional expectations which merely depend linearly on the predictors.
In situations where linearity can be reasonably assumed for every outcome, BayesSUR would remain a viable alternative to suBART.

The main focus of this paper is the application of suBART within the context of cost-effectiveness analysis, a setting in healthcare research where it is of interest to jointly estimate the healthcare costs and the health-related quality of life associated with two or more treatment options. Throughout Section \ref{sec:case_study}, we find some remarkable differences, depending on the particular method used for analysing the TTCM data. This applies both to the additional simulation experiment in Section \ref{sec:TTCM_sim} and the main application in Section \ref{sec:TTCM_real_V2}. It is of course expected to see large differences between linear SUR and the two BART-based models, given the very different model assumptions. The significant differences between suBART and mvBART are arguably even more interesting. To reiterate, suBART assigns each outcome its own tree ensemble, while the trees for all outcomes are the same for mvBART. The large differences in results suggest to us that the assumption of a shared tree structure is a significant one, which can strongly impact the results. Unlike mvBART, suBART can accommodate data where the dependence on $\mathbf{X}$ is very different for different components of the outcome vector. In the CEA context, this applies particularly in situations where factors which govern the costs are unrelated to the quality of life, and \emph{vice versa}. Such situations do occur in practise: when investigating the effect of total knee replacement, \citet{dakin2012rationing} found that the patients' sex was a strong predictor of healthcare costs yet had no measurable relationship with quality of life, while the exact opposite was true for the patients' age. On the other hand, a shared tree structure may lead to estimates which are more precise, in the sense of having a smaller variance, since there are more data available to inform the tree structure. \citet{um2023bayesian} claim this as an advantage of their method. Our results are somewhat inconclusive in this respect: when estimating the MATE on the costs, mvBART exhibits slightly smaller standard deviation than suBART, while both methods are similar in this respect for the quality of life and the INB. 

According to our simulation experiment in Section \ref{sec:TTCM_sim}, the ps-suBART model demonstrates superior performance over competing methodologies. Recall that this extension includes propensity scores estimated via BART as an additional predictor in suBART. As suBART in particular benefits greatly from the incorporation of propensity scores, and given that this experiment is explicitly based on the TTCM data, we are inclined to believe that the ps-suBART results are the most reliable for the real TTCM application. Overall, we consider the ps-suBART model to be a natural adaptation of the univariate BART approach and expect it to be a very useful tool in the analysis of cost-effectiveness data.

Despite the extensive array of comparisons and scenarios and the additional insights gleaned by suBART in the CEA setting, there remains ample opportunity for further exploration of various extensions to suBART. We delineate some of these possibilities below in light of limitations identified in the simulation experiments and real data application.
\begin{itemize}
    \item In the context of univariate outcomes, various extensions to the basic BART model have been proposed in the literature . It may be of interest to extend suBART in similar ways, for other applications beyond the TTCM data considered here. For example, methods designed to address the inherent discontinuity and lack of smoothness of regression trees based on sums of piecewise constants \citep[e.g.][]{linero2018bayesian, prado2021bayesian, maia2024gp} could be adapted to suBART. Another extension, which we did not adopt here due to the relatively small number of covariates in the TTCM data, is the sparsity-inducing Dirichlet prior on the splitting probabilities \citep{linero2018highdim}, which could allow suBART to be used in high-dimensional settings. However, it remains to be seen how such extensions would perform in causal settings. Furthermore, embedding these and other BART extensions in the suBART framework could present some computational challenges.
    \item 
    Our simulation experiments comprehensively demonstrated suBART's superiority in capturing non-linear relationships. However, BayesSUR was slightly preferable when the response was generated by a simple linear function. In light of this, we can also envision a hybrid approach, whereby some outcomes would be modelled nonparametrically via BART, and others parametrically via linear models. We stress that this would differ from existing approaches referred to in the literature as semi-parametric BART \citep[e.g.][]{zeldow2019semiparametric,deshpande2020,prado2021accounting}, where each response is modelled by a combination of parametric and non-parametric forms. While such semi-parametric extensions of suBART would also be of interest, we are describing here a model which would instead assign either a BART model or a linear predictor to each outcome. In situations where linearity holds for one outcome but not for others (such as the one we present in Section \ref{sec:TTCM_sim}), such a mixed linear-BART model may be preferable to both suBART and linear SUR. 
    In practical settings, a hybrid model of this kind would require strong beliefs \emph{a priori} about which outcomes should explicitly be specified as having a simpler, linear functional form. While this would reduce the computational burden, the need for pre-specification could present challenges to users of a practical implementation. At present, suBART is automatic insofar as it obviates the need for such pre-specifications, albeit at the expense of minor efficiency losses when the true conditional expectation is linear.
    \item In the classic SUR model context, accounting for heteroscedasticity has been a common challenge in the literature \citep{afolayan2018efficiency}. As suBART represents an effective alternative to the traditional linear SUR, the extension proposed by \citet{pratola2020heteroscedastic} to accommodate heteroscedasticity within BART could be adapted to the suBART setting to address cases where the homoscedasticity assumption is invalid. However, it is important to note that the approach of \citet{pratola2020heteroscedastic} has yet to be extended beyond scalar variance estimation.
    \item The proposed suBART accommodates two types of multivariate outcomes: all continuous and all binary. We stress however that this flexibility is exclusive and does not apply to both types simultaneously. Previous works in the literature, such as those by \citet{flexible2014papa}, \citet{zhang2015bayesprobit}, and \citet{bayesianmix20216tony}, have adapted a Bayesian framework for handling responses of mixed type. These approaches could potentially be adapted to the suBART framework, providing greater generalisation of the approach, particularly given that trading off treatment costs against binary outcomes (e.g., cancer remission) is often of interest in other healthcare applications.
    \item The ps-suBART model we used in our analysis of the TTCM data builds on the ps-BART model of \citet{hahn2020bayesian}, who also propose another related method, the Bayesian causal forest (BCF). The authors and their discussants found the performance of ps-BART and BCF to be similar overall, with each method improving on the other in some settings. One of the advantages of BCF is the ability to directly specify a prior on the amount of treatment effect heterogeneity, while this prior is left implicit in ps-BART. The drawback of this flexibility is that prior specification is more complex in BCF, and it is harder to find reasonable default choices which are appropriate in a variety of settings. Additionally, the computational demands of BCF are greater than those of standard BART. Both of these issues would become even more challenging in the SUR setting with multivariate outcomes. However, given that a multivariate generalisation of BCF has recently been proposed by \citet{mcjames2024bayesian}, which is analogous to the multivariate generalisation of BART developed by \citet{um2023bayesian}, there remains scope for embedding BCF in the seemingly unrelated framework as an alternative multivariate approach for conducting cost-effectiveness analyses.
    \item CEAs are sometimes conducted with more than two treatment options of interest. Conceptually this is similar to the two-treatments setting presented in the TTCM application. For details, we refer interested readers to \citet{drummond2015methods}, Section 4.4. The ps-suBART method could be easily extended to settings with more than two treatments, by estimating propensity scores using multinomial probit BART \citep{kindo2016multinomial}. With more than one treatment effect for each outcome, and more than one INB, each corresponding to a pairwise comparison of two treatments, the treatment which has a non-negative INB in all comparisons would be considered cost-effective.
    \item When presenting the TTCM data, we noted that the original dataset had missing responses for survey questions related to the outcome variables. We bypassed this by working instead with an artificial complete dataset. Ideally, we would prefer to keep the missing values and incorporate the imputation directly into the posterior computation, in order to get coherent estimates of posterior uncertainty. Assuming that the observations are missing at random (i.e., the probability of missingness is independent of the true values of the missing observations, conditional on the observed data), it is straightforward to incorporate imputation into our Gibbs sampler: we would simply draw the missing $y^{(j)}_i$ values from their conditional distribution, given in Equation \eqref{eq:marginal_normal}. Imputing outcomes with the probit suBART model could be done in a similar manner. However, imputation of missing outcomes for the present application would be further complicated by the fact that the outcomes were calculated after imputing the constituent survey responses by \citet{wiertsema2019cost}. Imputing missing \emph{covariate} values, on the other hand, is a totally different matter: our models are formulated conditionally on $\mathbf{X}$, and hence impose none of the distributional assumptions on $\mathbf{X}$ that would be required for imputation tasks. That being said, missing covariate values could be handled by simply adapting the approach of \citet{kapelner2016bartmachine}.
\end{itemize}
We hope to incorporate these advancements into our forthcoming research plans and anticipate suBART's adoption in other CEA settings to inspire further developments.

\begin{acks}[Acknowledgments]
\small
Mateus Maia's work was supported by Science Foundation Ireland Career Development Award grant number 17/CDA/4695 and SFI research centre award 12/RC/2289P2. Andrew Parnell's work was supported by: a Science Foundation Ireland Career Development Award (17/CDA/4695); an investigator award (16/IA/4520); a Marine Research Programme funded by the Irish Government, co-financed by the European Regional Development Fund (Grant-Aid Agreement No. PBA/CC/18/01); European Union's Horizon 2020 research and innovation programme under grant agreement No. 818144; SFI Centre for Research Training 18CRT/6049, and SFI Research Centre awards 16/RC/3872 and 12/RC/2289P2. For the purpose of Open Access, the author has applied a CC BY public copyright licence to any Author Accepted Manuscript version arising from this submission.

We thank the authors of \citet{wiertsema2019cost} for making part of the TTCM data available to us, and allowing us to present the results of our analysis in this paper. We are also grateful to Benedikt Schwab and Linus N{\"{u}}sing for alerting us to minor bugs in the computational implementation on our \texttt{subart} GitHub repository, following an earlier draft of this article.
\end{acks}

\interlinepenalty=1000
\bibliographystyle{ba}
\bibliography{sample}



\clearpage
\setcounter{subsection}{0}
\renewcommand\thesubsection{\Alph{subsection}}
\renewcommand\thefigure{\thesubsection.\arabic{figure}}
\renewcommand\thetable{\thesubsection.\arabic{table}}

\section*{Appendices}\label{sec:appendix}

\subsection{Comparison to independent BART models}
\setcounter{figure}{0}
\setcounter{table}{0}
\label{app:indBART}
 
In Section \ref{sec:application}, we discussed the importance of jointly modelling $c_i$ and $q_i$, so as not to enforce the unrealistic assumption that the treatment effects $\Delta_c$ and $\Delta_q$ are independent. This raises two potential questions; namely, what do we lose if we use suBART but the errors are independent, and what do we lose if the errors are not independent, but we assume that they are?

We address both questions by performing additional simulation experiments. We repeat the experiment based on the TTCM data from Section \ref{sec:TTCM_sim}, using two models to analyse the simulated datasets, for simplicity: ps-suBART, and independent BART models equipped with propensity scores as an additional covariate, with analogous priors as we do for the suBART model (i.e. independent half-$t$ priors on the standard deviations of the error terms). We refer to the latter model as ps-indBART. For the sake of fairness of the comparison, we use our own \texttt{subart} implementation for both models, with the code modified to accommodate the independent BART models as appropriate. 

We use three different choices for the correlation coefficient $\rho\colon 0, -0.25$, and $-0.5$, where the first value corresponds to the case of independent errors and the second value of $-0.25$ has already been considered for ps-suBART in Section \ref{sec:TTCM_sim}. 
From Table \ref{tab:indbart0}, we observe that the two models produce essentially the same results when $\rho = 0$. Hence, we do not lose anything from using suBART when the true correlation is zero. On the other hand, when the correlation is non-zero, as in Table \ref{tab:indbart025} and Table \ref{tab:indbart05}, ps-suBART produces smaller bias and standard deviations, as well as improved coverage rates for all estimands. 
The differences in the results become more pronounced in each successive table, as the strength of the correlation increases.
\begin{table}[H]
\caption{Performance metrics with $\rho = 0$.\label{tab:indbart0}}
\setlength{\tabcolsep}{3.75pt}
\begin{tabular}{ll|rrrrr}
\textbf{Estimand} & \textbf{Model} & Bias & SD & RMSE & CI coverage & CI width \\
\hline
\multirow{2}{*}{$\Delta_c$} & \textbf{ps-suBART} & $-71$& $117$& $137$& $0.432$& $159$ \\
&\textbf{ps-indBART}& $-71$& $117$& $137$& $0.333$& $159$ \\
\multirow{2}{*}{$\Delta_q$} & \textbf{ps-suBART} & $-0.0068$& $0.0168$& $0.0181$& $0.436$& $0.0208$ \\
& \textbf{ps-indBART} & $-0.0167$& $0.0169$& $0.0181$& $0.419$& $0.0208$\\
\multirow{2}{*}{$\text{INB}_{\mathrm{20k}}$} & \textbf{ps-suBART} & $-65$& $353$& $360$& $0.468$& $445$ \\
& \textbf{ps-indBART} & $-63$& $355$& $361$& $0.460$& $446$ \\
\multirow{2}{*}{$\text{INB}_{\mathrm{50k}}$} & \textbf{ps-suBART} & $-269$ & $845$& $887$& $0.456$& $1051$ \\
& \textbf{ps-indBART} & $-265$& $849$& $889$& $0.426$& $1051$
\end{tabular}
\end{table}
\begin{table}[H]
\caption{Performance metrics with $\rho = -0.25$.\label{tab:indbart025}}
\setlength{\tabcolsep}{3.75pt}
\begin{tabular}{ll|rrrrr}
\textbf{Estimand} & \textbf{Model} & Bias & SD & RMSE & CI coverage & CI width \\
\hline
\multirow{2}{*}{$\Delta_c$} & \textbf{ps-suBART} & $-60$& $118$& $132$& $0.423$& $159$ \\
&\textbf{ps-indBART} & $-66$& $118$& $135$& $0.425$& $159$ \\
\multirow{2}{*}{$\Delta_q$} & \textbf{ps-suBART} & $-0.0041$& $0.0160$& $0.0166$& $0.475$& $0.0207$ \\
& \textbf{ps-indBART} &$-0.0060$& $0.0162$& $0.0173$& $0.450$& $0.0208$\\
\multirow{2}{*}{$\text{INB}_{\mathrm{20k}}$} & \textbf{ps-suBART} &$-22$& $353$& $354$& $0.466$& $467$ \\
& \textbf{ps-indBART} & $-53$& $362$& $366$& $0.452$& $447$ \\
\multirow{2}{*}{$\text{INB}_{\mathrm{50k}}$} & \textbf{ps-suBART} & $-146$ & $823$& $836$& $0.473$& $1071$ \\
& \textbf{ps-indBART} &$-232$& $837$& $869$& $0.447$& $1054$
\end{tabular}
\end{table}

\begin{table}[H]
\caption{Performance metrics with $\rho = -0.5$.\label{tab:indbart05}}
\setlength{\tabcolsep}{3.75pt}
\begin{tabular}{ll|rrrrr}
\textbf{Estimand} & \textbf{Model} & Bias & SD & RMSE & CI coverage & CI width \\
\hline
\multirow{2}{*}{$\Delta_c$} & \textbf{ps-suBART} & $-54$& $116$& $128$& $0.447$& $158$ \\
 & \textbf{ps-indBART} & $-67$& $116$& $134$& $0.423$& $159$ \\
\multirow{2}{*}{$\Delta_q$} & \textbf{ps-suBART} &$-0.0017$& $0.0159$& $0.0160$& $0.464$& $0.0200$ \\
 & \textbf{ps-indBART} & $-0.0057$& $0.0165$& $0.0174$& $0.433$& $0.0208$\\
\multirow{2}{*}{$\text{INB}_{\mathrm{20k}}$} & \textbf{ps-suBART} & $20$& $379$& $379$& $0.440$& $479$ \\
& \textbf{ps-indBART} & $-47$& $395$& $398$& $0.412$& $446$ \\
\multirow{2}{*}{$\text{INB}_{\mathrm{50k}}$} & \textbf{ps-suBART} & $-31$ & $847$& $847$& $0.451$& $1067$ \\
& \textbf{ps-indBART} & $-217$& $881$& $907$& $0.435$& $1053$
\end{tabular}
\end{table}
In light of these results, we note that the presence of the covariance term in Equation \eqref{eq:Pr_Inb}, which gives the formula for $\Pr\left(\text{INB}_{\lambda} > 0\right)$,  shows how the independence assumption is likely to yield an inappropriate posterior value of $\Pr\left(\text{INB}_{\lambda} > 0\right)$. In the same way, separate BART models are likely to give inaccurate posterior quantiles for $\text{INB}_{\lambda}$ itself. This might be a cause of the lower coverage rates from ps-indBART in Tables \ref{tab:indbart025} and \ref{tab:indbart05}.
In this context, we draw attention to the average width of the credible intervals for $\text{INB}_{\mathbf{20k}}$ and $\text{INB}_{\mathbf{50k}}$: when $\rho$ is negative, the intervals from suBART are consistently wider than those from indBART. This is indicative of a larger posterior variance, and theoretically consistent with our specification of negative correlations. 
To see this, observe that $\text{Var}( \text{INB}_{\lambda}) = \lambda^2 \text{Var}(\Delta_q) + \text{Var}(\Delta_c) - 2 \lambda \text{Cov}\lbrack \Delta_q, \Delta_c\rbrack$.
A negative residual correlation between the observed $q_i$ and $c_i$ values, conditional on the covariates, is likely to cause the posterior covariance $\text{Cov}\lbrack \Delta_q, \Delta_c\rbrack$ to be negative too. Hence, the posterior variance of $\text{INB}_{\lambda}$ will increase, along with the width of the credible intervals. 

\setcounter{figure}{0}
\setcounter{table}{0}
\subsection{Convergence assessments}\label{app:convergence}

Throughout this paper, we adopted the consistent setting of 5000 MCMC iterations, with 1000 iterations discarded as burn-in, for our comparative experiments and application with continuous responses. Convergence appears satisfactory for all methods. However, 
due to the inefficiencies of the PX-MH algorithm for updating $\boldsymbol{\Sigma}$ under the probit version of suBART, we require longer chains and careful specification of the proposal tuning parameter $\nu_{\text{prop}}$ for these experiments. In particular, higher values $\nu_{\text{prop}}$ increase the acceptance rate at the expense of inducing significant autocorrelation in the chains. See \citet{zhang2006sampling} for a discussion of the difficulties of optimising this trade-off. In order to maintain an acceptance rate in the range of 20-30\% as the dimension and sample size varied, we found $\nu_{\text{prop}}=n_{\text{train}}/10$ (when $d=2$), $\nu_{\text{prop}}=n_{\text{train}}/2$ (when $d=3$), $N_{\text{MCMC}}=10000$, and $N_{\text{burn-in}}=2000$ to strike an appropriate balance between these two concerns and the runtime of the experiments. We hence adopted these MCMC settings for all applications of the probit version of our model and, for fairness, all other competing methods in the presence of binary responses. 

\begin{figure}[H]
\centering
\begin{subfigure}[t]{0.49\textwidth}
    \centering
    \includegraphics[width=\linewidth]{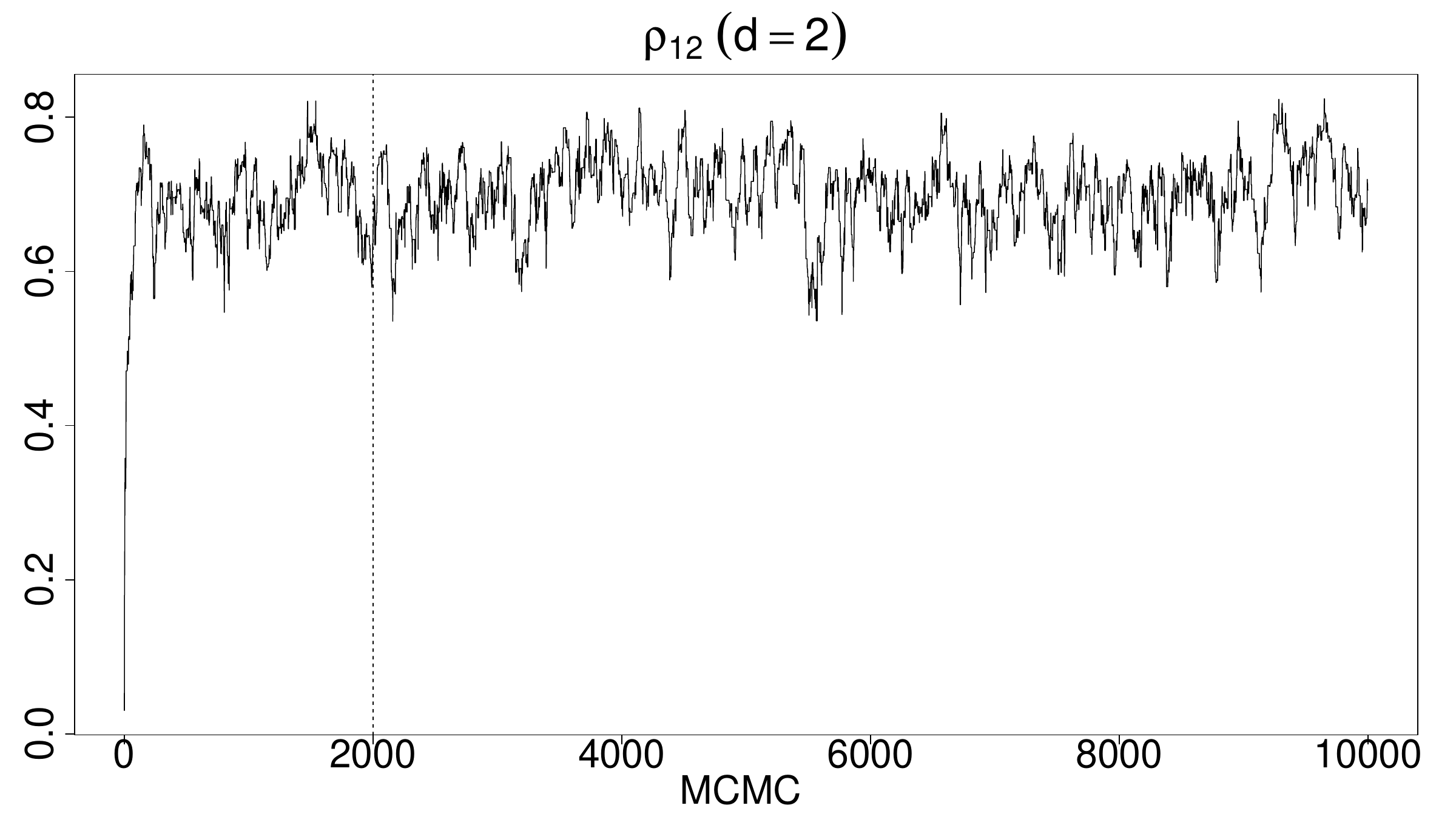}
\end{subfigure}%
\begin{subfigure}[t]{0.49\textwidth}
    \centering
    \includegraphics[width=\linewidth]{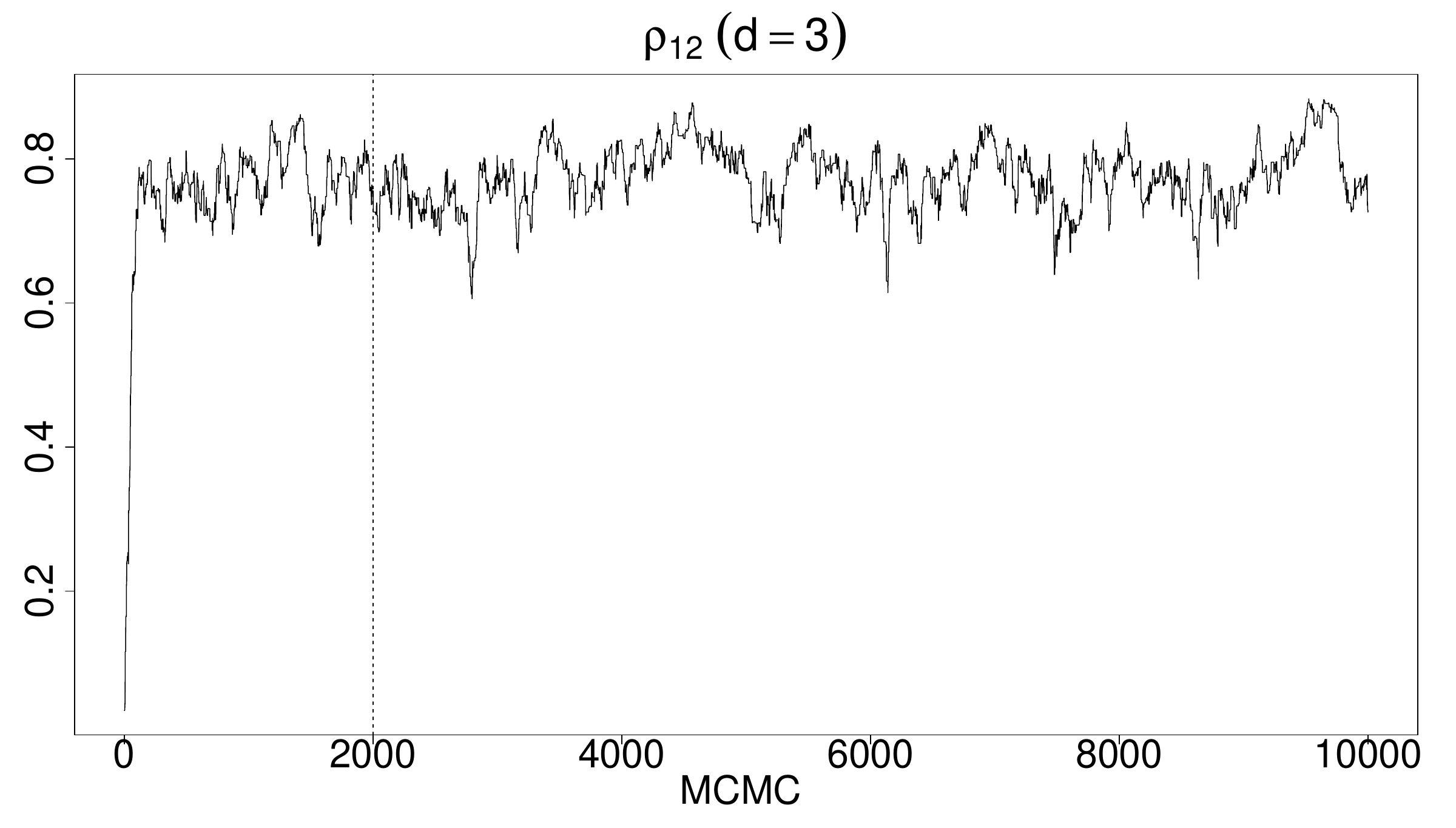}
\end{subfigure}
\caption{Traceplots of the posterior samples of $\rho_{12}$ considering the Friedman \#2 simulation scenario with $n_{\text{train}}=1000$ and $d=2$ (left) and $d=3$ (right). The dashed vertical line represents the number of iterations used as burn-in.}\label{fig:traceplot_d_23_rho_classification}
\end{figure}

\begin{figure}[H]
\centering
\begin{subfigure}[t]{0.49\textwidth}
    \centering
    \includegraphics[width=\linewidth]{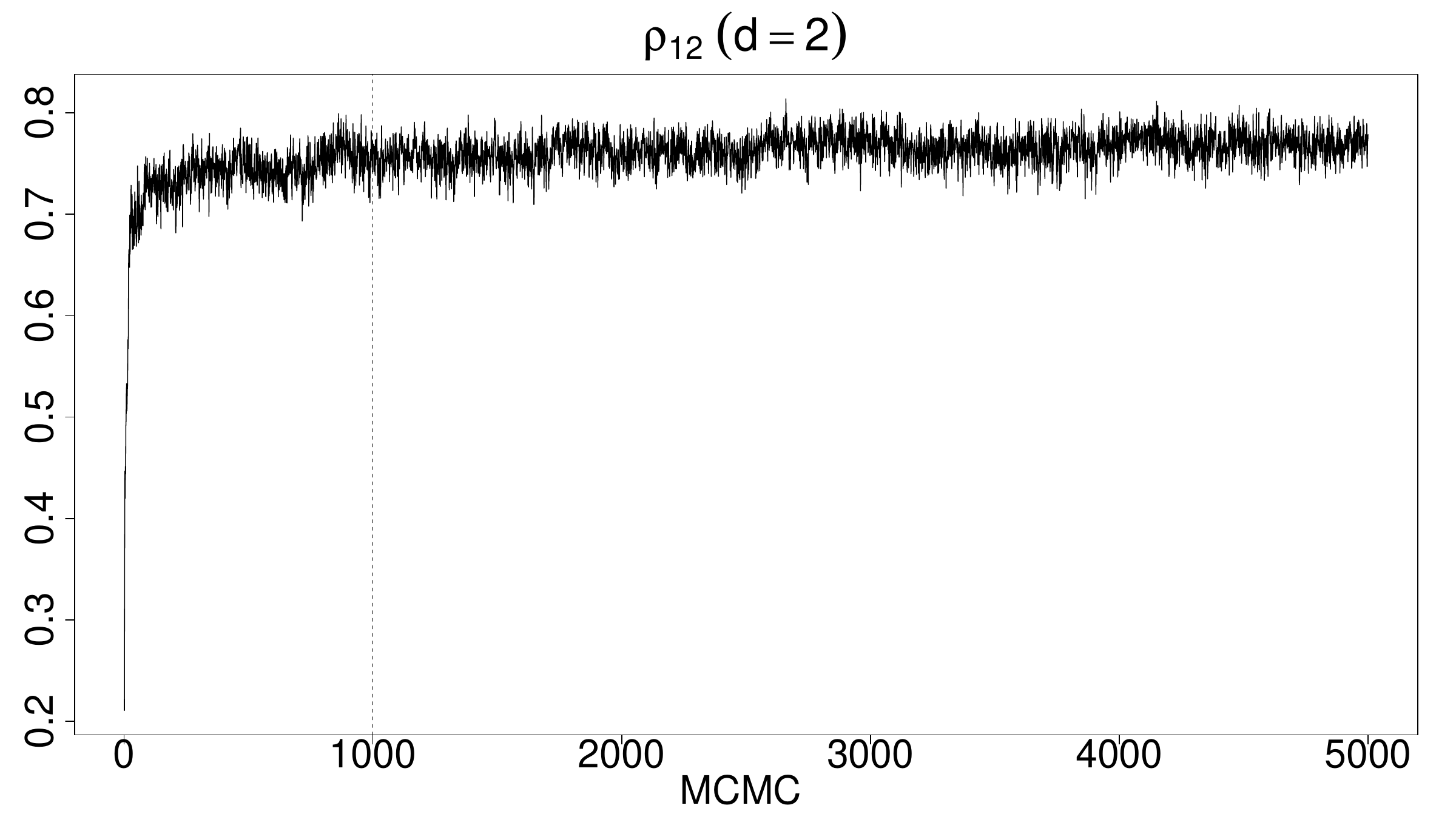}
\end{subfigure}%
\begin{subfigure}[t]{0.49\textwidth}
    \centering
    \includegraphics[width=\linewidth]{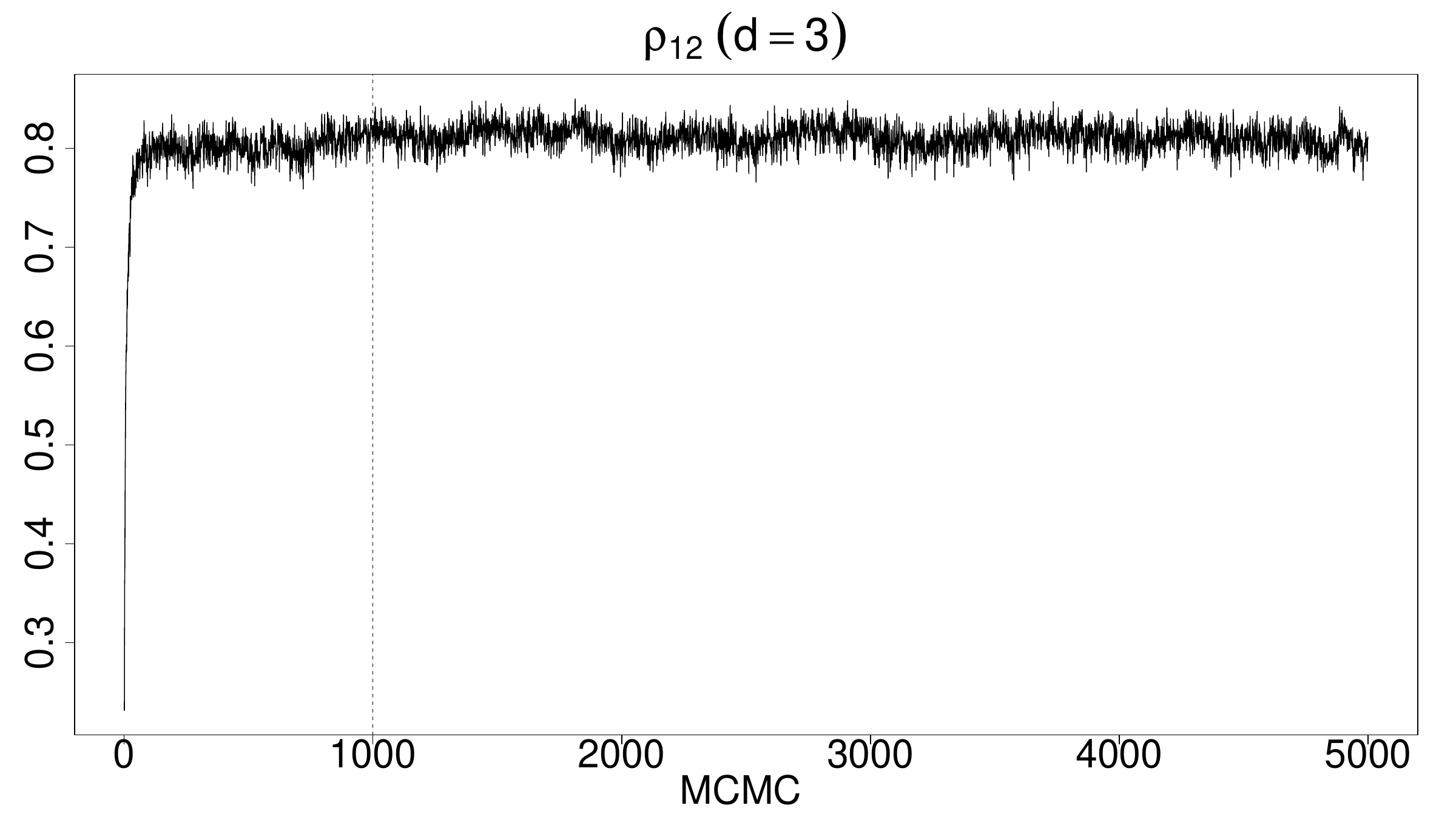}
\end{subfigure}
\caption{Traceplots of the posterior samples of $\rho_{12}$ considering the Friedman \#1 simulation scenario with $n_{\text{train}}=1000$ and $d=2$ (left) and $d=3$ (right). The dashed vertical line represents the number of iterations used as burn-in.}\label{fig:traceplot_d_23_rho_regression}
\end{figure}

In order to demonstrate the satisfactory convergence of the probit version of our model under these tuning parameter and MCMC settings, the MCMC samples for entries of $\boldsymbol{\Sigma}$ under one randomly chosen replicate for the Friedman \#2 experiment are shown in Figure \ref{fig:traceplot_d_23_rho_classification}. Specifically, these traceplots depict the posterior samples of $\rho_{12}$ under the $d=2$ (left) and $d=3$ (right) simulation settings, with $n_{\text{train}}=1000$. In each case, the samples concentrate around their respective true values of $0.75$ and $0.8$ after the burn-in period elapses. The chosen values of $\nu_{\text{prop}}=n_{\text{train}}/10$ and $\nu_{\text{prop}}=n_{\text{train}}/2$ for the $d=2$ and $d=3$ cases achieve acceptance rates of $0.28$ and $0.25$, respectively. Similar convergence behaviour and acceptance rates are achieved for the other entries of $\boldsymbol{\Sigma}$, across all evaluated sample sizes and replicates. For brevity, we omit the corresponding traceplots.
\vspace{-1em}%
For completeness, we also show the convergence behaviour of the continuous version of our model for the same $\rho_{12}$ under a randomly chosen replicate of the Friedman \#1 scenario, with $d=\{2,3\}$ and $n_{\text{train}}=1000$, in Figure \ref{fig:traceplot_d_23_rho_regression}. There are evidently no convergence issues in this scenario. Indeed, further examination of traceplots across all relevant simulation experiments and configurations of sample size and dimensionality suggest that our sampler is adequate in the case of continuous responses.

\subsection{Additional performance evaluation on simulation experiments}
\setcounter{figure}{0}
\setcounter{table}{0}
\label{app:friedman}

In order to evaluate the performance of suBART against its competitors, we conducted experiments outlined in Section \ref{sec:simstudies}, where we varied $n_{\text{train}} = n_{\text{test}} = \{250,500,1000\}$ and $d = \{2,3\}$. This appendix summarises the remaining results omitted from the main paper, for different configurations of sample size and dimensionality of the outcome. 
Recall that mvBART is excluded wherever $d=3$, due to the limitations of the \texttt{skewBART} package.
Overall, the findings illustrated in the boxplots below align with the conclusions drawn in Section \ref{sec:simstudies}. In particular, suBART exhibits reasonable performance metrics for both continuous (Friedman \#1) and binary (Friedman \#2) outcome scenarios, either matching or surpassing its tree-based model counterparts. Notably, suBART outperforms the linear BayesSUR across all scenarios, with the exception of the responses whose dependence on the predictors is linear. Additionally, we provide a summary of the estimated correlation parameters through tables displaying~the RMSE and 50\% CI coverage for all $\rho_{kj}$ where $j\neq k$. Across all scenarios, suBART consistently approaches the true values and provides credible intervals with proper coverage rates.
\begin{figure}[H]
    \centering
    \includegraphics[width=.85\textwidth]{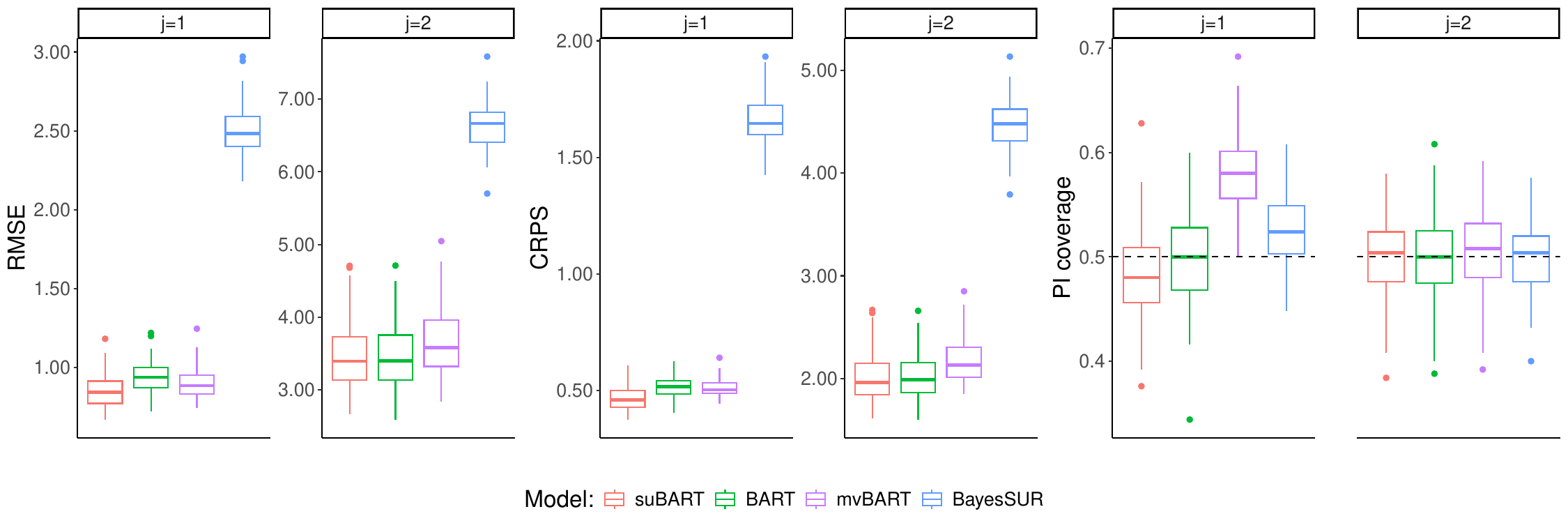}
    \caption{Simulation results for Friedman \#1 under the suBART, BART, mvBART, and BayesSUR models (from left to right) with $n_{\text{train}} = n_{\text{test}}=250 $ and $d = 2$.}
    \label{fig:sim_reg_friedman_one_250_2d}
\end{figure}\vspace{-2.25em}
\begin{figure}[H]
    \centering
    \includegraphics[width=.85\textwidth]{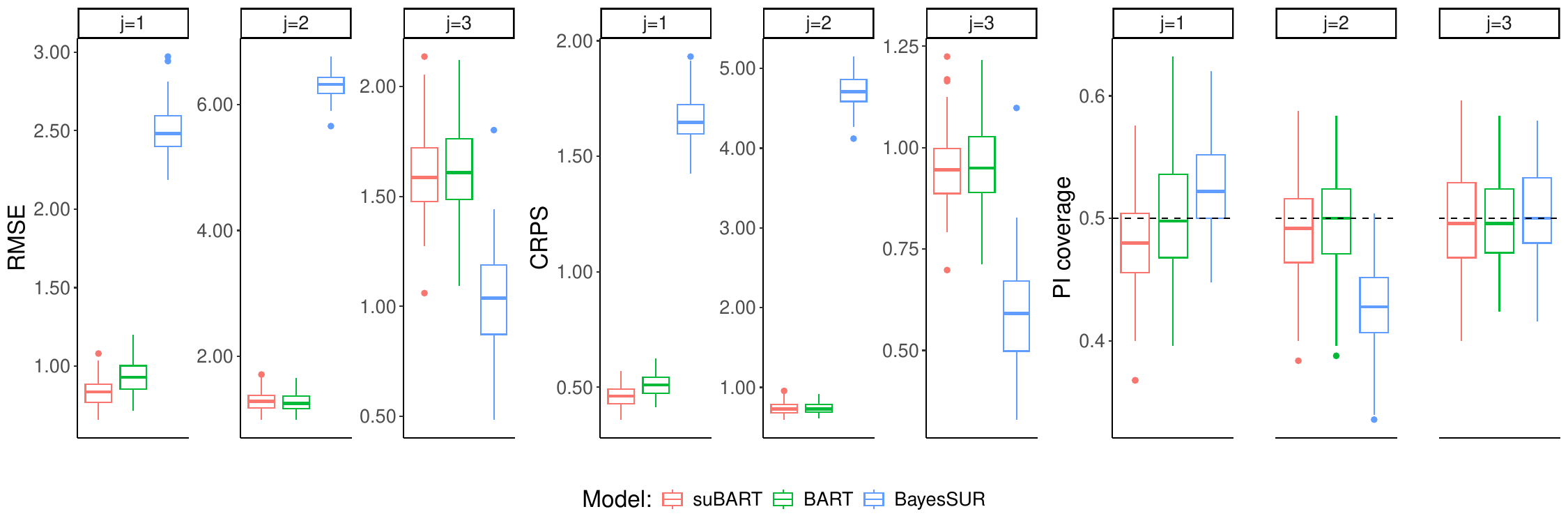}
    \caption{Simulation results for Friedman \#1 under the suBART, BART, and BayesSUR models (from left to right) with $n_{\text{train}} = n_{\text{test}}=250 $ and $d = 3$.}
    \label{fig:sim_reg_friedman_one_250_3d}
\end{figure}\vspace{-2.25em}
\begin{figure}[H]
    \centering
    \includegraphics[width=.85\textwidth]{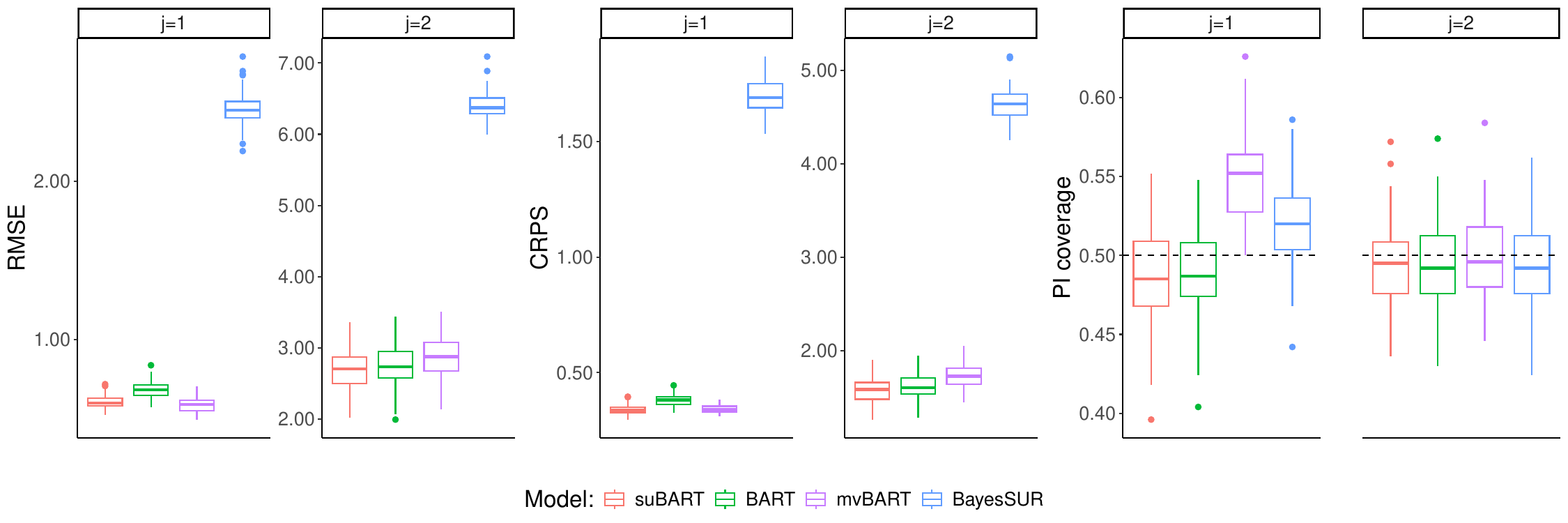}
    \caption{Simulation results for Friedman \#1 under the suBART, BART, mvBART, and BayesSUR models (from left to right) with $n_{\text{train}} = n_{\text{test}}=500 $ and $d = 2$.}
    \label{fig:sim_reg_friedman_one_500_2d}
\end{figure}\vspace{-2.25em}
\begin{figure}[H]
    \centering
    \includegraphics[width=.85\textwidth]{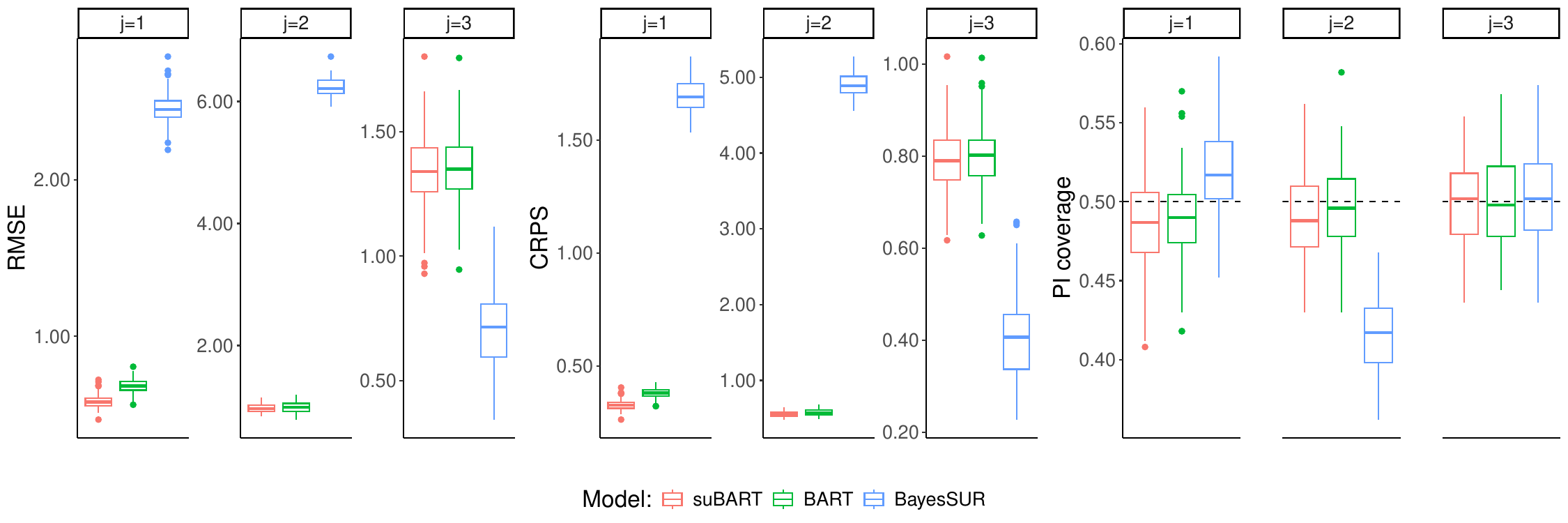}
    \caption{Simulation results for Friedman \#1 under the suBART, BART, and BayesSUR models (from left to right) with $n_{\text{train}} = n_{\text{test}}=500 $ and $d = 3$.}
    \label{fig:sim_reg_friedman_one_500_3d}
\end{figure}\vspace{-2.25em}
\begin{figure}[H]
    \centering
    \includegraphics[width=.85\textwidth]{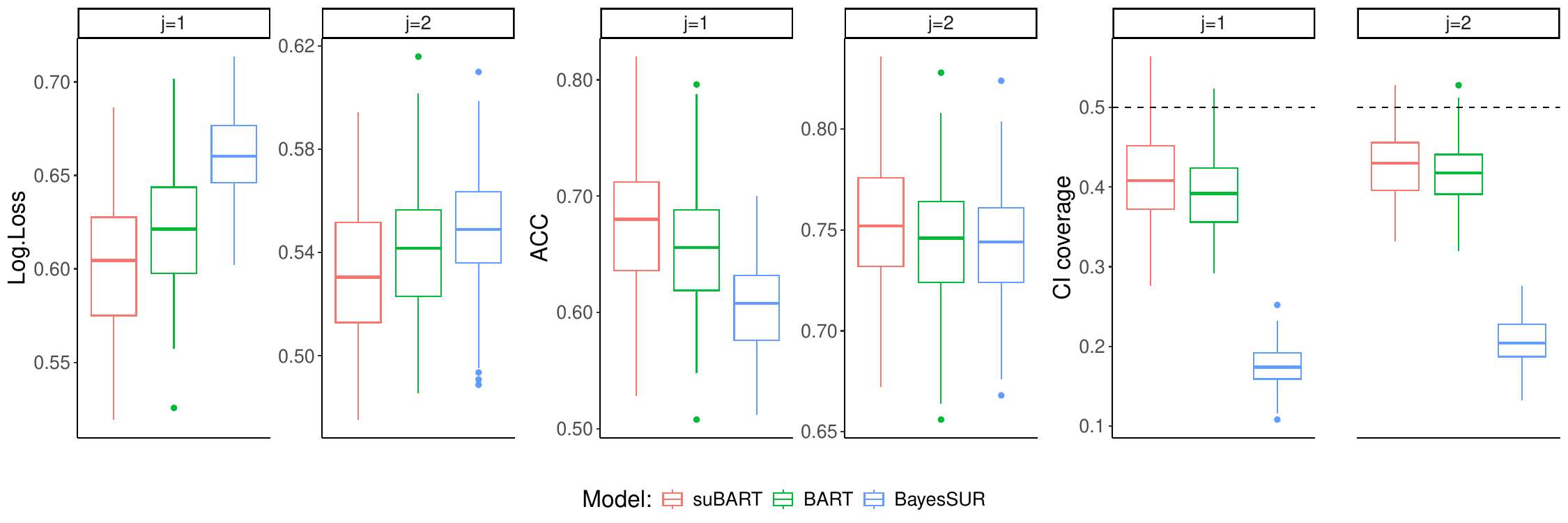}
    \caption{Simulation results for Friedman \#2 under the suBART, BART, and BayesSUR models (from left to right) with $n_{\text{train}} = n_{\text{test}}=250 $ and $d = 2$.}
    \label{fig:sim_class_friedman_one_250_2d}
\end{figure}\vspace{-2.25em}
\begin{figure}[H]
    \centering
    \includegraphics[width=.85\textwidth]{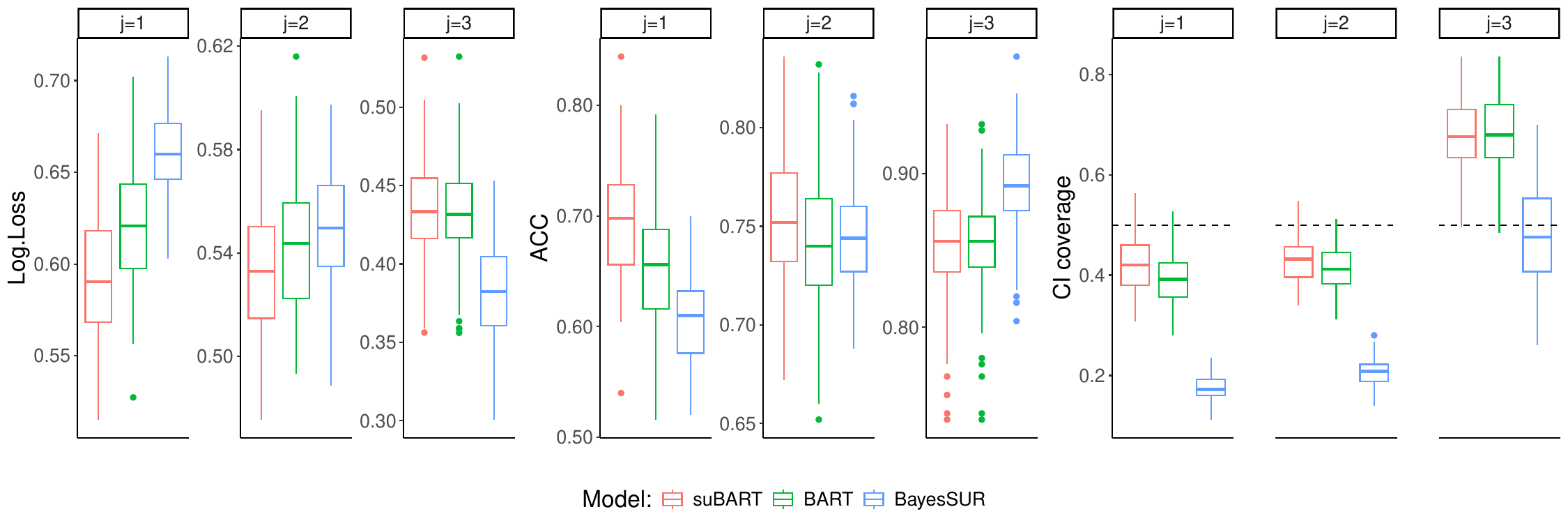}
    \caption{Simulation results for Friedman \#2 under the suBART, BART, and BayesSUR models (from left to right) with $n_{\text{train}} = n_{\text{test}}=250 $ and $d = 3$.}
    \label{fig:sim_class_friedman_one_250_3d}
\end{figure}\vspace{-2.25em}
\begin{figure}[H]
    \centering
    \includegraphics[width=.85\textwidth]{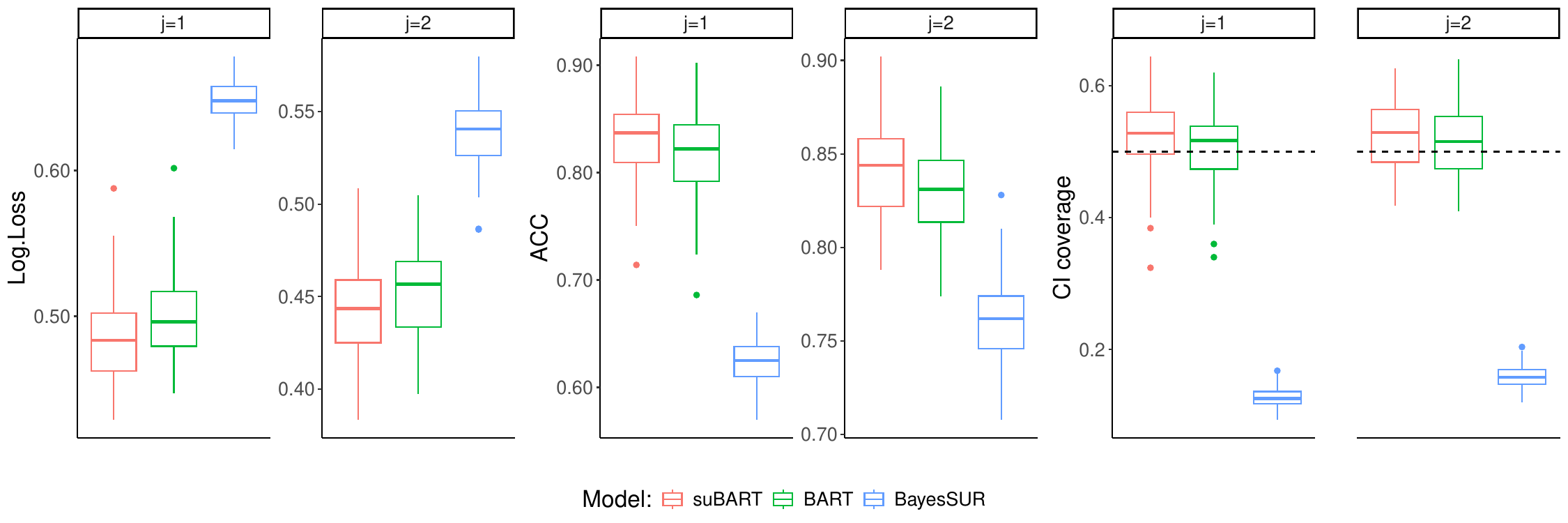}
    \caption{Simulation results for Friedman \#2 under the suBART, BART, and BayesSUR models (from left to right) with $n_{\text{train}} = n_{\text{test}}=500 $ and $d = 2$.}
    \label{fig:sim_class_friedman_one_500_2d}
\end{figure}\vspace{-2.25em}
\begin{figure}[H]
    \centering
    \includegraphics[width=.85\textwidth]{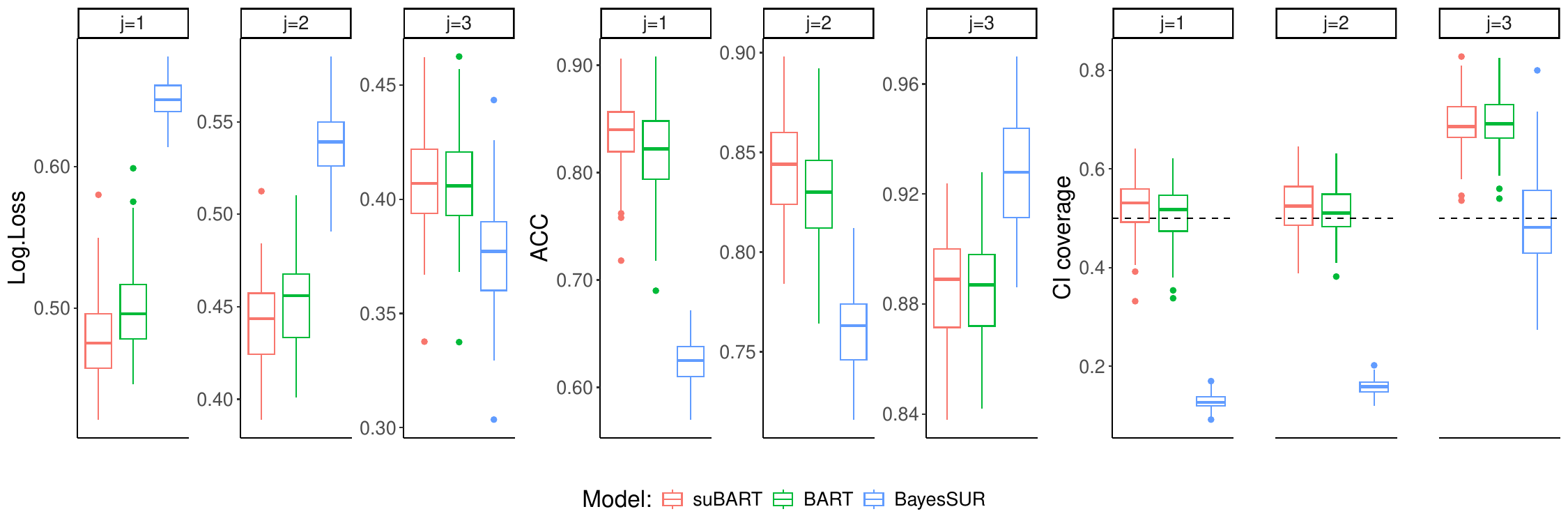}
    \caption{Simulation results for Friedman \#2 under the suBART, BART, and BayesSUR models (from left to right) with $n_{\text{train}} = n_{\text{test}}=500 $ and $d = 3$.}
    \label{fig:sim_class_friedman_one_500_3d}
\end{figure}\vspace{-2.25em}
\begin{figure}[H]
    \centering
    \includegraphics[width=.85\textwidth]{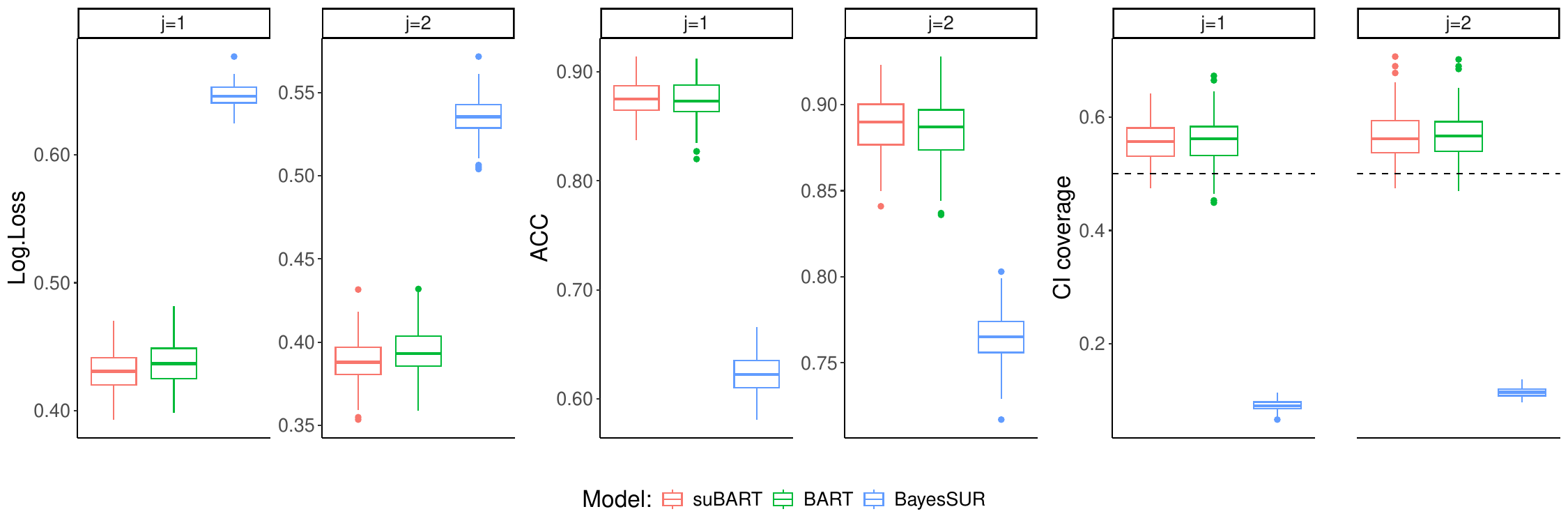}
    \caption{Simulation results for Friedman \#2 under the suBART, BART, and BayesSUR models (from left to right) with $n_{\text{train}} = n_{\text{test}}=1000 $ and $d = 2$.}
    \label{fig:sim_class_friedman_one_1000_2d}
\end{figure}\vspace{-2.25em}
\begin{table}[H]
\centering
\caption{RMSE and coverage of a 50\% CI for $\sigma_j$ and ${\rho}_{jk}$  for Friedman \#1 with $n_{\text{train}} = n_{\text{test}} =250$.}
\label{tab:sim_corr_fried_one_combined_250}
\begin{tabular}{lcccccccc}
\hline
\small
& \multicolumn{4}{c}{\textbf{RMSE}} & \multicolumn{4}{c}{\textbf{CI coverage}} \\
 & \textbf{suBART} & \textbf{BART} & \textbf{mvBART} & \textbf{BayesSUR} & \textbf{suBART} & \textbf{BART} & \textbf{mvBART} & \textbf{BayesSUR} \\
\hline
$d=2$ & & & & & & & & \\
\hline
{$\sigma_1$} & $0.08$ & $0.09$ & $0.08$ & $1.62$ & $0.28$ & $0.38$ & $0.46$ & $0.00$ \\
{$\sigma_2$} & $0.58$ & $0.63$ & $0.82$ & $1.87$ & $0.30$ & $0.26$ & $0.15$ & $0.00$ \\
{$\rho_{12}$} & $0.06$ & --- & $0.04$ & $0.31$ & $0.28$ & --- & $0.59$ & $0.00$ \\ 
\hline
$d=3$ & & & & & & & & \\
\hline
{$\sigma_1$} & $0.06$ & $0.08$ & --- & $1.62$ & $0.46$ & $0.41$ & --- & $0.00$ \\
{$\sigma_2$} & $0.26$ & $0.32$ & --- & $4.18$ & $0.10$ & $0.03$ & --- & $0.00$ \\
{$\sigma_3$} & $0.35$ & $0.35$ & --- & $0.20$ & $0.27$ & $0.25$ & --- & $0.47$ \\
{$\rho_{12}$} & $0.06$ & --- & --- & $0.35$ & $0.16$ & --- & --- & $0.00$ \\ 
{$\rho_{13}$} & $0.06$ & --- & --- & $0.31$ & $0.40$ & --- & --- & $0.00$ \\
{$\rho_{23}$} & $0.06$ & --- & --- & $0.16$ & $0.47$ & --- & --- & $0.03$ \\ 
\end{tabular}
\end{table}\vspace{-2em}
\begin{table}[H]
\centering
\caption{RMSE and coverage of a 50\% CI for $\sigma_j$ and ${\rho}_{jk}$  for Friedman \#1 with $n_{\text{train}} = n_{\text{test}} =500$.}
\label{tab:sim_corr_fried_one_combined_500}
\begin{tabular}{lcccccccc}
\hline
\small
& \multicolumn{4}{c}{\textbf{RMSE}} & \multicolumn{4}{c}{\textbf{CI coverage}} \\
 & \textbf{suBART} & \textbf{BART} & \textbf{mvBART} & \textbf{BayesSUR} & \textbf{suBART} & \textbf{BART} & \textbf{mvBART} & \textbf{BayesSUR} \\
\hline
$d=2$ & & & & & & & & \\
\hline
{$\sigma_1$} & $0.05$ & $0.07$ & $0.06$ & $1.61$ & $0.35$ & $0.09$ & $0.20$ & $0.00$ \\
{$\sigma_2$} & $0.44$ & $0.48$ & $0.56$ & $1.75$ & $0.30$ & $0.28$ & $0.18$ & $0.00$ \\
{$\rho_{12}$} & $0.03$ & --- & $0.02$ & $0.32$ & $0.38$ & --- & $0.58$ & $0.00$ \\ 
\hline
$d=3$ & & & & & & & & \\
\hline
{$\sigma_1$} & $0.04$ & $0.07$ & --- & $1.61$ & $0.41$ & $0.12$ & --- & $0.00$ \\
{$\sigma_2$} & $0.14$ & $0.20$ & --- & $4.15$ & $0.18$ & $0.08$ & --- & $0.00$ \\
{$\sigma_3$} & $0.25$ & $0.26$ & --- & $0.12$ & $0.17$ & $0.17$ & --- & $0.58$ \\
{$\rho_{12}$} & $0.03$ & --- & --- & $0.35$ & $0.30$ & --- & --- & $0.00$ \\ 
{$\rho_{13}$} & $0.03$ & --- & --- & $0.31$ & $0.50$ & --- & --- & $0.00$ \\
{$\rho_{23}$} & $0.04$ & --- & --- & $0.16$ & $0.58$ & --- & --- & $0.00$ \\ 
\end{tabular}
\end{table}
\vspace{-2em}
\begin{table}[H]
\centering
\caption{RMSE and coverage of a 50\% CI for ${\rho}_{jk}$ for Friedman \#2 with $n_{\text{train}} = n_{\text{test}} = 250$.}
\label{tab:sim_corr_class_fried_one_combined_250}
\begin{tabular}{lcccccc}
\hline
\small
 & \multicolumn{3}{c}{\textbf{RMSE}} & \multicolumn{3}{c}{\textbf{CI coverage}} \\
 & \textbf{suBART} & \textbf{BayesSUR} & \textbf{suBART} & \textbf{BayesSUR} \\
\hline
$d=2$ & & & & \\
\hline
{$\rho_{12}$} & $0.12$ & $0.57$ & $0.29$ & $0.00$ \\ 
\hline
$d=3$ & & & & \\
\hline
{$\rho_{12}$} & $0.13$ & $0.61$ & $0.19$ & $0.00$ \\
{$\rho_{13}$} & $0.09$ & $0.41$ & $0.46$ & $0.00$ \\
{$\rho_{23}$} & $0.09$ & $0.38$ & $0.52$ & $0.02$ \\
\hline
\end{tabular}
\end{table}\vspace{-2em}
\begin{table}[H]
\centering
\caption{RMSE and coverage of a 50\% CI for ${\rho}_{jk}$ for Friedman \#2 with $n_{\text{train}} = n_{\text{test}} = 500$.}
\label{tab:sim_corr_class_fried_one_combined_500}
\begin{tabular}{lcccccc}
\hline
\small
 & \multicolumn{3}{c}{\textbf{RMSE}} & \multicolumn{3}{c}{\textbf{CI coverage}} \\
 & \textbf{suBART} & \textbf{BayesSUR} & \textbf{suBART} & \textbf{BayesSUR} \\
\hline
$d=2$ & & & & \\
\hline
{$\rho_{12}$} & $0.06$ & $0.27$ & $0.48$ & $0.00$ \\ 
\hline
$d=3$ & & & & \\
\hline
{$\rho_{12}$} & $0.05$ & $0.28$ & $0.46$ & $0.00$ \\
{$\rho_{13}$} & $0.06$ & $0.11$ & $0.51$ & $0.12$ \\
{$\rho_{23}$} & $0.08$ & $0.08$ & $0.49$ & $0.41$ \\
\hline
\end{tabular}
\end{table}

\setcounter{figure}{0}
\setcounter{table}{0}
\subsection{Additional TTCM application results}\label{app:extraTTCM}

In Appendices \ref{app:CATE} and \ref{app:var_imp}, we present some additional insights gleaned from the TTCM application by our flagship ps-suBART model. In the former, we graphically explore the heterogeneity in the estimated causal effects, showing how they vary for different values of the covariates and propensity scores. In the latter, we investigate variable importance scores using the ideas of \citet{chipman2010bart}. However, it is important to emphasise that we do so with reference to each individual outcome, given that the covariates (or propensity scores) may be strongly related to one outcome but not necessarily to another. This is not possible under mvBART, which enforces a common tree structure and does not permit variation in splitting rules across each response.

\subsubsection{Conditional treatment effects and heterogeneity}
\label{app:CATE}
In our analyses of the TTCM data, we focussed on the MATE, which measures the expected treatment effect averaged over the observed sample. We did this because the interest in cost-effectiveness analysis lies primarily in average effects, not individual ones. Nonetheless, we may sometimes also want to analyse individual treatment effects, and in particular how they depend on the patients' baseline characteristics. We thus present here some ideas on how to do this, using the TTCM data. For brevity, we restrict attention to the results obtained from the flagship ps-suBART model.

We first recall the following definition, which was made in Section \ref{sec:TTCM_sim}:
$$
    \tau_c(\mathbf{x}_i) = \mathbb{E}\left\lbrack c_i\given t_i = 1, \mathbf{x}_i\right\rbrack - \mathbb{E}\left\lbrack c_i\given t_i = 0, \mathbf{x}_i\right\rbrack.
$$
This quantity is called the \emph{conditional average treatment effect} (CATE), because it depends on $\mathbf{x}_i$, unlike the MATE $\Delta_c$ defined in Equation \eqref{eq:li_MATE}. We also define $\tau_q(\mathbf{x}_i)$ in the same way.
In general, $\tau_c(\mathbf{x}_i)$ is different for different $i$, unless the dependence of $\mathbb{E}\left\lbrack c_i\given t_i, \mathbf{x}_i\right\rbrack$ on $t_i$ and $\mathbf{x}_i$ is linear. See \citet{li2023bayesian} for details. We now further define the \emph{conditional incremental net benefit} (CINB)\footnote{There is no standard terminology for this estimand in the literature. This acronym is our own invention.} by 
\begin{equation}\label{eq:CNB}
    \text{CINB}_{\lambda}(\mathbf{x}_i) = \lambda \tau_q(\mathbf{x}_i) - \tau_c(\mathbf{x}_i).
\end{equation}
Again, the CINB is the counterpart of the INB defined earlier, where the former depends on $\mathbf{x}_i$ and the latter does not.

\begin{figure}[H]
    \centering
    \includegraphics[width=1\textwidth]{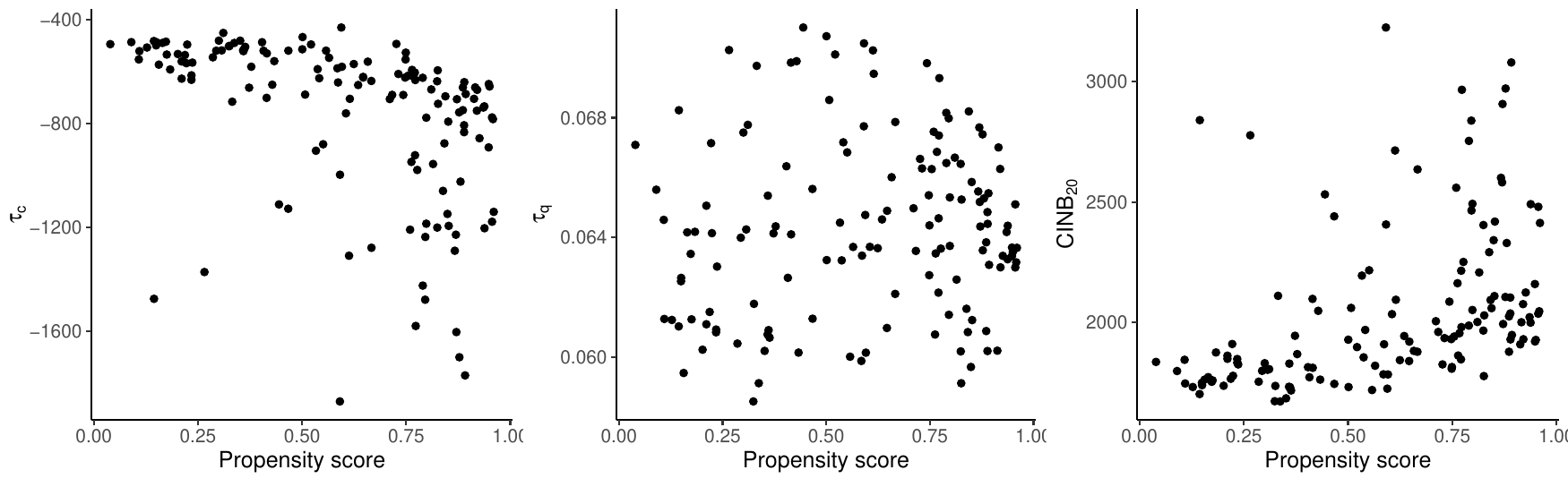}
    \caption{Plots of the estimated propensity scores supplied as a predictor to ps-suBART against the estimated CATE for costs (left), quality (middle), and conditional incremental net benefit (right).}
    \label{fig:CATEplot_ps}
\end{figure}

In Figure \ref{fig:CATEplot_ps} we plot the estimated propensity scores against the posterior means of $\tau_c(\mathbf{x}_i)$, $\tau_q(\mathbf{x}_i)$, and $\text{CINB}_{20}(\mathbf{x}_i)$, respectively. The plots indicate some considerable heterogeneity. In particular, we see that, while the effect on costs is negative for all patients, there seems to be a further negative relationship with the propensity scores: patients who are more likely to receive the treatment also benefit more from it. For the effect on $q_i$ there is no clear relationship with the propensity scores. Finally, the relationship between the CINB and the propensity scores is positive, which makes sense in light of the definition in Equation \eqref{eq:CNB}.

\begin{figure}[H]
    \centering
    \includegraphics[width=1\textwidth]{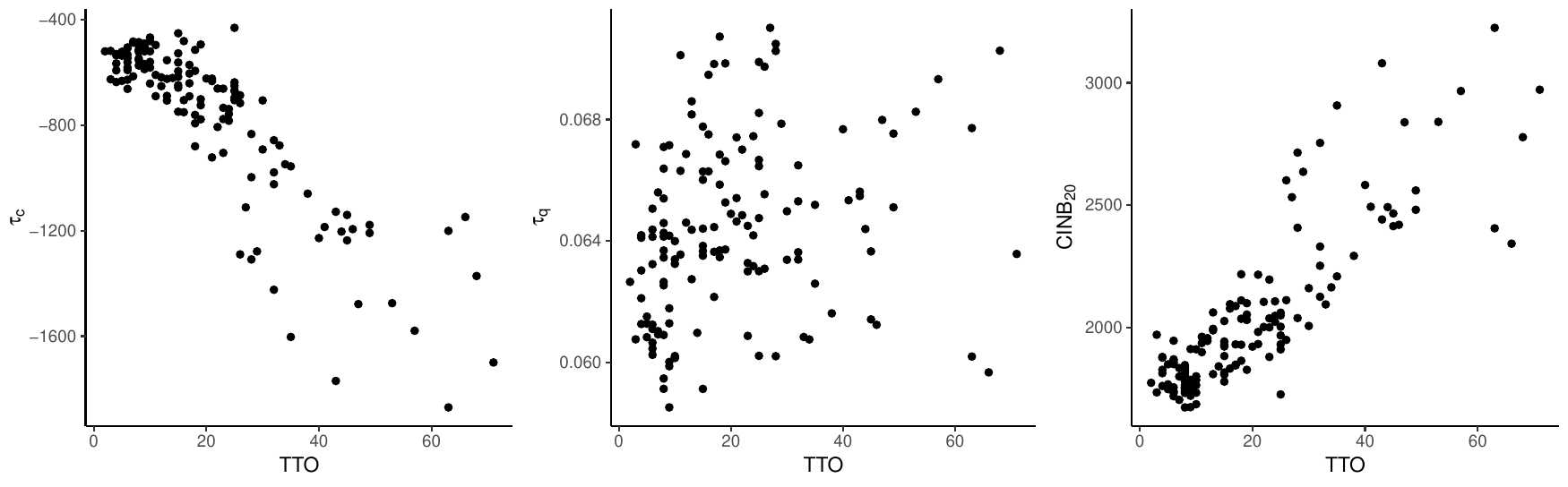}
    \caption{Plots of the TTO predictor against the estimated CATE for costs (left), quality (middle), and conditional incremental net benefit (right).}
    \label{fig:CATEplot_TTO}
\end{figure}

\begin{figure}[H]
    \centering
    \includegraphics[width=1\textwidth]{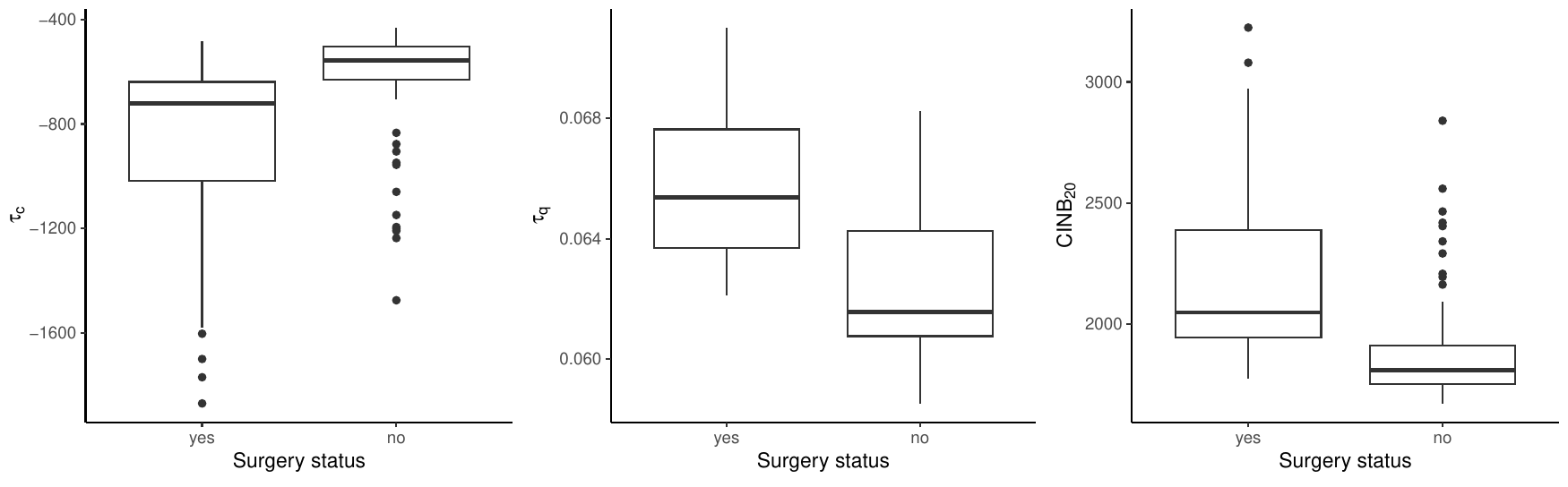}
    \caption{Box plots of the surgery predictor against the estimated CATE for costs (left), quality (middle), and conditional incremental net benefit (right).}
    \label{fig:CATEplot_surgery}
\end{figure}

In addition to the propensity scores, such plots can also be produced for any individual covariate. We show similar plots against the continuous `TTO' variable and the categorical `surgery' variable from the TTCM dataset in Figure \ref{fig:CATEplot_TTO} and Figure \ref{fig:CATEplot_surgery}, where strong relationships and similar levels of heterogeneity are evident. Interestingly, Figure \ref{fig:CATEplot_surgery} shows that, for patients who receive surgery, the treatment is both more effective and more cost-saving.
\vspace{-2em}%
\vspace{-2em}%

Finally, it is also worth noting that the observed heterogeneity in the above plots of $\tau_c(\mathbf{x}_i)$ against various predictors is indicative of non-linearity in $\mathbb{E}\left\lbrack c_i\given t_i, \mathbf{x}_i\right\rbrack$. The same applies to $\tau_q(\mathbf{x}_i)$ and $\text{CINB}_{20}(\mathbf{x}_i)$. Thus, there is evidence to support the flexibility of ps-suBART in capturing non-linear relationships.

\subsubsection{Visualising variable importance}
\label{app:var_imp}

\begin{figure}[H]
    \centering
    \includegraphics[width=\textwidth]{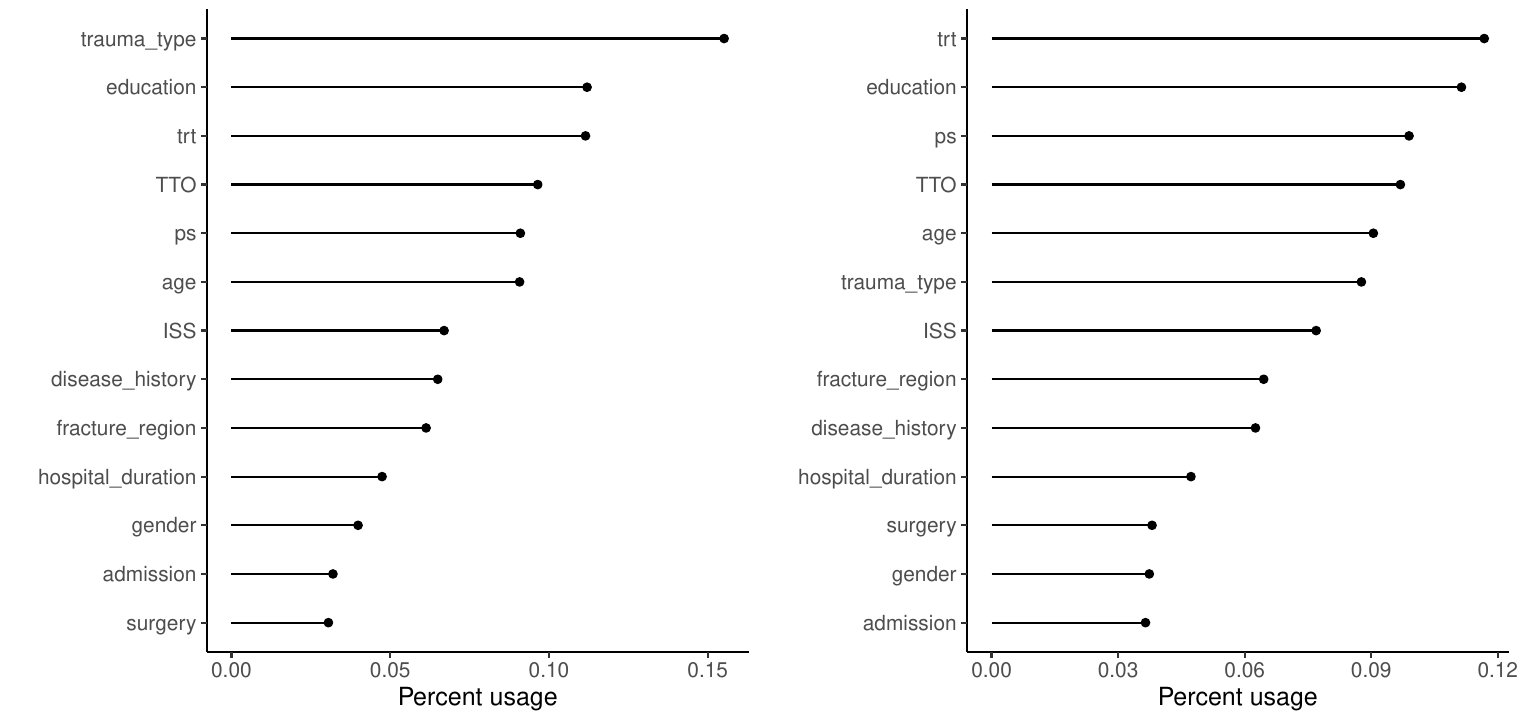}
    \caption{Average usage of each variable in the TTCM dataset in the 100 trees of the ps-suBART model. The left side shows usage for the outcome $c_i$, the right for $q_i$.}
    \label{fig:varimp_plot}
\end{figure}

Figure \ref{fig:varimp_plot} shows the percentage of times, averaged across all trees in the outcome-specific ensemble, that a given predictor was used to form a splitting rule for a given outcome in the TTCM application. Similar plots were shown to demonstrate the quantification of variable importance in the original univariate BART setting by \citet{chipman2010bart}. Here, the variable importance scores in Figure \ref{fig:varimp_plot} illustrate the practical implications of a defining feature of suBART, namely that the relative importance of the predictors differs in relation to each outcome. Crucially, this behaviour is achieved without having to pre-specify the functional form of the relationships between the cost and quality outcomes and the available predictors, as a result of BART's intrinsic variable-selection properties. Some interesting findings are worth briefly commenting on. Firstly, we have variables which are important to both outcomes, such as the treatment indicator. Secondly, we have variables which are relatively unimportant for both outcomes, with the hospital admission indicator being the second-least and least used variable, respectively. Thirdly, we have variables which differ in the extent of their importance for each outcome: trauma type is the most used variable for the $c_i$ outcome, while being number six in the ranking for $q_i$. Finally, we note that tailoring the suBART framework to causal inference objectives by supplying propensity scores as a predictor has indeed led to the propensity scores being frequently used to form splitting rules for both outcomes.

\end{document}